\documentclass[prd,reprint,superscriptaddress,altaffillsymbol,amsmath,nofootinbib,longbibliography]{revtex4-1}
\pdfoutput=1

\usepackage{ifthen}
\usepackage{epsfig}
\usepackage{booktabs} 
\usepackage{natbib}
\usepackage{wasysym}
\usepackage{slashed} 
\usepackage{bbm}
\usepackage{mathrsfs}
\usepackage{amssymb}
\usepackage{dsfont}
\usepackage[export]{adjustbox}

\usepackage{color}

\definecolor{OurBlue}{rgb}{0.384,0.616,0.784}
\definecolor{OurRed}{rgb}{0.878,0.388,0.212}

\usepackage[
colorlinks=true, 
linkcolor=OurBlue,
citecolor=OurRed,
urlcolor=OurBlue,
]{hyperref}

\allowdisplaybreaks

\newcommand{\erf}{\mathop{\mathrm{erf}}}
\newcommand{\beqra}{\begin{eqnarray}}
\newcommand{\eeqra}{\end{eqnarray}}
\newcommand{\beq}{\begin{equation}}
\newcommand{\eeq}{\end{equation}}
\newcommand{\dd}{\mathrm{d}}

\renewcommand{\epsilon}{\varepsilon}

\renewcommand{\vec}[1]{\mathbf{#1}}
\renewcommand{\bar}{\overline}

\newcommand{\vPerpEl}{\mathbf{v}_{\rm el}^\perp}

\begin{document}

\title{\boldmath Crystal responses to general dark matter-electron interactions}

\author{Riccardo Catena}
\email{catena@chalmers.se}
\affiliation{Chalmers University of Technology, Department of Physics, SE-412 96 G\"oteborg, Sweden}

\author{Timon Emken}
\email{timon.emken@fysik.su.se}
\affiliation{The Oskar Klein Centre, Department of Physics, Stockholm University, AlbaNova, SE-10691 Stockholm, Sweden}
\affiliation{Chalmers University of Technology, Department of Physics, SE-412 96 G\"oteborg, Sweden}

\author{Marek Matas}
\email{marek.matas@mat.ethz.ch}
\affiliation{Department of Materials, ETH Z\"urich, CH-8093 Z\"urich, Switzerland}

\author{Nicola A. Spaldin}
\email{nicola.spaldin@mat.ethz.ch}
\affiliation{Department of Materials, ETH Z\"urich, CH-8093 Z\"urich, Switzerland}

\author{Einar Urdshals}
\email{urdshals@chalmers.se}
\affiliation{Chalmers University of Technology, Department of Physics, SE-412 96 G\"oteborg, Sweden}

\begin{abstract}

We develop a formalism to describe the scattering of dark matter (DM) particles by electrons bound in crystals for a general form of the underlying DM-electron interaction.  Such a description is relevant for direct-detection experiments of DM particles lighter than a nucleon, which might be observed in operating DM experiments via electron excitations in semiconductor crystal detectors.  Our formalism is based on an effective theory approach to general non-relativistic DM-electron interactions, including the anapole, and magnetic and electric dipole couplings, combined with crystal response functions defined in terms of electron wave function overlap integrals. Our main finding is that, for the usual simplification of the velocity integral, the rate of DM-induced electronic transitions in a semiconductor material depends on at most five independent crystal response functions, four of which were not known previously. We identify these crystal responses, and evaluate them using density functional theory for crystalline silicon and germanium, which are used in operating DM direct detection experiments. Our calculations allow us to set 90\% confidence level limits on the strength of DM-electron interactions from data reported by the SENSEI and EDELWEISS experiments. The novel crystal response functions discovered in this work encode properties of crystalline solids that do not interact with conventional experimental probes, suggesting the use of the DM wind as a probe to reveal new kinds of hidden order in materials.

\end{abstract}

\maketitle

\section{Introduction}
An invisible and unidentified mass component, dark matter (DM), is the leading form of matter in galaxies, galaxy clusters and the large scale structures we observe in the Cosmos~\cite{Bertone:2016nfn}. It is responsible for the bending of light emitted by distant luminous sources and provides the seeds for the gravitational collapse modern cosmology predicts to be at the origin of all objects we see in the Universe~\cite{Peebles2017Mar}. And yet, the nature of this essential, but elusive cosmological component remains unidentified. In the leading paradigm of astroparticle physics, DM is made of hypothetical, as yet undetected particles that the Standard Model of particle physics cannot account for~\cite{Profumo2017Feb}. While different methods have been proposed in the past decades to unveil the identity of DM, only a direct detection of the hypothetical particles forming the Milky Way DM component will likely prove the microscopic nature of DM conclusively~\cite{Bertone:2004pz}. 

DM direct detection experiments play a central role in this context~\cite{Drukier:1983gj,goodman:1984dc}. Typically, they operate ultra-low background detectors located deep underground to search for signals of interactions between DM particles from our galaxy and nuclei forming the detector material~\cite{Undagoitia_2015}. So far, this experimental technique has mostly focused on the search for nuclear recoils induced by Weakly Interacting Massive Particles (WIMPs) -- a class of DM candidates with interactions at about the scale of weak interactions and mass ranging from a few GeV to a few hundreds of TeV~\cite{Arcadi:2017kky}. The upper bound arises from the requirement that the WIMP annihilation cross section is unitary~\cite{Griest1990}, whereas the lower bound guarantees that the predicted density of WIMPs in the present Universe matches that measured via cosmic microwave background (CMB) observations~\cite{Lee1977Jul}.
The possibility of explaining the present DM cosmological density in terms of particle masses and coupling constants is one of the defining features of the WIMP paradigm and is based upon the so-called WIMP chemical decoupling mechanisms, which occurs when the rate of WIMP annihilations in the early Universe equals the rate of expansion of the Universe~\cite{Steigman2012Jul}.

In spite of about four decades of searches at DM direct detection experiments, WIMPs have so far escaped detection~\cite{Schumann2019Aug}. While WIMPs remain a central element of modern cosmology, the fact that they have not been detected has recently motivated a critical reconsideration of the assumptions underlying the WIMP paradigm, and in particular the DM direct detection technique~\cite{Essig:2011nj}. One important example is the restricted mass range within which the search for DM particles has so far been performed. Indeed, DM particles of mass smaller than about 1 GeV would be too light to induce an observable nuclear recoil, and this might explain why experiments searching for WIMPs via nuclear recoils have so far not been able to report an unambiguous discovery. On the other hand, a DM particle of mass in the MeV - GeV range might have enough kinetic energy to induce an observable electronic transition in a target material -- a possibility which has recently attracted a great deal of attention~\cite{Battaglieri:2017aum}.

From the experimental side, a number of different approaches has been pushed forward to search for sub-GeV DM particles. These involve the use of dual-phase argon~\cite{Agnes:2018oej} and xenon~\cite{Essig:2012yx,Essig:2017kqs,Aprile:2019xxb} targets, silicon and germanium semiconductors~\cite{Essig:2011nj,Graham:2012su,Lee:2015qva,Essig:2015cda,Crisler:2018gci,Agnese:2018col,Abramoff:2019dfb,Aguilar-Arevalo:2019wdi,Amaral:2020ryn,Andersson:2020uwc}, sodium iodide crystals~\cite{Roberts:2016xfw}, graphene~\cite{Hochberg:2016ntt,Geilhufe:2018gry}, 3D Dirac materials~\cite{Hochberg:2017wce,Geilhufe:2019ndy}, polar crystals~\cite{Knapen:2017ekk}, scintillators~\cite{Derenzo:2016fse,Blanco:2019lrf} and superconductors~\cite{Hochberg:2015pha,Hochberg:2015fth,Hochberg:2019cyy}.

From the theoretical side, the search for sub-GeV DM particles involves the modelling of DM-electron interactions in detector materials, e.g.~\cite{Kopp:2009et,Essig:2011nj,Essig:2015cda}. 
One generic feature of models for DM-electron interactions is that they involve a new mediator particle, in addition to the DM candidate~~\cite{Battaglieri:2017aum}. A new mediator is needed to reconcile the chemical decoupling mechanism with current observations on the present DM density. In this context, the most extensively investigated model extends the Standard Model by an additional $U(1)$ gauge group, under which only the DM particle is charged~\cite{Holdom:1985ag}. The associated gauge boson is referred to as the ``dark photon''. Interactions between the DM particle and the electrically charged particles in the Standard Model arise from a ``kinetic mixing'' between ordinary and dark photons. While this framework allows to interpret the results of present and future DM direct detection experiments, it is also rather restrictive, as it a priori excludes the possibility that the amplitude for DM-electron scattering depends on momenta different from the transferred momentum~\cite{Catena:2019gfa}. Scenarios in which the amplitude for DM-electron scattering depends on the initial electron momentum, in addition to the transferred momentum, include models where the DM-electron interaction is generated by an anapole moment or a magnetic and electric dipole~\cite{Kavanagh:2018xeh,Catena:2020tbv}. 

In this work, we extend the formalism that we developed in Ref.~\cite{Catena:2019gfa} for the scattering of DM particles by electrons without making any restrictive assumption on the form of the underlying DM-electron interaction, to the case of electrons bound in crystals used in operating DM direct detection experiments. Our formalism is based on an effective theory approach to non-relativistic DM-electron interactions, and a set of crystal response functions defined in terms of electron wave function overlap integrals. Effective theory methods have previously been used in the scattering of DM particles by nuclei~\cite{Fan:2010gt,Fitzpatrick:2012ix}, in modelling collective excitations in DM direct detection experiments~\cite{trickle2020effective} and in a study of the DM scattering by bound electrons in isolated atomic systems~\cite{Catena:2019gfa}. This latter work introduced the notion of ``atomic response'' to DM-electron interactions that we here extend to the case of semiconductor crystals. 
By applying our new formalism to the study of DM-electron scattering in crystals, we discover that, under standard assumptions for the local DM velocity distribution, the rate of DM-induced electronic transitions in a semiconductor material depends on at most five independent ``crystal response'' functions. We express these response functions in terms of electron wave function overlap integrals and evaluate them numerically using \texttt{QEdark-EFT}~\cite{QEdark-EFT}, our extension of the \texttt{QEdark} code~\cite{Essig:2015cda}, which relies on the integrated suite of open-source computer codes, \texttt{QuantumEspresso}~\cite{Giannozzi2009Sep}. Leveraging on this finding, within our effective theory framework we are able to set 90\% confidence level exclusion limits on the strength with which DM can couple to electrons, for general models. As illustrative examples we explicitly give the limits for specific DM models that were not tractable before our work, including the anapole, magnetic and electric DM-electron interaction models. Four of the five crystal response functions computed in this work were not known previously, and have explicitly been identified for the first time here. From a practical point of view, these can be used to compute the rate of DM-induced electron excitations in virtually all DM models where the free amplitude for DM-electron interactions does not explicitly depend on the mass of the particle that mediates the underlying interaction. By promoting the coupling constant to a general function of the momentum transfer our formalism could in principle be applied to any interaction model. At a more speculative level, the novel crystal response functions discovered in this work encode properties of crystals that have so far remained hidden, and that could be revealed if it becomes possible to use the DM particles that form our Milky Way as a probe in scattering experiments with semiconductor targets. 

This work is organised as follows. In Sec.~\ref{sec:theory}, we introduce our formalism to model the scattering of DM particles by electrons in semiconductor crystals. In Sec.~\ref{sec:crysres}, we introduce and numerically evaluate the novel crystal response functions that our formalism predicts. In Sec.~\ref{sec:results}, we present the 90\% confidence level exclusion limits on the strength with which DM can couple to electrons, both within our effective theory framework and within specific DM models. We summarise and conclude in Sec.~\ref{sec:summary}. Finally, details  underlying our analytical and numerical calculations are presented in the appendices.

\section{Dark matter-induced electronic excitations in crystals}
\label{sec:theory}
In this section, we extend the formalism of DM-electron scattering in crystals to general non-relativistic DM-electron interactions, including the anapole, magnetic and electric dipole couplings.~We first introduce an expression for the rate of DM-induced electronic excitations that applies to arbitrary target materials and interactions, Sec.~\ref{sec:rategen}.~Then, we narrow it down to the case of crystals, Sec.~\ref{sec:ratecrys}, and general non-relativistic DM-electron interactions, Sec.~\ref{sec:nr}.

\subsection{General rate of electronic excitations}
\label{sec:rategen}

For a generic target material and arbitrary DM-electron interactions, the rate of DM-induced electronic excitations from an initial electron bound state $|\mathbf{e}_1\rangle$ to a final state $|\mathbf{e}_2\rangle$ is given by~\cite{Catena:2019gfa}
\begin{align}
\mathscr{R}_{1\rightarrow 2}&=\frac{n_{\chi}}{16 m^2_{\chi} m^2_e} \,
\int \frac{{\rm d}^3 q}{(2 \pi)^3} \int {\rm d}^3 v f_{\chi}(\mathbf{v}) (2\pi) \delta(E_f-E_i) \nonumber \\
&\times \overline{\left| \mathcal{M}_{1\rightarrow 2}\right|^2}\, , 
\label{eq:transition rate}
\end{align}
where the initial and final states, $|\mathbf{e}_1\rangle$, and $|\mathbf{e}_2\rangle$, are energy, not momentum eigenstates, and $E_i$ ($E_f$) is the initial (final) energy of the DM-electron system.~In Eq.~(\ref{eq:transition rate}), the squared electron transition amplitude, $\overline{\left| \mathcal{M}_{1\rightarrow 2}\right|^2}$, is defined in terms of the initial and final momentum-space electron wave functions, $\widetilde{\psi}_1$ and $\widetilde{\psi}_2$, and of the amplitude for DM scattering by free electrons, $\mathcal{M}$, as
\begin{align}
    \overline{\left| \mathcal{M}_{1\rightarrow 2}\right|^2}\equiv \overline{\left|\int  \frac{{\rm d}^3 \ell}{(2 \pi)^3} \, \widetilde{\psi}_2^*(\boldsymbol{\ell}+\mathbf{q})  
\mathcal{M}(\boldsymbol{\ell},\mathbf{p},\mathbf{q})
\widetilde{\psi}_1(\boldsymbol{\ell}) \right|^2}\, ,
\label{eq:transition amplitude}
\end{align}
where a bar denotes an average (sum) over initial (final) spin states and we integrate over the electron momenta, $\boldsymbol{\ell}$.~Here, $\mathbf{q}=\mathbf{p}-\mathbf{p}^\prime$ is the momentum transferred, $\mathbf{p}'$ is the outgoing DM particle momentum, and $\mathbf{p}=m_\chi\mathbf{v}$, where $\mathbf{v}$ is the initial DM particle velocity, while $m_\chi$ and $m_e$ are the DM and electron mass, respectively.~Eq.~(\ref{eq:transition rate})
also depends on the DM number density, $n_\chi$, and the local DM velocity distribution in the detector rest frame, $f_\chi$.~For the latter, we assume
\begin{align}
    f_\chi(\mathbf{v})&= \frac{1}{N_{\rm esc}\pi^{3/2}v_0^3}\exp\left[-\frac{(\mathbf{v}+\mathbf{v}_\oplus)^2}{v_0^2} \right]
    \nonumber\\
    &\times \Theta\left(v_{\rm esc}-|\mathbf{v}+\mathbf{v}_\oplus|\right)\,,
\end{align}
where 
\begin{align}
N_{\rm esc}\equiv \erf(v_{\rm esc}/v_0)-2 (v_{\rm esc}/v_0)\exp(-v_{\rm esc}^2/v_0^2)/\sqrt{\pi}
\end{align}
With this definition for $N_{\rm esc}$, $f_{\chi}(\mathbf{v})$ is normalised to one.~In all numerical applications, we set the most probable speed to the value of the local standard of rest, $v_0=220~\text{km sec}^{-1}$~\cite{Kerr:1986hz}, the local galactic escape velocity to $v_{\rm esc} = 544~\text{km sec}^{-1}$~\cite{Smith:2006ym}, the speed of the Earth in the galactic reference frame (where the mean DM particle velocity is zero) to~$v_\oplus= 244~\text{km sec}^{-1}$ and the local DM number density to $n_\chi=~0.4\text{ GeV/cm}^{3}/m_\chi$ \cite{Catena:2009mf}.

\subsection{Rate of electronic excitations in crystals}
\label{sec:ratecrys}
In the case of crystalline materials, we label the initial (final) electron state by a band index $i$ ($i'$) and a wavevector in the first Brillouin Zone (BZ) $\mathbf{k}$ ($\mathbf{k}'$), i.e. in the notation of Eq.~(\ref{eq:transition rate}), $1\equiv\{i\mathbf{k}\}$ and $2\equiv\{i'\mathbf{k}'\}$.~Furthermore, we express the initial electron position space wave function at $\mathbf{x}$ in the Bloch form 
\begin{equation}
\psi_{i\mathbf{k}}(\mathbf{x})=\frac{1}{\sqrt{V}}\sum_\mathbf{G}{u_i(\mathbf{k}+\mathbf{G})e^{i(\mathbf{k}+\mathbf{G})\cdot \mathbf{x}}}\, ,
\label{eq:psi}
\end{equation}
and similarly for $\psi_{i'\mathbf{k}'}$.~Here, $V=N_{\rm cell} V_{\rm cell}$ is the volume of the crystal, $V_{\rm cell}$ is the volume of a single unit cell,  $N_{\rm cell} = M_{\rm target}/M_{\rm cell}$ is the number of unit cells in the crystal, $M_{\rm target}$ is the detector target mass, while $M_{\rm cell}= 2m_{\rm Ge}=135.33\text{ GeV}$ for germanium and $M_{\rm cell}= 2m_{\rm Si}=52.33\text{ GeV}$ for silicon.~The $u_i$ coefficients in Eq.~(\ref{eq:psi}) fulfil $\sum_{\mathbf{G}}|u_i(\mathbf{k}+\mathbf{G})|^2=1$, where the sum runs over reciprocal lattice vectors $\mathbf{G}$.~On evaluating Eq.~(\ref{eq:transition rate}) using wave functions $\psi_{i\mathbf{k}}$ and $\psi_{i'\mathbf{k}'}$ of the type in Eq.~(\ref{eq:psi}), we denote the corresponding rate of DM-induced electronic excitation by $\mathscr{R}_{i\mathbf{k} \rightarrow i'\mathbf{k}'}$.

In the case of DM direct detection experiments using semiconducting crystals such as silicon and germanium as target materials, the observable quantity is the total rate of valence to conduction band electron excitations in the whole crystal,
$\mathscr{R}_{\rm crystal}$, which is given by
\begin{align}
\mathscr{R}_{\rm crystal}=2 \sum_{i i'} \int_{\rm BZ} \frac{V {\rm d}^3 k}{(2\pi)^3} \int_{\rm BZ} \frac{V {\rm d}^3 k'}{(2\pi)^3} \,
\mathscr{R}_{i\mathbf{k} \rightarrow i'\mathbf{k}'} \,.
\label{eq:Rcrys}
\end{align}
The factor of 2 is a result of the spin degeneracy and consequent double occupation of each crystal orbital.~In order to evaluate Eq.~(\ref{eq:Rcrys}), we expand the free scattering amplitude, $\mathcal{M}$, in the small electron momentum to electron mass ratio.~As we will see in detail below, this non-relativistic expansion allows us to identify the response of crystals to general DM-electron interactions in a model-independent manner.
Note that, in principle, one should obtain this total rate by summing Eq.~(\ref{eq:transition rate}) over all unfilled conduction bands and all filled valence bands.~However, in practical applications one has to truncate the number of conduction bands included in the sum to a manageable number, as we will see in Sec.~\ref{sec:num}.~

\begin{table}[t]
    \centering
    \begin{tabular*}{\columnwidth}{@{\extracolsep{\fill}}ll@{}}
    \toprule
      $\mathcal{O}_1 = \mathds{1}_{\chi e}$ & $\mathcal{O}_9 = i\mathbf{S}_\chi\cdot\left(\mathbf{S}_e\times\frac{ \mathbf{q}}{m_e}\right)$  \\
        $\mathcal{O}_3 = i\mathbf{S}_e\cdot\left(\frac{ \mathbf{q}}{m_e}\times \mathbf{v}^{\perp}_{\rm el}\right)$ &   $\mathcal{O}_{10} = i\mathbf{S}_e\cdot\frac{ \mathbf{q}}{m_e}$   \\
        $\mathcal{O}_4 = \mathbf{S}_{\chi}\cdot \mathbf{S}_e$ &   $\mathcal{O}_{11} = i\mathbf{S}_\chi\cdot\frac{ \mathbf{q}}{m_e}$   \\                                                                             
        $\mathcal{O}_5 = i\mathbf{S}_\chi\cdot\left(\frac{ \mathbf{q}}{m_e}\times \mathbf{v}^{\perp}_{\rm el}\right)$ &  $\mathcal{O}_{12} = \mathbf{S}_{\chi}\cdot \left(\mathbf{S}_e \times \mathbf{v}^{\perp}_{\rm el} \right)$ \\                                                                                                                 
        $\mathcal{O}_6 = \left(\mathbf{S}_\chi\cdot\frac{ \mathbf{q}}{m_e}\right) \left(\mathbf{S}_e\cdot\frac{\hat{{\bf{q}}}}{m_e}\right)$ &  $\mathcal{O}_{13} =i \left(\mathbf{S}_{\chi}\cdot  \mathbf{v}^{\perp}_{\rm el}\right)\left(\mathbf{S}_e\cdot \frac{ \mathbf{q}}{m_e}\right)$ \\   
        $\mathcal{O}_7 = \mathbf{S}_e\cdot  \mathbf{v}^{\perp}_{\rm el}$ &  $\mathcal{O}_{14} = i\left(\mathbf{S}_{\chi}\cdot \frac{ \mathbf{q}}{m_e}\right)\left(\mathbf{S}_e\cdot  \mathbf{v}^{\perp}_{\rm el}\right)$  \\
        $\mathcal{O}_8 = \mathbf{S}_{\chi}\cdot  \mathbf{v}^{\perp}_{\rm el}$  & $\mathcal{O}_{15} = i\mathcal{O}_{11}\left[ \left(\mathbf{S}_e\times  \mathbf{v}^{\perp}_{\rm el} \right) \cdot \frac{ \mathbf{q}}{m_e}\right] $ \\       
    \bottomrule
    \end{tabular*}
    \caption{Interaction operators defining the non-relativistic effective theory of spin 1/2 DM-electron interactions~\cite{Fan:2010gt,Fitzpatrick:2012ix,Catena:2019gfa}.~$\mathbf{S}_e$ ($\mathbf{S}_\chi$) is the electron (DM) spin, $\mathbf{v}_{\rm el}^\perp=\mathbf{v}-\boldsymbol{\ell}/m_e-\mathbf{q}/(2 \mu_{\chi e})$, where $\mu_{\chi e}$ is the DM-electron reduced mass, $\mathbf{v}_{\rm el}^\perp$ is the transverse relative velocity and $\mathds{1}_{\chi e}$ is the identity in the DM-electron spin space.}
\label{tab:operators}
\end{table}

\subsection{Non-relativistic expansion}
\label{sec:nr}
Assuming that both initial and final electron and DM particle move at a non-relativistic speed, the free scattering amplitude, $\mathcal{M}$, can in general be expressed as a function of two momenta only~\cite{Fitzpatrick:2012ix}:~the transferred momentum, $\mathbf{q}$, and the transverse relative velocity $\mathbf{v}_{\rm el}^\perp$, i.e.~the component of $\mathbf{v}$ that in the case of elastic DM-electron scattering is perpendicular to $\mathbf{q}$.~Namely, $\mathcal{M}=\mathcal{M}(\mathbf{q},\mathbf{v}_{\rm el}^\perp)$, where $\mathbf{v}_{\rm el}^\perp=\mathbf{v}-\boldsymbol{\ell}/m_e-\mathbf{q}/(2 \mu_{\chi e})$ and $\mu_{\chi e}$ is the DM-electron reduced mass.~By expanding $\mathcal{M}$ in the electron momentum to electron mass ratio, $|\boldsymbol{\ell}|/m_e\ll 1$, we find
\begin{align}
    \label{eq:Mnr2}
    \mathcal{M}(\mathbf{q},\mathbf{v}_{\rm el}^\perp) &\simeq \mathcal{M}(\mathbf{q},\mathbf{v}_{\rm el}^\perp)_{\boldsymbol{\ell}=0} \nonumber\\
    &+ \left(\frac{\boldsymbol{\ell}}{m_e}\right)\cdot m_e\nabla_{\boldsymbol{\ell}} \mathcal{M}(\mathbf{q},\mathbf{v}_{\rm el}^\perp)_{\boldsymbol{\ell}=0}\,.
\end{align}
While Eq.~(\ref{eq:Mnr2}) applies to any model for DM-electron interactions as a first order expansion in $|\boldsymbol{\ell}|/m_e\ll 1$, it is an exact equation in the case of the so-called non-relativistic effective theory of DM-electron interactions, where the free scattering amplitude is by construction expressed as a sum of interaction operators in the DM and electron spin space that are at most linear in $\mathbf{v}_{\rm el}^\perp$,
\begin{equation}
 \label{eq:Mnr}
\mathcal{M}(\mathbf{q},\mathbf{v}_{\rm el}^\perp) = \sum_i \left(c_i^s +c^\ell_i \frac{q_{\rm ref}^2}{|\mathbf{q}|^2} \right) \,\langle \mathcal{O}_i \rangle  \,,
\end{equation}
where the interaction operators $\mathcal{O}_i$ for spin 1/2 DM are defined in Tab.~\ref{tab:operators}, $q_{\rm ref}$ is a reference momentum given by $q_{\rm ref}\equiv \alpha m_e$ and $\alpha$ is the fine structure constant.~Angle brackets in Eq.~(\ref{eq:Mnr}) denote matrix elements between the two-component spinors $\xi^\lambda$ and $\xi^s$ ($\xi^{\lambda'}$ and $\xi^{s'}$) associated with the initial (final) state electron and DM particle, respectively.~For example, $\langle \mathcal{O}_1 \rangle \equiv \xi^{s'} \xi^{s} \xi^{\lambda'} \xi^\lambda$.~Finally, the dimensionless coefficients $c_i^s$ and $c_i^\ell$ in Eq.~(\ref{eq:Mnr}) are the coupling constants of the interaction operators in Tab.~\ref{tab:operators}.~When $c^s_i\neq 0$ and $c^\ell=0$, we refer to the interactions in Tab.~\ref{tab:operators} as of contact type; we refer to them as of long-range type when $c^s_i=0$ and $c^\ell\neq0$.

In the non-relativistic limit, almost any model for DM-electron interactions in crystals can be matched onto the free scattering amplitude in Eq.~(\ref{eq:Mnr}).
~By substituting Eq.~(\ref{eq:Mnr}) into Eq.~(\ref{eq:transition amplitude}), we find
\begin{align}
\overline{\left|\mathcal{M}_{i\mathbf{k}\rightarrow i'\mathbf{k}'}\right|^2} &= \overline{\left|\mathcal{M}\right|^2}_{\boldsymbol{\ell}=0}\left|f_{i\mathbf{k}\rightarrow i^\prime\mathbf{k}^\prime}\right|^2 \nonumber\\
&+ 2 m_e \overline{\Re\left[\mathcal{M}f_{i\mathbf{k}\rightarrow i^\prime \mathbf{k}^\prime}(\nabla_{\boldsymbol{\ell}}\mathcal{M}^*)_{\boldsymbol{\ell}=0}\cdot\left(\mathbf{f}_{i\mathbf{k}\rightarrow i^\prime \mathbf{k}^\prime}\right)^*\right]} 
\nonumber\\
&+  m_e^2 \overline{\left|(\nabla_{\boldsymbol{\ell}}\mathcal{M})_{\boldsymbol{\ell}=0}\cdot \mathbf{f}_{i\mathbf{k}\rightarrow i^\prime \mathbf{k}^\prime}\right|^2}
\label{eq:M2}
\end{align}
where
\begin{align}
f_{i\mathbf{k} \rightarrow i' \mathbf{k}'}(\mathbf{q})&=\int \mathrm{d}^3x \, \psi_{i' \mathbf{k}'}^*(\mathbf{x})\,e^{i\mathbf{x}\cdot\mathbf{q}} \,\psi_{i\mathbf{k}}(\mathbf{x}) \label{eq:f}\\
\mathbf{f}_{i\mathbf{k}\rightarrow i' \mathbf{k}'}(\mathbf{q})&=\int \mathrm{d}^3x \, \psi_{i' \mathbf{k}'}^*(\mathbf{x}) \,e^{i\mathbf{x}\cdot\mathbf{q}}\, \frac{i\nabla_\mathbf{x}}{m_e} \psi_{i\mathbf{k}}(\mathbf{x})\, ,
\label{eq:fvec}
\end{align}
are electron wave function overlap integrals. ~The first one (Eq.~(\ref{eq:f})) arises within the standard treatment of DM-electron interactions in crystals~\cite{Essig:2015cda}.~The second one (Eq.~(\ref{eq:fvec})) is responsible for the novel crystal responses we define below in Sec.~\ref{sec:crysres}.~An explicit calculation performed in Appendix~\ref{sec:R}
shows that from each term in Eq.~(\ref{eq:M2}), we can collect a Dirac delta function and rewrite Eq.~(\ref{eq:M2}) as follows
\begin{align}
\overline{\left|\mathcal{M}_{i\mathbf{k}\rightarrow i'\mathbf{k}'}\right|^2} &=\sum_{\Delta \mathbf{G}}\frac{(2\pi)^3\delta^3(\mathbf{k}+\mathbf{q}-\mathbf{k}^\prime-\Delta\mathbf{G})}{V} \nonumber\\
&\times\Bigg[\overline{\left|\mathcal{M}\right|^2}\left|f_{i\mathbf{k}\rightarrow i^\prime, \mathbf{k}^\prime}^\prime\right|^2 \nonumber\\
&+ 2 m_e \overline{\Re\left[\mathcal{M}f_{i\mathbf{k}\rightarrow i^\prime \mathbf{k}^\prime}^\prime(\nabla_{\boldsymbol{\ell}}\mathcal{M}^*)_{\boldsymbol{\ell}=0}\cdot\left(\mathbf{f}_{i\mathbf{k}\rightarrow i^\prime \mathbf{k}^\prime}^\prime\right)^*\right]} 
\nonumber\\
&+ m_e^2 \overline{\left|(\nabla_{\boldsymbol{\ell}}\mathcal{M})_{\boldsymbol{\ell}=0}\cdot \mathbf{f}_{i\mathbf{k}\rightarrow i^\prime \mathbf{k}^\prime}^\prime\right|^2} \Bigg]
\label{eq:M2_prime}
\end{align}
where $\Delta \mathbf{G}\equiv \mathbf{G}^\prime-\mathbf{G}$ and 
\begin{align}
 f_{i,\mathbf{k}\rightarrow i^\prime, \mathbf{k}^\prime}^\prime &\equiv \sum_{\mathbf{G}}u_{i^\prime}^*\left(\mathbf{k}^\prime+\mathbf{G}+\Delta\mathbf{G}\right) u_i\left(\mathbf{k}+\mathbf{G}\right)\label{eq:f_prime} \\
    \mathbf{f}_{i,\mathbf{k}\rightarrow i^\prime, \mathbf{k}^\prime}^\prime &\equiv -\frac{1}{m_e}\sum_{\mathbf{G}}u_{i^\prime}^*\left(\mathbf{k}^\prime+\mathbf{G}+\Delta\mathbf{G}\right)\left(\mathbf{k}+\mathbf{G}\right)
    \nonumber\\
    &\times u_i\left(\mathbf{k}+\mathbf{G}\right)\, .
\label{eq:fvec_prime}
\end{align}
Consequently,
\begin{align}
\overline{\left|\mathcal{M}_{i\mathbf{k}\rightarrow i'\mathbf{k}'}\right|^2} &\equiv  \sum_{\Delta\mathbf{G}}\frac{(2\pi)^3\delta^3(\mathbf{k}+\mathbf{q}-\mathbf{k}^\prime-\Delta\mathbf{G})}{V} \,
\nonumber\\
&\times\overline{\left|\mathcal{M}_{i\mathbf{k}\rightarrow i^\prime \mathbf{k}^\prime}^\prime\right|^2}\,,
\label{eq:M2final}
\end{align}
where $\overline{\left|\mathcal{M}_{i\mathbf{k}\rightarrow i^\prime \mathbf{k}^\prime}^\prime\right|^2}$ is defined through the comparison of Eq.~(\ref{eq:M2}) and Eq.~(\ref{Eq:Crystal_matrixelement}).~In the non-relativistic limit, the initial and final DM-electron energies, $E_i$ and $E_f$, respectively, are given by
\begin{align}
    E_i &= m_\chi + m_e + \frac{m_\chi}{2}v^2 + E_{i\mathbf{k}}\, , \label{eq: energy initial}\\
    E_f &= m_\chi + m_e + \frac{|m_\chi\vec{v}-\mathbf{q}|^2}{2m_\chi} + E_{i'\mathbf{k}'}\,, \label{eq: energy final}
\end{align}
where $v=|\mathbf{v}|$ is the initial DM~particle speed and $E_{i\mathbf{k}}$ ($E_{i'\mathbf{k}'}$) is the energy of the initial (final) electron bound state.
~Defining $\Delta E_{i\mathbf{k} \rightarrow i'\mathbf{k}'}\equiv E_{i'\mathbf{k}'}-E_{i\mathbf{k}}$ allows us to write 
$E_f-E_i$ as follows
\begin{align}
E_f-E_i &= \Delta E_{i\mathbf{k} \rightarrow i'\mathbf{k}'} + \frac{q^2}{2m_\chi}-q v \cos \theta\,,
\label{eq:E}
\end{align}
where $q=|\mathbf{q}|$ and $\theta$ is the angle between the momentum transfer $\mathbf{q}$ and the initial DM particle velocity $\mathbf{v}$.~By replacing Eq.~(\ref{eq:E}), Eq.~(\ref{eq:M2final}), Eq.~(\ref{eq:psi}) and Eq.~(\ref{eq:transition rate}) in Eq.~(\ref{eq:Rcrys}), for the total rate we find

\begin{align}
\mathscr{R}_{\textrm{crystal}}&=\frac{2 n_\chi V}{16m_\chi^2 m_e^2}\int\mathrm{d}^3q \int \mathrm{d}^3 v \,f_\chi(\mathbf{v}) 
\nonumber \\
&\times\sum_{\Delta\mathbf{G} ii^\prime}\int_{\textrm{BZ}}\frac{\mathrm{d}^3k }{(2\pi)^3}\int_{\textrm{BZ}}\frac{\mathrm{d}^3k^\prime}{(2\pi)^3}
\nonumber \\
&\times\delta^3(\mathbf{k}+\mathbf{q}-\mathbf{k}^\prime-\Delta\mathbf{G})\, 
\nonumber\\
&\times 
 (2\pi) \delta\left( \Delta E_{i\mathbf{k} \rightarrow i'\mathbf{k}'} + \frac{q^2}{2m_\chi}-q v \cos \theta\right)
 \nonumber \\
&\times\overline{\left|\mathcal{M}^\prime_{i,\mathbf{k}\rightarrow i^\prime, \mathbf{k}^\prime}\right|^2}
\label{eq:Rv2}
\end{align}

We can now use the Dirac delta function in Eq.~(\ref{eq:Rv2}) to perform the integration over the polar angle $\theta$ in the velocity dependent part of the total excitation rate $\mathscr{R}_{\rm crystal}$ that we here denote by
\begin{align}
\Gamma^{ii'}_{\mathbf{k}\mathbf{k}'}(q)&=\int \mathrm{d} v v^2  \int_{0}^{2\pi} {\rm d}\phi \int_{-1}^{+1} {\rm d}\cos\theta \,\frac{f_\chi(\mathbf{v})}{v}
\nonumber\\
&\times
\delta\left(\cos\theta- \xi
\right) \overline{\left|\mathcal{M}^\prime_{i,\mathbf{k}\rightarrow i^\prime, \mathbf{k}^\prime}\right|^2}, 
\end{align}
where
\begin{align}
\xi
= \frac{q}{2 m_\chi v} + \frac{\Delta E_{i\mathbf{k} \rightarrow i'\mathbf{k}'} }{q v} \,.
\end{align}
We find
\begin{align}
\Gamma^{ii'}_{\mathbf{k}\mathbf{k}'}(q)&\simeq 
\int_{|\mathbf{v}|\ge v_{\rm min}}
{\rm d}v\,
\frac{ v^2f_\chi(v)}{v}\nonumber\\
&\times 
\int_{0}^{2\pi} {\rm d}\phi \, \overline{\left|\mathcal{M}^\prime_{i,\mathbf{k}\rightarrow i^\prime, \mathbf{k}^\prime}\right|^2}_{\cos\theta=\xi}
 \nonumber\\
&= 
\int_{|\mathbf{v}|\ge v_{\rm min}}
{\rm d}v\,
 \frac{2\pi v^2f_\chi(v)}{v} 
 \overline{\overline{\left|\mathcal{M}^\prime_{i,\mathbf{k}\rightarrow i^\prime, \mathbf{k}^\prime}\right|^2}}_{\cos\theta=\xi}
\nonumber\\
& \simeq \frac{1}{2} \widehat{\eta}(q, \Delta E_{i\mathbf{k} \rightarrow i'\mathbf{k}'} ) \left[ \overline{\overline{\left|\mathcal{M}^\prime_{i,\mathbf{k}\rightarrow i^\prime, \mathbf{k}^\prime}\right|^2}}_{\cos\theta=\xi}
\right],
\label{eq:velint}
\end{align}
where
\begin{align}
v_{\rm min}=\frac{q}{2 m_\chi } + \frac{\Delta E_{i\mathbf{k} \rightarrow i'\mathbf{k}'} }{q } \,.
\label{eq:vmin}
\end{align}
In the first step of Eq.~(\ref{eq:velint}) we follow Essig {\rm et al.}~\cite{Essig:2015cda} and use the simplification $f_\chi(\mathbf{v}) = f_\chi(v)$, performing the integration over $\theta$ by using the Dirac delta function.~In the second step of Eq.~(\ref{eq:velint}), we introduce the azimuthal-angle-averaged squared transition amplitude, defined as
 \begin{align}
\overline{\overline{\left|\mathcal{M}^\prime_{i,\mathbf{k}\rightarrow i^\prime, \mathbf{k}^\prime}\right|^2}}_{\cos\theta=\xi
} \equiv \frac{1}{2\pi}\int_0^{2\pi} {\rm d}\phi \,\overline{\left|\mathcal{M}^\prime_{i,\mathbf{k}\rightarrow i^\prime, \mathbf{k}^\prime}\right|^2}_{\cos\theta=\xi
}\,.
\label{eq:Mazi}
\end{align}
Again following~\cite{Essig:2015cda}, in the third step of Eq.~(\ref{eq:velint}) we restore the angular dependence of $f_{\chi}(v)=f_\chi(\mathbf{v})$ and introduce the linear operator $\widehat{\eta}(q, \Delta E_{i\mathbf{k} \rightarrow i'\mathbf{k}'} )$.~Acting on $C g(\mathbf{v})$ where $C$ is a constant and $g(\mathbf{v})$ a function of the DM particle velocity in the detector rest frame, it gives
\begin{align}
\widehat{\eta}(q, \Delta E_{i\mathbf{k} \rightarrow i'\mathbf{k}'} ) \left[C g(\mathbf{v}) \right] &= C \int_{|\mathbf{v}|\ge v_{\rm min}
}
{\rm d}^3v \, g(\mathbf{v})\nonumber\\
&\times\frac{f_\chi(\mathbf{v})}{v} \,.
\end{align}
In terms of $\widehat{\eta}$ and $\overline{\overline{\left|\mathcal{M}^\prime_{i,\mathbf{k}\rightarrow i^\prime, \mathbf{k}^\prime}\right|^2}}_{\theta=\bar{\theta}(q, \Delta E_{i\mathbf{k} \rightarrow i'\mathbf{k}'} )}$, we can finally express the total excitation rate in a crystal as
\begin{align}
\mathscr{R}_{\textrm{crystal}}&=\frac{\pi n_\chi V}{8m_\chi^2 m_e^2} \sum_{\Delta\mathbf{G} ii^\prime}\int_{\textrm{BZ}}\frac{\mathrm{d}^3k }{(2\pi)^3}\int_{\textrm{BZ}}\frac{\mathrm{d}^3k^\prime}{(2\pi)^3}
\nonumber\\
&\times\int\mathrm{d}^3q \frac{1}{q} \,\widehat{\eta}\left(q, \Delta E_{i\mathbf{k} \rightarrow i'\mathbf{k}'}\right)\delta^3(\mathbf{k}+\mathbf{q}-\mathbf{k}^\prime-\Delta\mathbf{G})\,\nonumber\\
&\times 
\overline{\overline{\left|\mathcal{M}^\prime_{i,\mathbf{k}\rightarrow i^\prime, \mathbf{k}^\prime}\right|^2}}_{\cos\theta=\xi}
\,.
\end{align}

An interesting class of models in the search for DM via electronic excitations is that in which DM couples to electrons via higher order moments in the multipole expansion of the electromagnetic field ~\cite{Catena:2019gfa,Catena:2020tbv}.~If $\chi$ ($\psi$) is a Majorana (Dirac) spinor describing the DM~particle, $g$ a dimensionless coupling constant and $\Lambda$ a mass scale, the anapole, magnetic dipole and electric dipole DM models are described by the following interaction Lagrangians~\cite{Catena:2019gfa},
\begin{align}
\mathscr{L}_{\rm anapole}&= \frac{g}{2\Lambda^2} \, \bar{\chi}\gamma^\mu\gamma^5\chi \, \partial^\nu F_{\mu\nu} \,, \\
\mathscr{L}_{\rm magnetic}&= \frac{g}{\Lambda} \, \bar{\psi}\sigma^{\mu\nu}\psi \, F_{\mu\nu}\,, \\
\mathscr{L}_{\rm electric}&= \frac{g}{\Lambda} \, i\bar{\psi} \sigma^{\mu\nu} \gamma^5 \psi \, F_{\mu\nu} \,,
\label{eq:L}
\end{align}
where $F_{\mu\nu} = \partial_\mu A_\nu - \partial_\nu A_\mu$ is the photon field strength tensor, and $A^\nu$ the photon field.~In the non-relativistic limit, the free electron scattering amplitudes associated with the Lagrangians in Eq.~(\ref{eq:L}) are~\cite{Catena:2019gfa}
\begin{align}
\mathscr{M}_{\rm anapole}&=\frac{4 e g}{\Lambda^2} m_\chi m_e \Bigg\{
    2 \left(\mathbf{v}_{\rm el}^\perp \cdot \xi^{\dagger s'} \mathbf{S}_\chi \xi^s \right) \delta^{\lambda' \lambda}
    \nonumber\\
    &+  
    g_e \left( \xi^{\dagger s'} \mathbf{S}_\chi \xi^s \right) \cdot \left( i\frac{\mathbf{q}}{m_e} \times \xi^{\dagger \lambda'} \mathbf{S}_e \xi^\lambda  \right)
    \Bigg\} \,, \label{eq: amplitude anapole}\\ 
\mathscr{M}_{\rm magnetic}&=
\frac{e g}{\Lambda}  \Bigg\{
    4m_e\delta^{s's}\delta^{\lambda'\lambda} \nonumber\\
    &+\frac{16m_\chi m_e}{q^2}  i\mathbf{q} \cdot \left(\mathbf{v}_{\rm el}^\perp \times \xi^{\dagger s'} \mathbf{S}_\chi \xi^s \right)\delta^{\lambda'\lambda} 
    \nonumber\\
    &-  
    \frac{8 g_em_\chi}{q^2} \Bigg[\left( \mathbf{q} \cdot \xi^{\dagger s'} \mathbf{S}_\chi \xi^s \right)\left( \mathbf{q} \cdot \xi^{\dagger \lambda'} \mathbf{S}_e \xi^\lambda \right)
    \nonumber\\
    &- q^2  \left( \xi^{\dagger s'} \mathbf{S}_\chi \xi^s \right)\cdot \left( \xi^{\dagger \lambda'} \mathbf{S}_e \xi^\lambda \right)
    \Bigg] \Bigg\}\,,\label{eq: amplitude magnetic dipole}\\
\mathscr{M}_{\rm electric}&= \frac{e g}{\Lambda} \frac{16 m_\chi m_e}{q^2} i\mathbf{q} \cdot \left( \xi^{\dagger s'} \mathbf{S}_\chi \xi^s \right)\delta^{\lambda' \lambda}\label{eq: amplitude electric dipole} \,.
\end{align}
where $g_e \simeq 2$ is the electron $g$-factor.~From the free electron scattering amplitude in Eq.~(\ref{eq: amplitude anapole}), we find that, in the case of anapole DM-electron interactions
\begin{subequations}
\label{eq: anapole effective couplings}
\begin{align}
    c_8^s &= 8 e m_e m_\chi\frac{g}{\Lambda^2}\, ,\\
    c_9^s &= -8 e m_e m_\chi\frac{g}{\Lambda^2}\, ,
\end{align}
\end{subequations}
are the only coupling constants different from zero.~From the amplitude in Eq.~(\ref{eq: amplitude magnetic dipole}), we find that the only coupling constants different from zero in the case of magnetic dipole DM-electron interactions are
\begin{subequations}
\label{eq: magnetic dipole effective couplings}
\begin{align}
    c_1^s &= 4 e m_e \frac{g}{\Lambda}\, ,\\
    c_4^s &= 16 e m_\chi\frac{g}{\Lambda}\, ,\\
    c_5^\ell &= \frac{16em_e^2 m_\chi}{q_{\rm ref}^2} \frac{g}{\Lambda}\, ,\\
    c_6^\ell &= -\frac{16em_e^2 m_\chi}{q_{\rm ref}^2} \frac{g}{\Lambda}\, .
\end{align}
\end{subequations}
Finally, from the amplitude in Eq.~(\ref{eq: amplitude electric dipole}), we find that in the case of electric dipole DM-electron interactions, one coupling constant only is different from zero,
\begin{align}
\label{eq: electric dipole effective couplings}
    c_{11}^\ell &= \frac{16 e m_\chi m_e^2}{q_{\rm ref}^2}\frac{g}{\Lambda}\, .
\end{align}

\section{Novel crystal responses}
\label{sec:crysres}
We now focus on the novel crystal responses that arise from the electron wave function overlap integral in Eqs.~(\ref{eq:f}) and (\ref{eq:fvec}).

\subsection{Excitation rate and crystal response functions}
As shown in the appendix, the azimuthal-angle-averaged squared transition amplitude, can be expressed as the sum of $r$ products between a DM response function $R_l(q,v)$ and a crystal response function $W_l(\mathbf{q},\Delta E)$, $l=1,\dots r$, where $\Delta E$ is defined implicitly via Eq.~(\ref{eq:W_scalar}) below.~As a result, the total electron excitation rate in crystals can be written as
\begin{align}
\mathscr{R}_{\textrm{crystal}}&=\frac{n_\chi N_\text{cell} }{128\pi m_\chi^2 m_e^2}\int \mathrm{d} (\ln\Delta E)\int \mathrm{d}^3 q 
 \frac{1}{q} \,
\widehat{\eta}\left(q, \Delta E
\right) \nonumber\\
&\times
\sum_{l=1}^r \Re\left(R_l^*(q,v) W_l(\mathbf{q},\Delta E)\right) \,,
\label{eq:R_crystal}
\end{align}
where the DM response functions $R_l(q,v)$ given in App.~\ref{App:DM_responses} depend on the coupling constants $c^s_i$ and $c^\ell_i$, in addition to $v$, $q$ and $\xi$, and
\begin{align}
W_l(\mathbf{q},\Delta E)&=(4\pi)^2V_\text{cell}\Delta E\sum_{\Delta\mathbf{G} ii^\prime}\int_{\textrm{BZ}}\frac{\mathrm{d}^3k }{(2\pi)^3}\int_{\textrm{BZ}}\frac{\mathrm{d}^3k^\prime}{(2\pi)^3}\,B_l \, \nonumber\\
&\times\delta^3(\mathbf{q}-\mathbf{k}^\prime-\Delta\mathbf{G}+\mathbf{k})  \nonumber\\
&\times\delta(\Delta E -E_{i\mathbf{k}}+E_{i'\mathbf{k}^\prime})
\label{eq:W_scalar}\,.
\end{align}
Here,
\begin{align}
B_1 =& \left| f_{i,\mathbf{k}\rightarrow i^\prime, \mathbf{k}^\prime}^\prime \right|^2 \\
B_2=&\frac{\mathbf{q}}{m_e} \cdot(f_{i,\mathbf{k}\rightarrow i^\prime, \mathbf{k}^\prime}^\prime) (\mathbf{f}_{i,\mathbf{k}\rightarrow i^\prime, \mathbf{k}^\prime}^\prime)^* \\
B_3=&\left| \mathbf{f}_{i,\mathbf{k}\rightarrow i^\prime, \mathbf{k}^\prime}^\prime \right|^2  \\
B_4=& \left|\frac{\mathbf{q}}{m_e} \cdot \mathbf{f}_{i,\mathbf{k}\rightarrow i^\prime, \mathbf{k}^\prime}^\prime \right|^2 \\
B_5=& i\frac{\mathbf{q}}{m_e} \cdot \left[\mathbf{f}_{i,\mathbf{k}\rightarrow i^\prime, \mathbf{k}^\prime}^\prime \times \left(\mathbf{f}_{i,\mathbf{k}\rightarrow i^\prime, \mathbf{k}^\prime}^\prime\right)^*\right]
\end{align}
This factorisation into DM and crystal responses of the total electron excitation rate in crystals is analogous to the factorisation of the total DM-induced ionisation rate in atoms we found in~\cite{Catena:2019gfa}. In App.~\ref{App:Velocity_averaging}, we show that within our simplified treatment of the velocity integral the two vectorial crystal responses, $\mathbf{W}_6$ and $\mathbf{W}_7$, arising from the overlap integrals,
\begin{align}
    \mathbf{B}_6=&f_{i,\mathbf{k}\rightarrow i^\prime, \mathbf{k}^\prime}^\prime \left(\mathbf{f}_{i,\mathbf{k}\rightarrow i^\prime, \mathbf{k}^\prime}^\prime\right)^*\\
\mathbf{B}_7=&\frac{\mathbf{q}}{m_e}\times f_{i,\mathbf{k}\rightarrow i^\prime, \mathbf{k}^\prime}^\prime \left(\mathbf{f}_{i,\mathbf{k}\rightarrow i^\prime, \mathbf{k}^\prime}^\prime\right)^*
\end{align}
are zero. For this reason, they are not included in Eq.~(\ref{eq:W_scalar}).
This is a good approximation for isotropic materials such as silicon and germanium, but will not suffice for anistropic materials such as graphene. For the rest of this paper we will focus on the $5$ responses relevant to silicon and germanium.

Because of the simplified treatment of the velocity integral discussed in the previous section~\cite{Essig:2015cda}, we can now perform the integral over the direction of the transfer momentum $\mathbf{q}$, finding 
\begin{align}
\mathscr{R}_{\textrm{crystal}}&=\frac{n_\chi N_\text{cell} }{128\pi m_\chi^2 m_e^2}\int \mathrm{d} (\ln\Delta E)\int \mathrm{d} q \, q \,\widehat{\eta}\left(q, \Delta E
\right)
\nonumber\\
&\times \sum_{l=1}^r \Re\left(R_l^*(q,v) \overline{W}_l(q,\Delta E)\right)\,,
\label{eq:R_crystal_2D}
\end{align}
where 
\begin{align}
    \overline{W}_l(q,\Delta E)=&\int \mathrm{d}\Omega_q\, W_l(\mathbf{q},\Delta E)\nonumber\\
    =&\frac{1}{q^2} \int \mathrm{d}^3 q^\prime\, W_l(\mathbf{q}^\prime,\Delta E)\delta(|\mathbf{q}^\prime|-q)\label{eq:W_scalar_2D}\,.
\end{align}
We can then use the Dirac delta function in Eq.~\eqref{eq:W_scalar} to perform the integral over ${\rm d^3} q^\prime$, obtaining
\begin{align}
    \overline{W}_l(q,\Delta E)&=(4\pi)^2V_\text{cell}\frac{\Delta E}{q^2}\sum_{\Delta \mathbf{G} ii^\prime}\int_{\textrm{BZ}}\frac{\mathrm{d}^3k }{(2\pi)^3}\int_{\textrm{BZ}}\frac{\mathrm{d}^3k^\prime}{(2\pi)^3} \,B_l \, \nonumber\\
    &\times\delta(\left|\mathbf{k}-\Delta\mathbf{G}-\mathbf{k}^\prime\right|-q) \nonumber\\
    &\times\delta(\Delta E -E_{i\mathbf{k}}+E_{i'\mathbf{k}^\prime})
    \label{eq:W_scalar_2D_2}\,.
\end{align}
which is our final expression for the novel crystal responses to general DM-electron interactions that we implement in the following.

\subsection{Numerical implementation}
\label{sec:num}
The numerical evaluation of the new crystal responses was implemented in \texttt{QEdark-EFT}~\cite{QEdark-EFT}, an extension to the \texttt{QEdark} package \cite{Essig:2015cda}, which interfaces with the plane-wave self-consistent field (PWscf) density functional theory (DFT) code, \texttt{QuantumEspresso v.5.1.2}~\cite{Giannozzi_2009, Giannozzi_2017, doi:10.1063/5.0005082}. 

\begin{figure}[h]
\centering
  \includegraphics[width=0.48\textwidth]{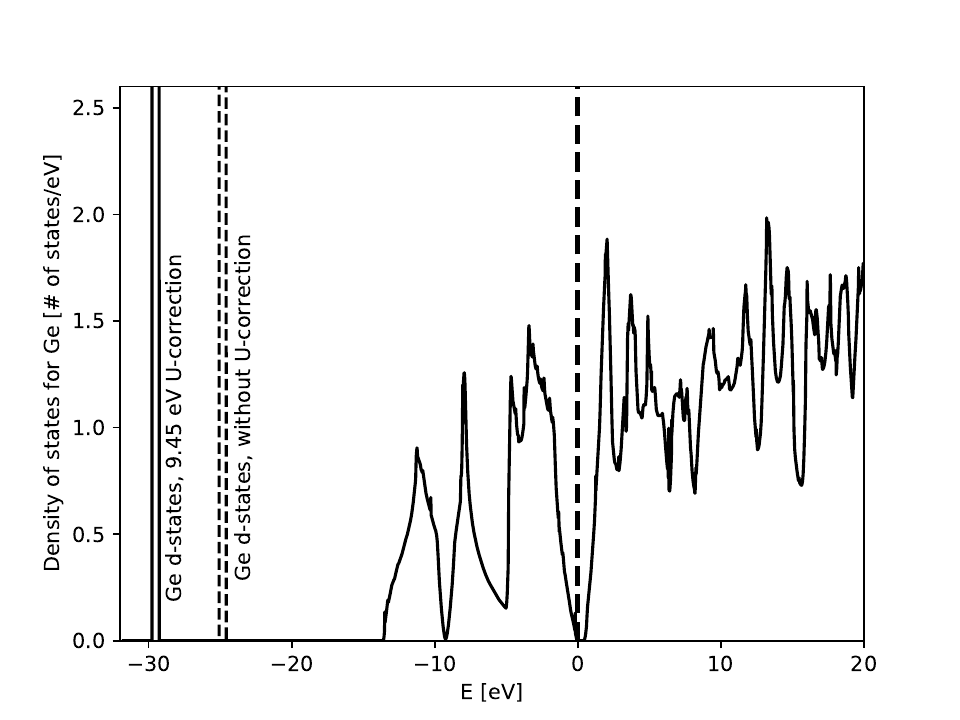}
    \includegraphics[width=0.48\textwidth]{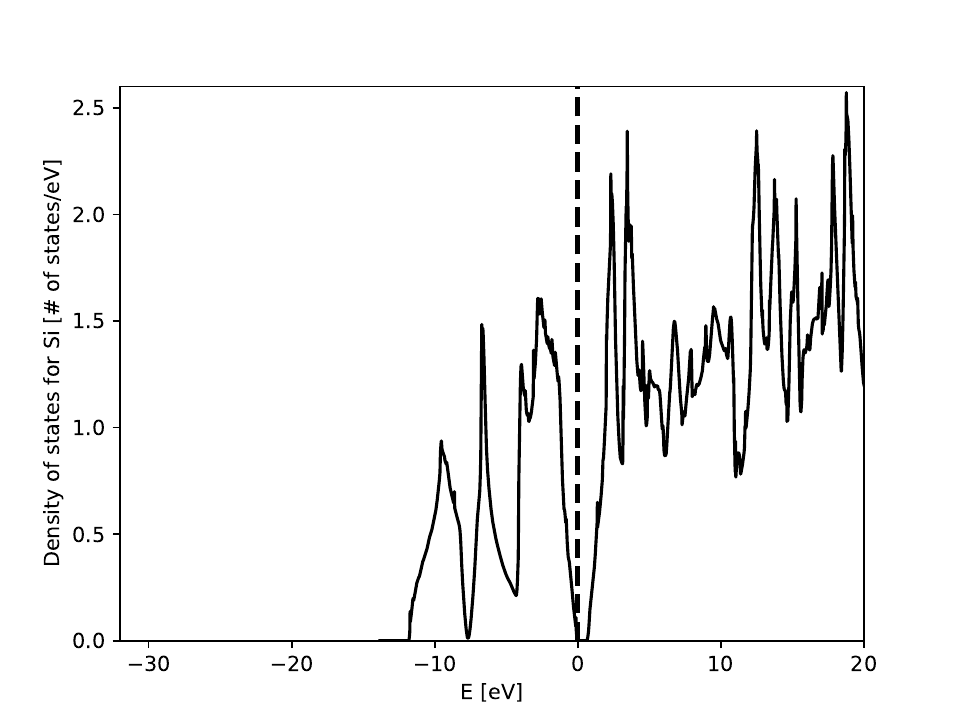}
\caption{Density of states for germanium (upper) and silicon (lower), with the $U$-correction of $9.45\,\mathrm{eV}$ applied to Ge $3d$ orbitals. The top of the valence band is set to $0 \,\mathrm{eV}$ in both cases, and is shown with a dashed vertical line.}
\label{fig:dos}
\end{figure}

For our self-consistent calculations for silicon and germanium we use the \texttt{Si.pbe-n-rrkjus\_psl.0.1.UPF} and \texttt{Ge.pbe-dn-rrkjus\_psl.0.2.2.UPF} pseudopotential provided with the \texttt{QuantumEspresso} package, which include the  $3s^2$, $3p^2$  electrons for silicon and $4s^2$, $4p^2$ and $3d^{10}$ electrons for germanium in the valence configuration. The electron-electron exchange and correlation are treated using the PBE functional~\cite{Perdew/Burke/Ernzerhof:1996}. We sample reciprocal space using a $6 \times 6 \times 6$ Monkhorst-Pack $k$-point grid, supplemented with additional $k$-points at and close to $\Gamma$ and half way to the zone boundary to give a total of 243 $k$ points; this grid was shown in Ref.~\cite{Essig:2015cda} to give good convergence of the calculated scattering cross sections. We take an energy cutoff, $E_\mathrm{cut}$ of $120$ Ry ($1.6$ keV) for silicon and $100$ Ry ($1.4$ keV) for germanium to expand the plane-wave basis, as explained below. We perform our calculations at the established experimental lattice constants of $a_{\mathrm{Si}}=10.3305~\mathrm{a.u.}$ and $a_{\mathrm{Ge}}=10.8171~\mathrm{a.u.}$. With the exception of $E_\mathrm{cut}$, the above parameters are the same as those used previously in Ref.~\cite{Essig:2015cda}. Since the spurious self-interaction within the PBE exchange-correlation functional causes the Ge $3d$ bands to lie $\sim5$ eV higher than observed experimentally~\cite{Gedstates}, we apply a Hubbard $U$ correction with a value of $U_\mathrm{eff} = 9.45$ eV to the Ge $3d$ orbitals using the approach of Cococcioni and de Gironcoli~\cite{PhysRevB.71.035105}. This has the effect of shifting the narrow Ge $3d$ band rigidly down in energy by $\sim \frac{U_\mathrm{eff}}{2}$ so that its position below the Fermi level ($\sim30$ eV) is consistent with experimental observations. Our resulting densities of states (Fig.~\ref{fig:dos}) show the usual DFT underestimation of the band gaps, which we correct in our response calculations using a scissors correction to set the band gap of silicon (germanium) to the experimental value of $1.2\,\mathrm{eV}$ ($0.67\,\mathrm{eV}$).

In order to numerically evaluate Eq.~(\ref{eq:W_scalar_2D_2}) we discretise it by introducing binning in $q$ and $\Delta E$,
\begin{align}
\overline{W}_l(q_n,\Delta E_m)&=\int_{q_n-\frac{1}{2}\delta q}^{q_n+\frac{1}{2}\delta q} \frac{\mathrm{d} q^\prime}{ \delta q} \int_{\Delta E_m-\frac{1}{2}\delta E}^{\Delta E_m + \frac{1}{2}\delta E} \frac{\mathrm{d} \Delta E^\prime}{\delta E}\nonumber\\
&\times
\overline{W}_l(q^\prime,\Delta E^\prime)\,,
\end{align}
where $q_n$ and $\Delta E_m$ are the central values of the $n^\mathrm{th}$ $q$-bin and $m^\mathrm{th}$ $\Delta E$-bin respectively. We use $2000$ bins in $q$ and $\Delta E$, letting $q$ take values between $0$ and $40.8\,\mathrm{keV}$ ($37.3\,\mathrm{keV}$) for silicon (germanium), and $\Delta E$ take values between $0$ and $85\,\mathrm{eV}$. From  Eq.~(\ref{eq:vmin}) we see that the minimum velocity required to cause excitation with these highest values of $\Delta E$ and $q$ is $625 \,\mathrm{km}/\mathrm{s}$ ($684 \,\mathrm{km}/\mathrm{s}$), well below the maximum dark matter velocity, $v_\mathrm{esc}+v_\oplus$. 
As in Ref.~\cite{Essig:2015cda}, we replace the $k$-integral with a discrete mesh and numerically evaluate 
\begin{align}
\overline{W}_l(q_n,E_m)&=\frac{2V_\mathrm{BZ}E_m}{\pi q_n^2\delta E\delta q }\sum_{\mathbf{k},\mathbf{k}^\prime} \sum_{i,i^\prime} \sum_{\Delta\mathbf{G}} \frac{w_{\mathbf{k}}}{2} \frac{w_{\mathbf{k}^\prime}}{2}B_l\nonumber \\
&\times
 \Theta \left(1 - \frac{\left|| \mathbf{k}^\prime -\mathbf{k} + \Delta\mathbf{G}| -q_{n}\right|}{\frac{1}{2}\delta q} \right)\nonumber \\
&\times \Theta \left(1 -\frac{\left| E_{i^\prime, \mathbf{k}^\prime} - E_{i, \mathbf{k}} -E_m \right|}{\frac{1}{2} \delta E} \right) \,,
\end{align}
where the $k$-sums go over the $243$ $k$-point grid mentioned above, each with weight $\omega_\mathbf{k}$, such that $\sum_k \omega_\mathbf{k}=2$. The lattice vectors $\Delta\mathbf{G}$ satisfy the cutoff relation 
\begin{equation}
    \frac{\left|\mathbf{k}+\mathbf{G}\right|^2}{2m_e}\leq E_\mathrm{cut}\,,
\end{equation}
causing $q$ to have a cutoff roughly at $\sqrt{2m_e E_\mathrm{cut}}$, where $E_\mathrm{cut}$ is our plane-wave energy cutoff described above. Our values of $E_\mathrm{cut}=1.6 \,\mathrm{keV}$ ($E_\mathrm{cut}=1.4 \,\mathrm{keV}$) for silicon (germanium) correspond to cutoffs in $q$ of $\sim 41\,\mathrm{keV}$ ($\sim 37\,\mathrm{keV}$). Note that these values are considerably larger than those used by \cite{Essig:2015cda}, which is necessary for two reasons: First, some of the new interactions that we consider here, in which the DM response functions $R_l(q,v)$ depend on positive powers of $q$, means that the integrand in the crystal excitation rate formula, Eq.~(\ref{eq:R_crystal}), is significantly different from zero for higher $q$-values than in the previously studied case of the dark photon model. And second, because we cover a larger range of deposited energies.
The higher energy cutoffs require in turn inclusion of a larger number of bands in our DFT calculation; we include 102 unoccupied bands, which again is almost twice as many as in Ref.~\cite{Essig:2015cda}.

Note that, as in Ref.~\cite{Essig:2015cda} our transition matrix elements are calculated between the Kohn-Sham pseudo-wavefunctions for the occupied and empty states. It will be an interesting future direction to evaluate the effect of the use of all-electron rather than pseudo wavefunctions on the calculated crystal excitation rates for our novel response functions, as was recently done in Ref.~\cite{Griffin:2021znd} for the previously known $W_1$ response.

\begin{figure*}
\centering
  \centering
  \includegraphics[width=0.48\textwidth]{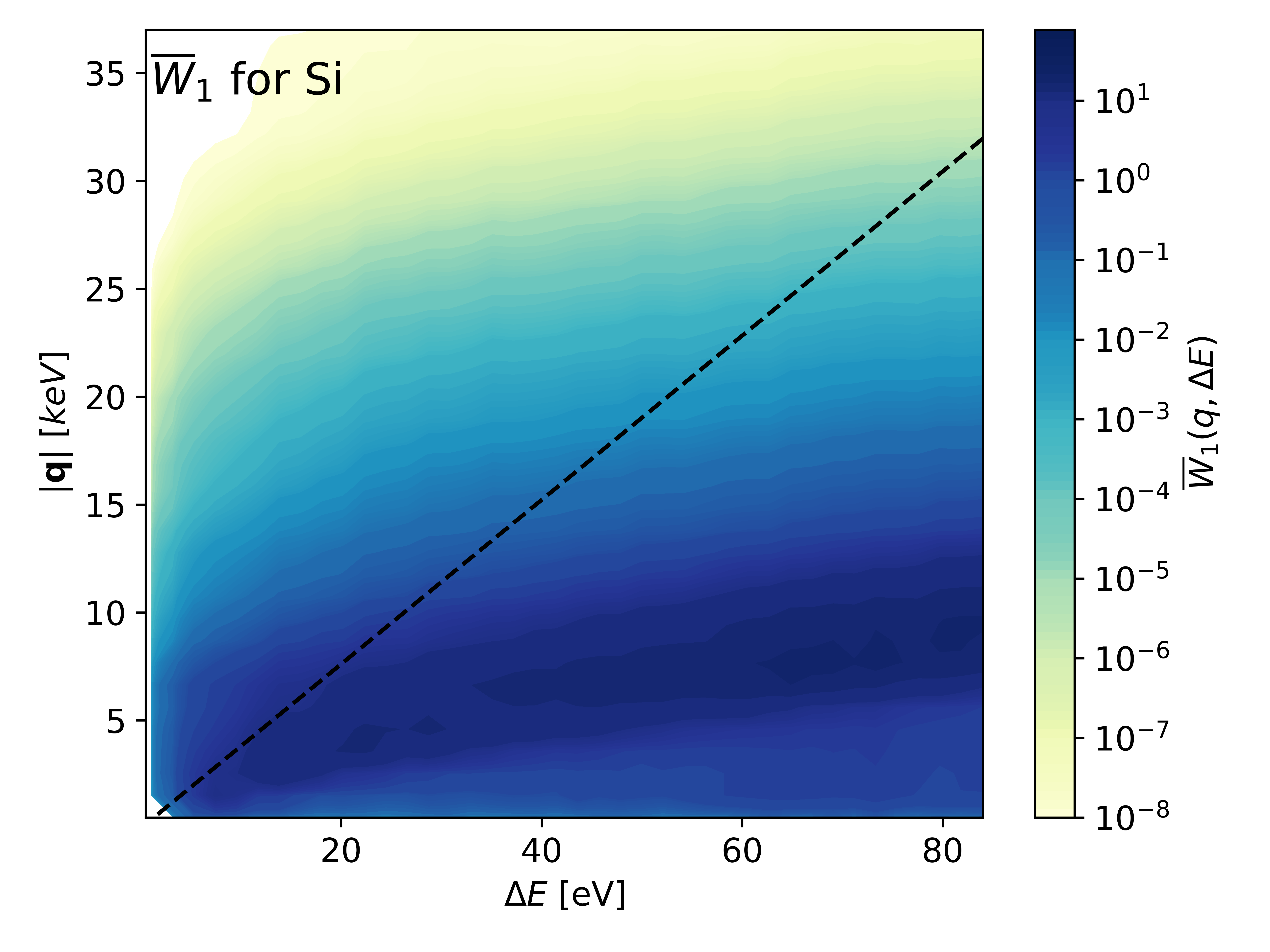}
  \includegraphics[width=0.48\textwidth]{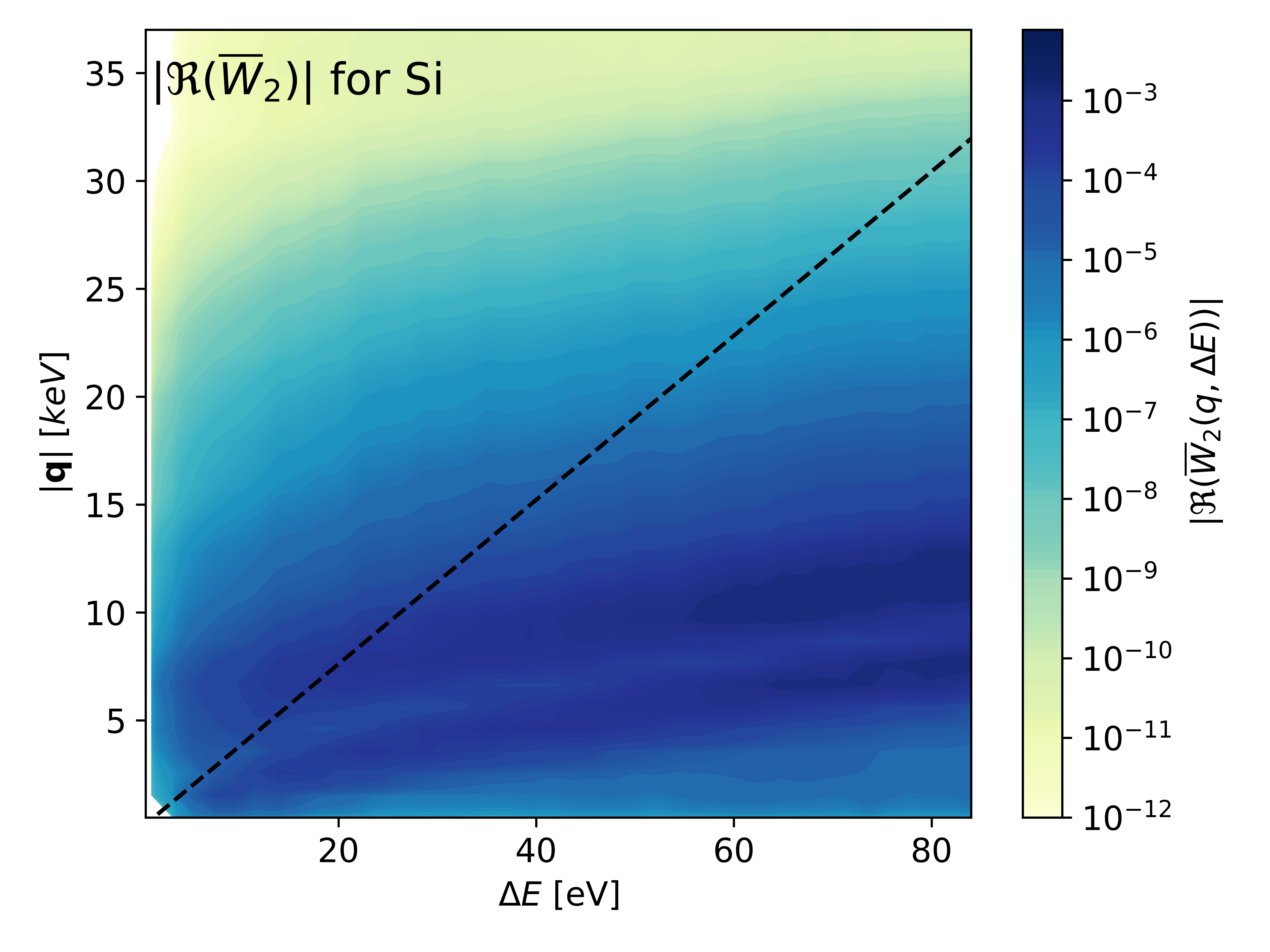}
  \includegraphics[width=0.48\textwidth]{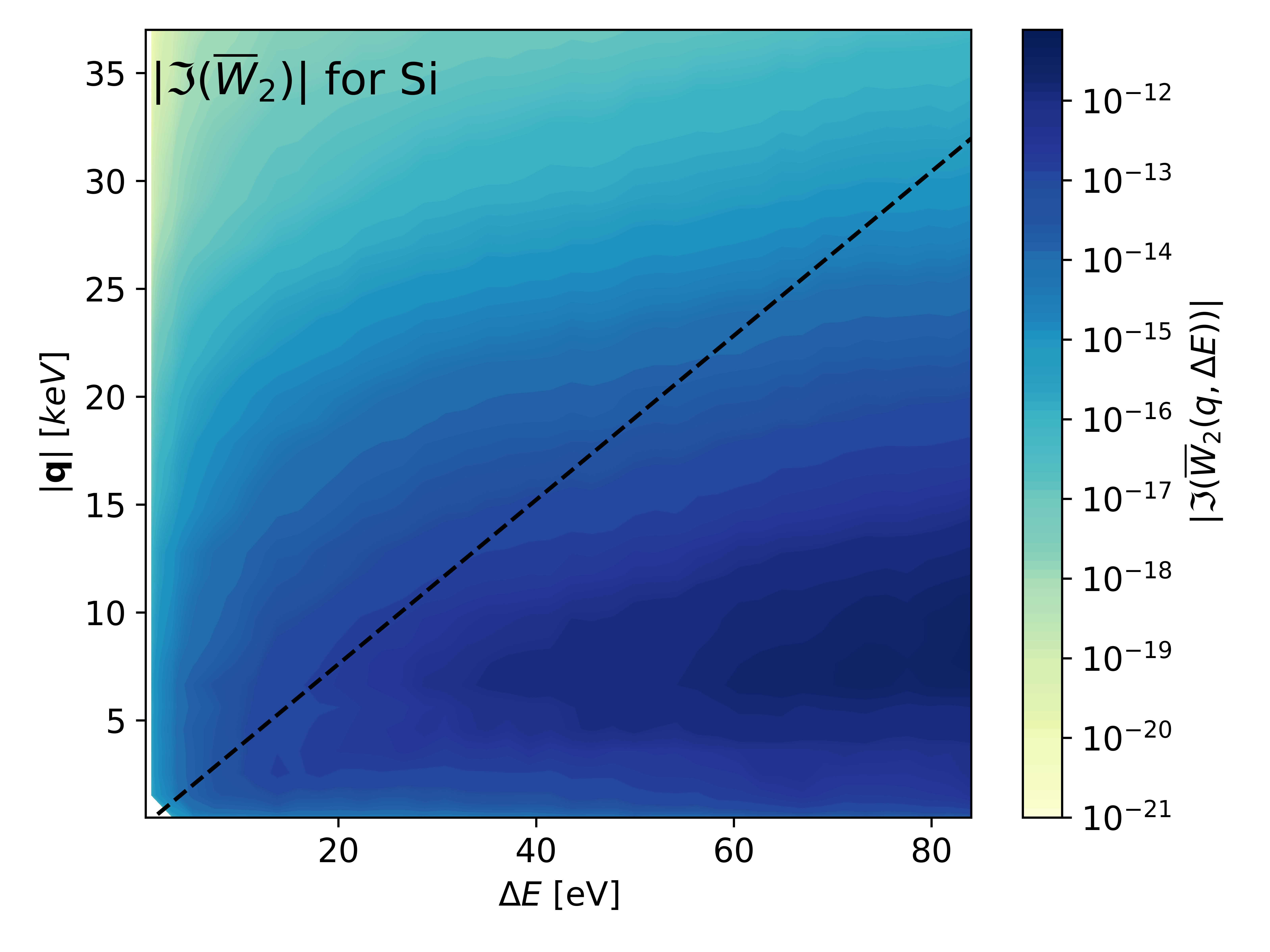}
  \includegraphics[width=0.48\textwidth]{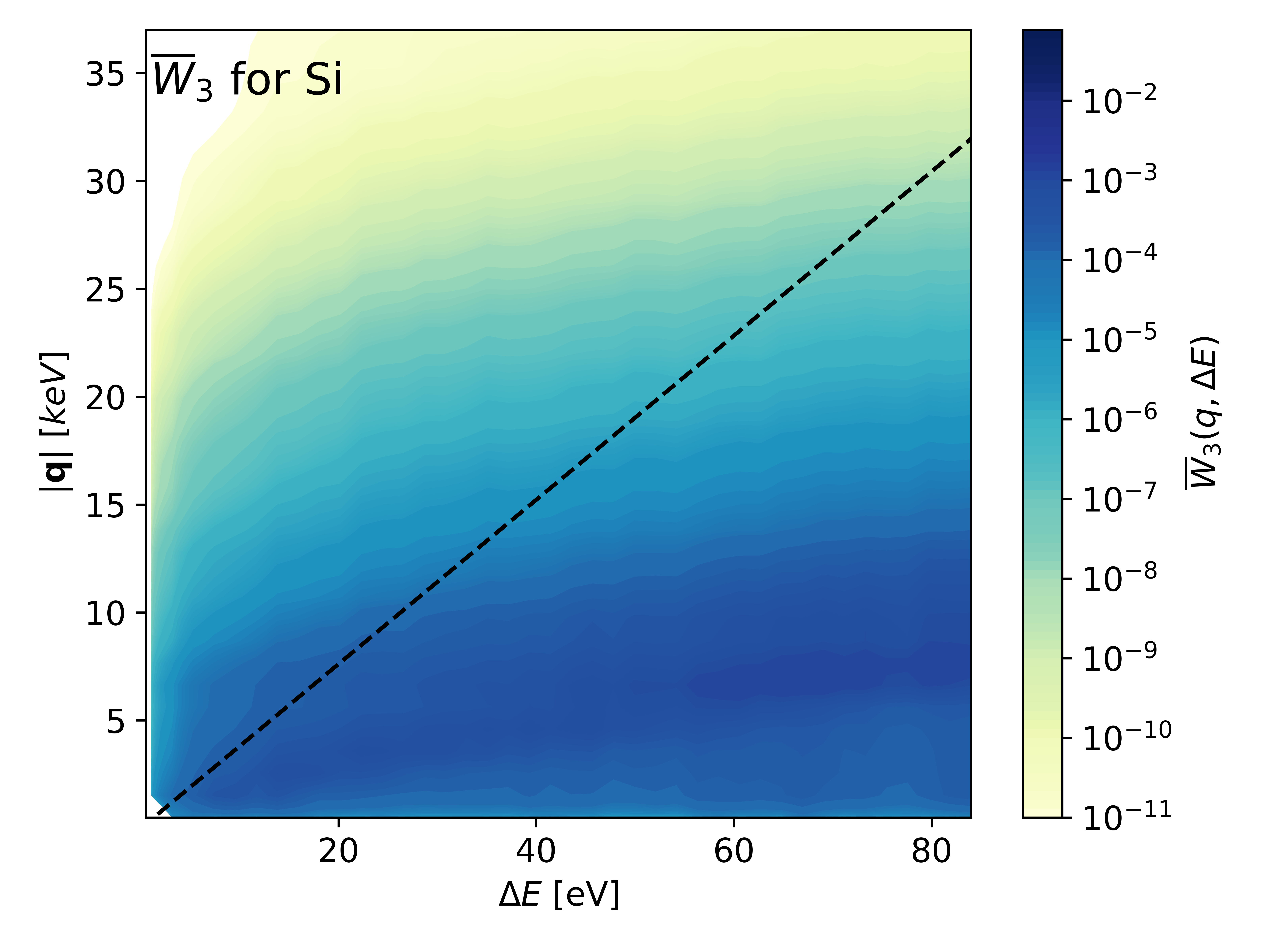}
  \includegraphics[width=0.48\textwidth]{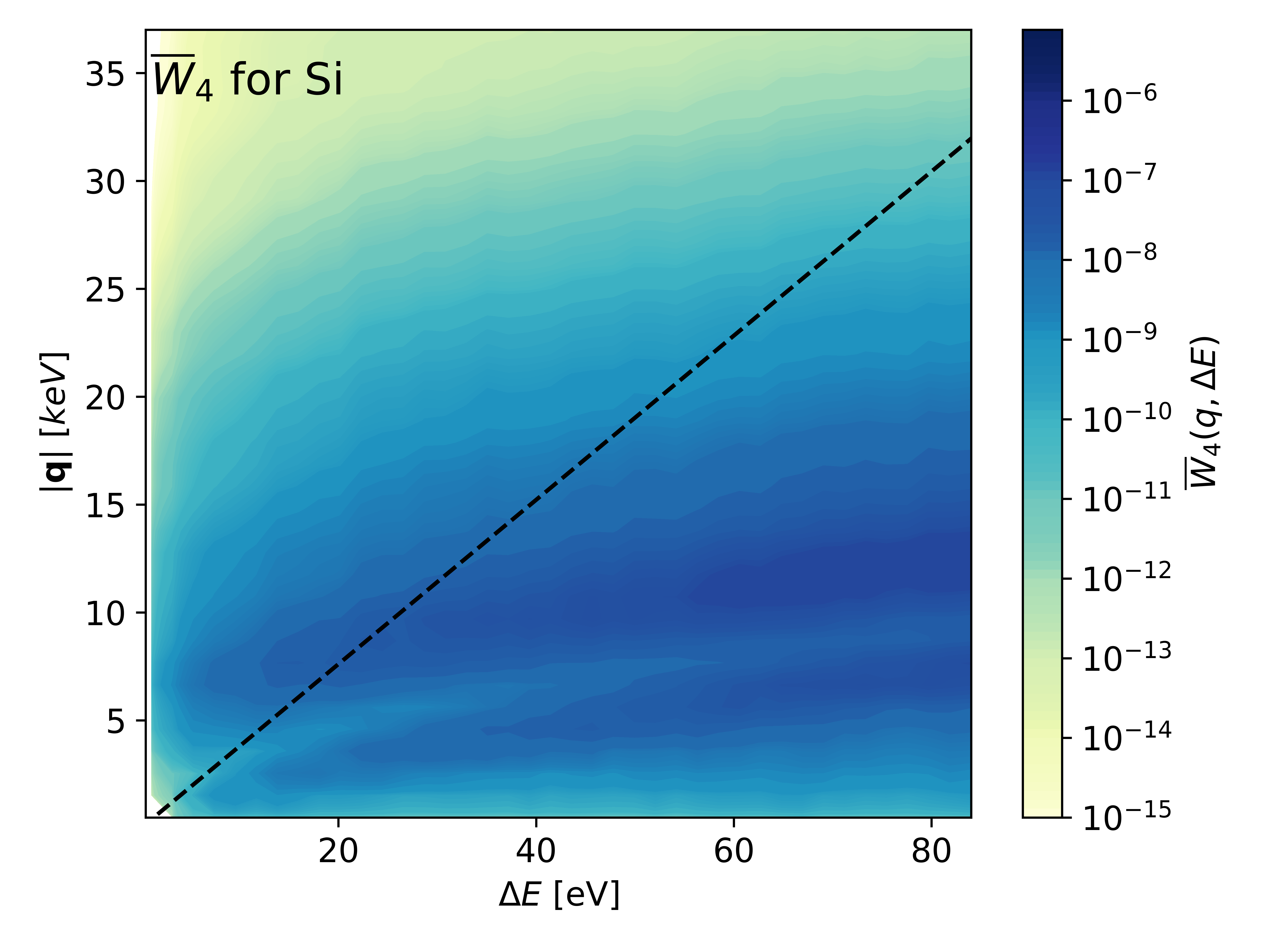}
  \includegraphics[width=0.48\textwidth]{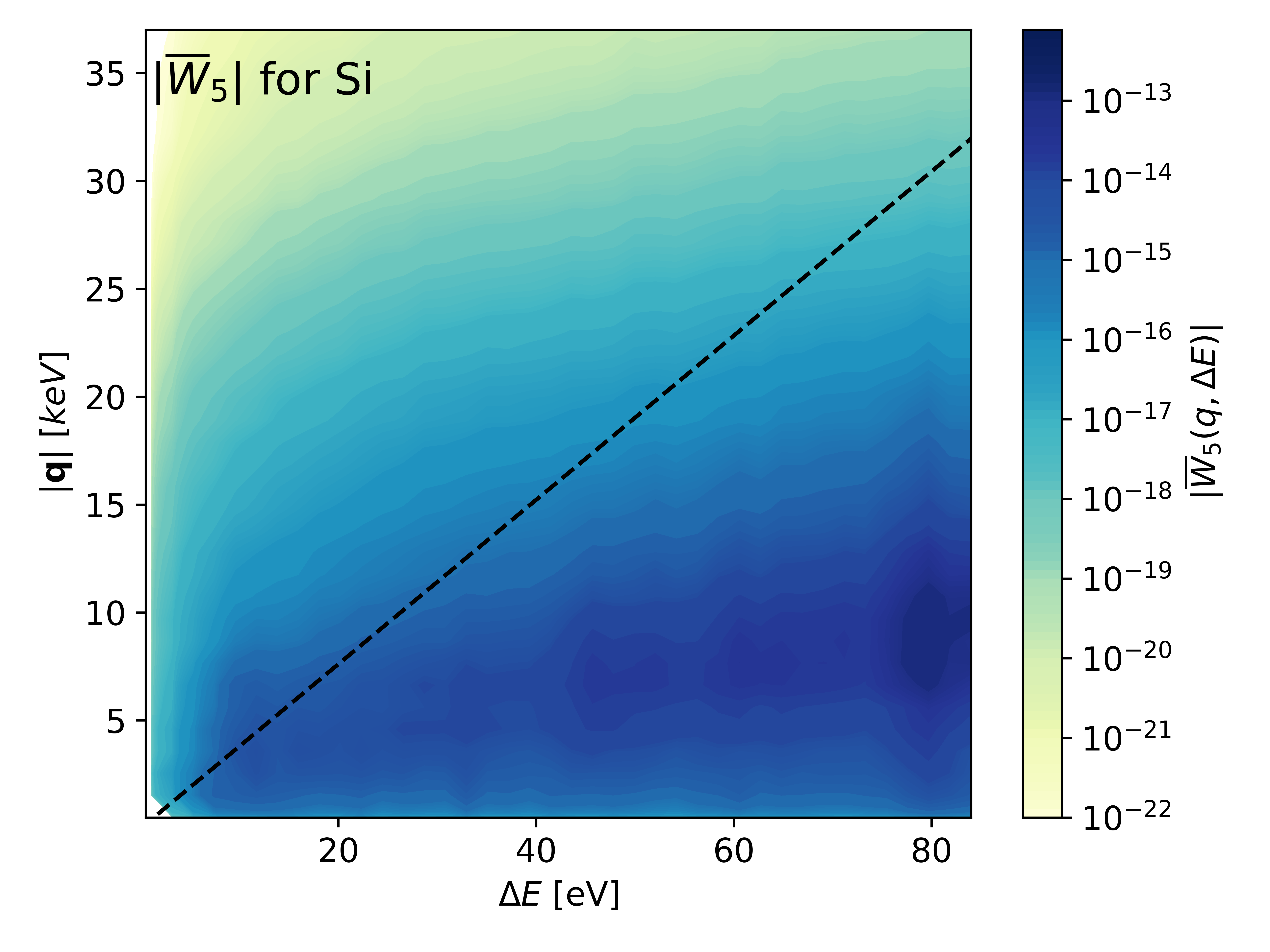}
\caption{Crystal response functions for silicon as a function of $q$ and $\Delta E$. Note that the color bars vary between the plots. $\overline{W}_2$ and $\overline{W}_5$ can take negative values, and we therefore show their absolute values. The region below the dashed line is kinematically inaccessible for $v_\mathrm{esc}=544\,\mathrm{km/s}$ and $v_E=244\,\mathrm{km/s}$.}
\label{fig:Silicon}
\end{figure*}

\begin{figure*}
\centering
  \includegraphics[width=0.48\textwidth]{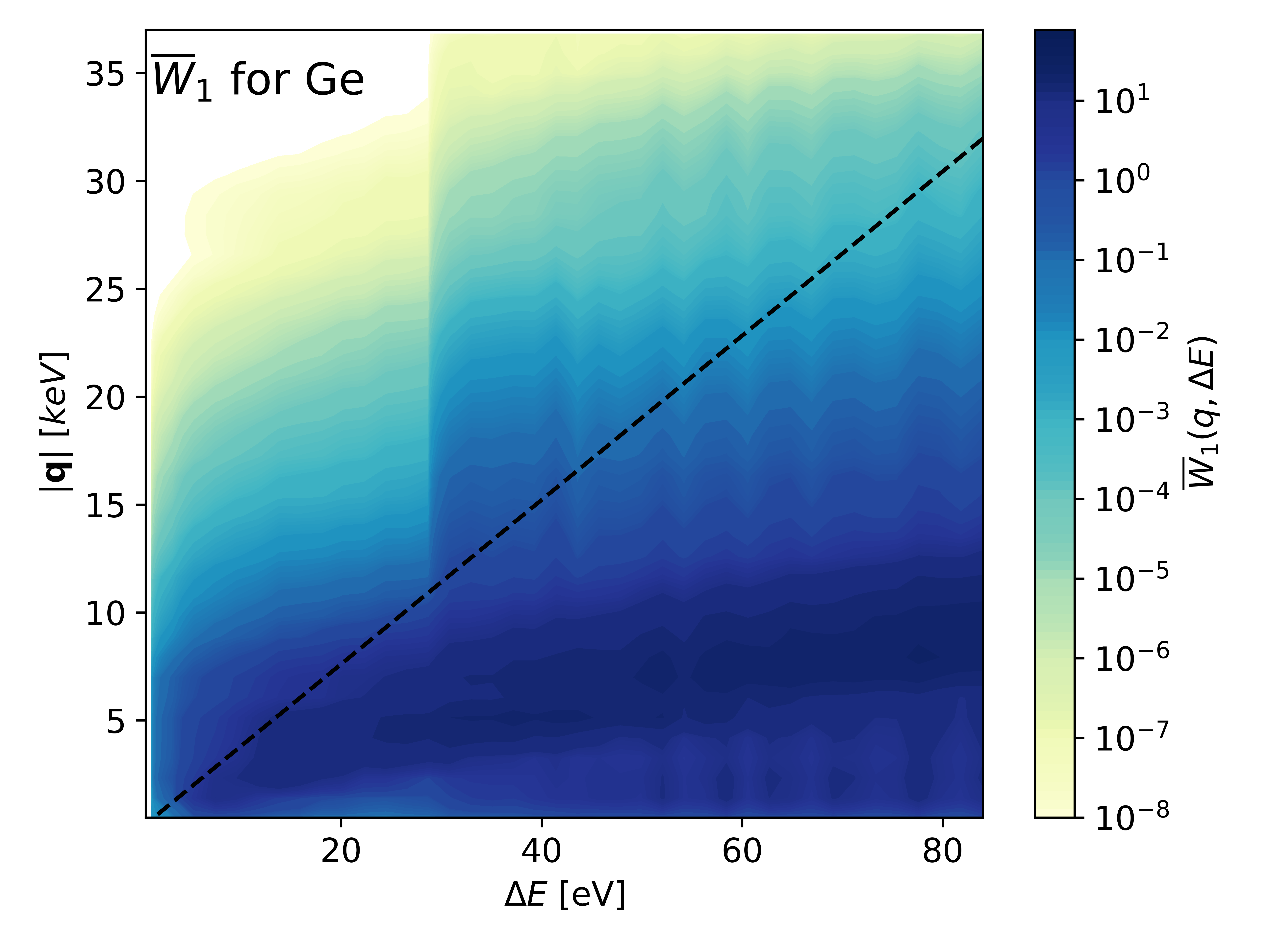}
  \includegraphics[width=0.48\textwidth]{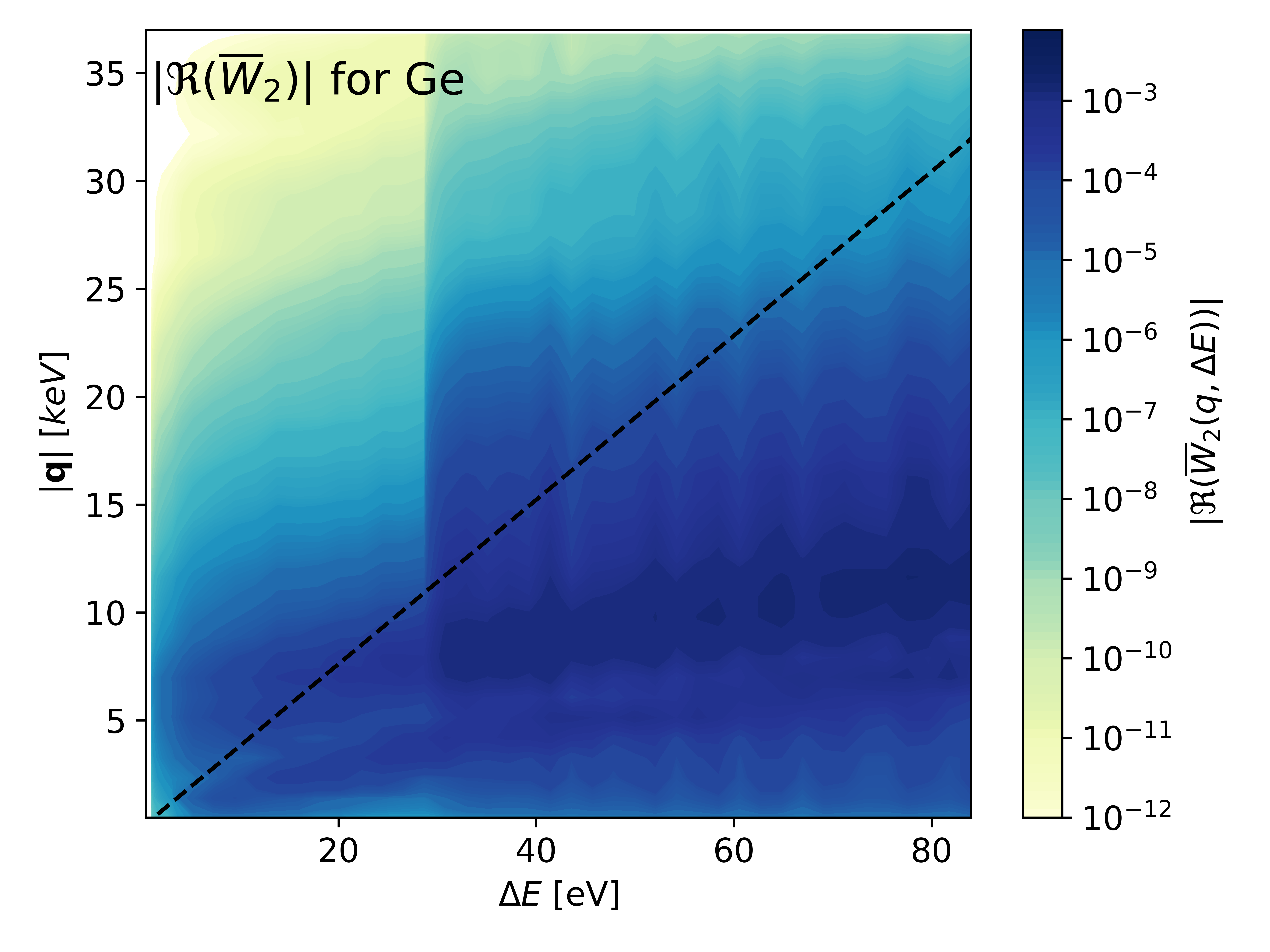}
  \includegraphics[width=0.48\textwidth]{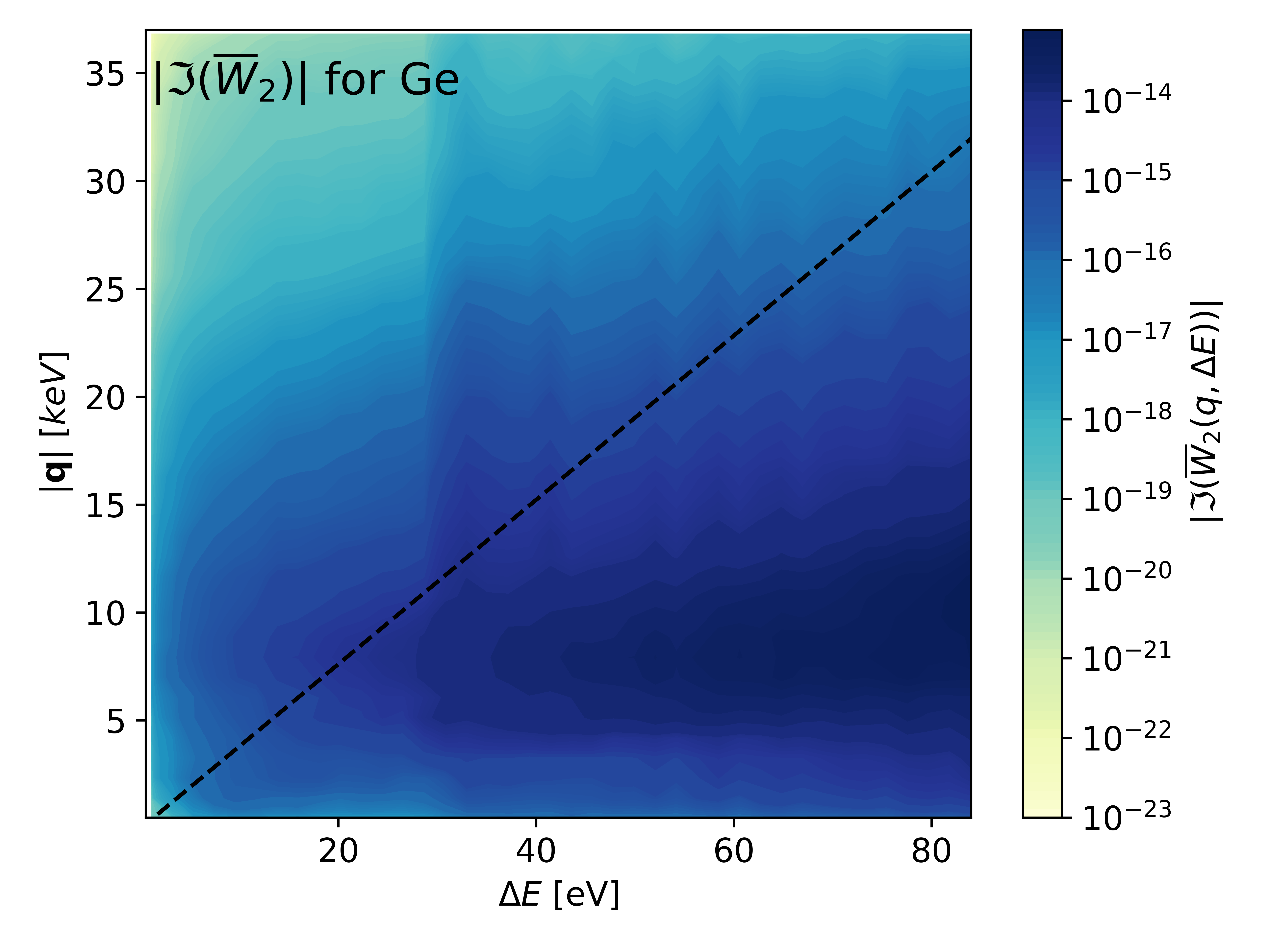}
\includegraphics[width=0.48\textwidth]{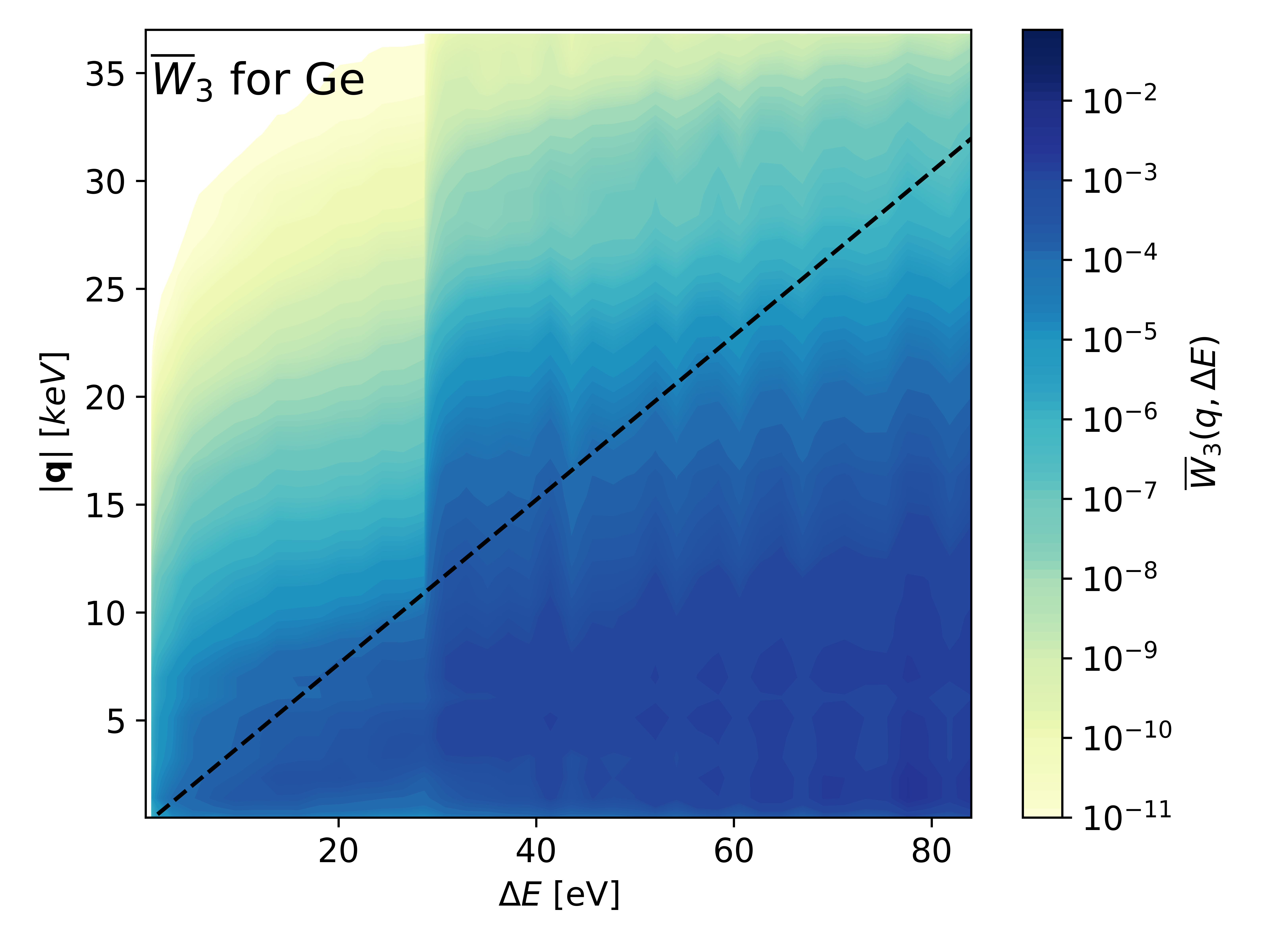}
  \includegraphics[width=0.48\textwidth]{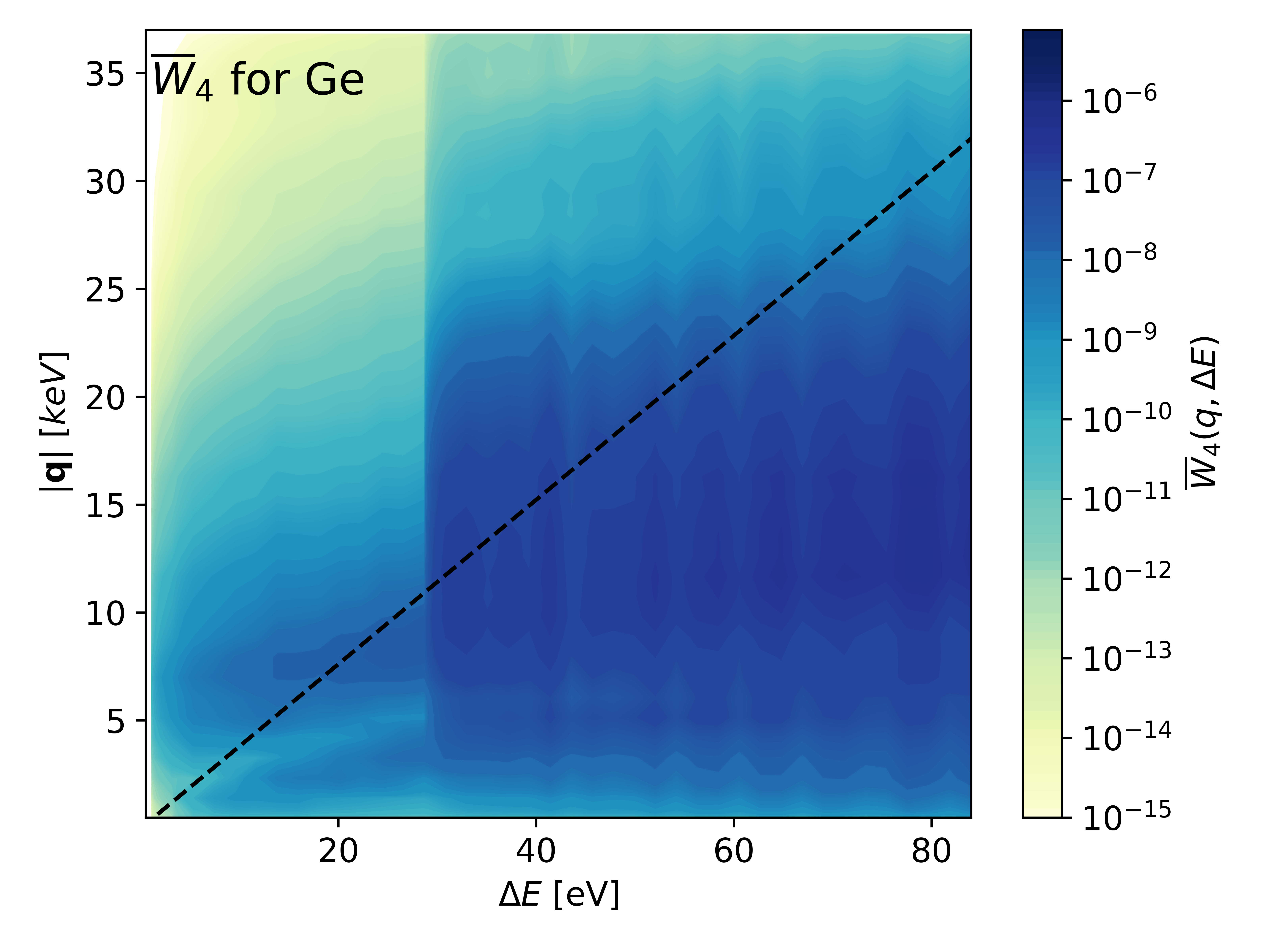}
  \includegraphics[width=0.48\textwidth]{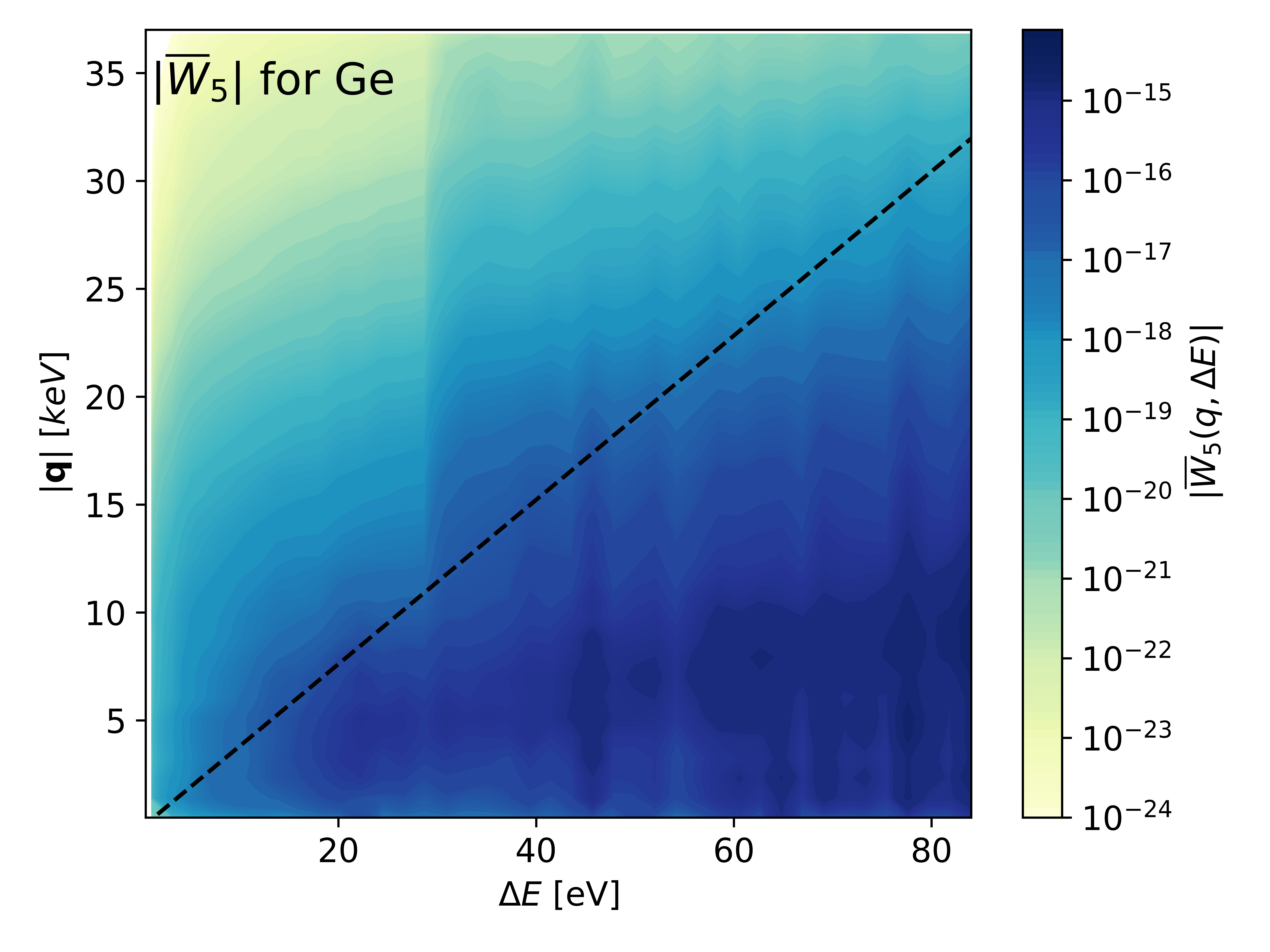}
\caption{Crystal response functions for germanium. For $\Delta E>30 \,\mathrm{eV}$ and $q>10\,\mathrm{keV}$ we see the influence of the $3d$ electrons.}
\label{fig:Germanium}
\end{figure*}

\begin{figure*}
    \centering
    \includegraphics[width=1\textwidth]{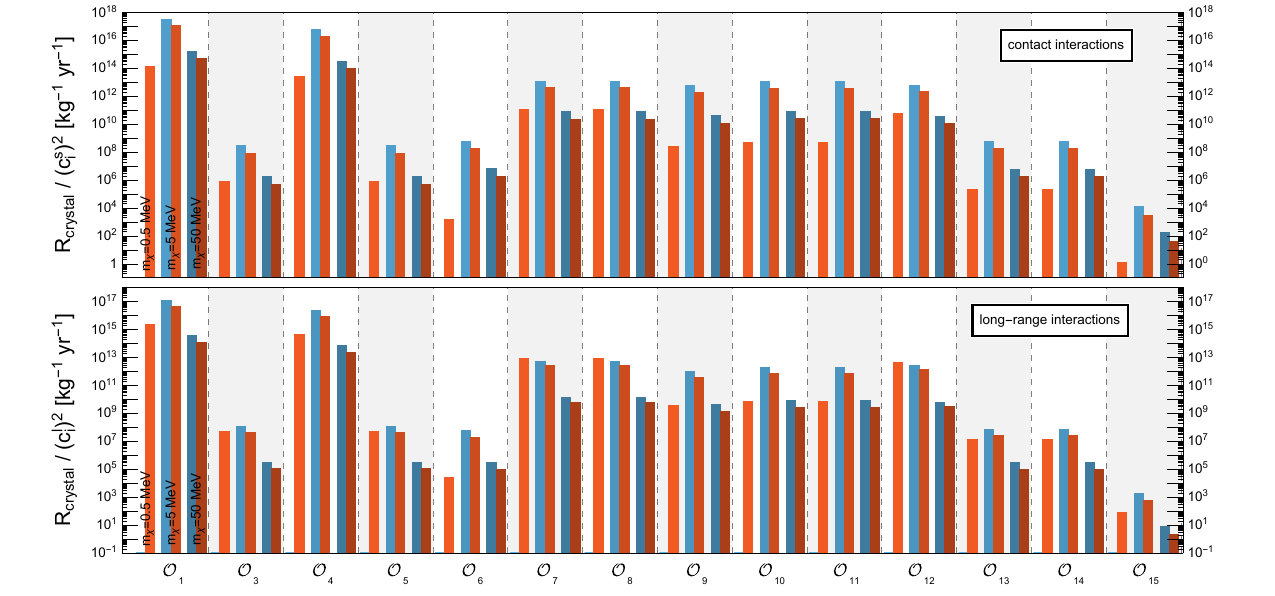}
    \caption{Rate contributions from operators $\mathcal{O}_1$ to $\mathcal{O}_{15}$ of Table~\ref{tab:operators} for silicon (blue) and germanium (red), with DM masses of $0.5\,\mathrm{MeV}$ (left), $5\,\mathrm{MeV}$ (middle) and $50\,\mathrm{MeV}$ (right). For $m_\chi=0.5\,\mathrm{MeV}$ the expected excitation rate in silicon is $0$ since the gravitationally bound dark matter particles have too low energy to overcome the larger band-gap.
    \label{fig:Total_Rate_Comparison}}
\end{figure*}

\begin{figure*}
\centering
  \includegraphics[width=0.48\textwidth]{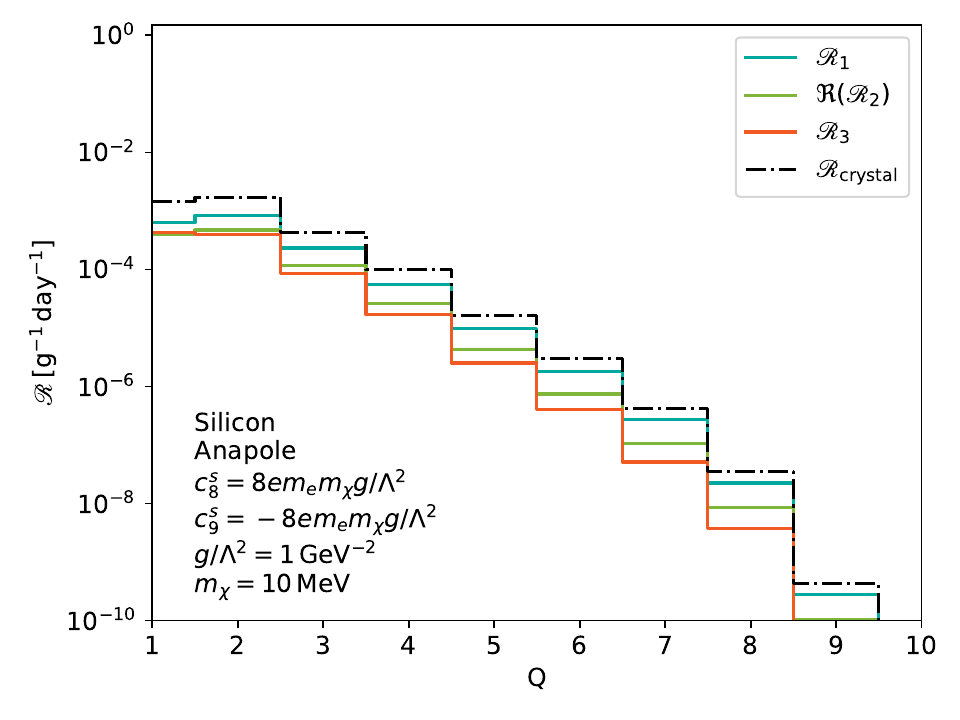}
  \includegraphics[width=0.48\textwidth]{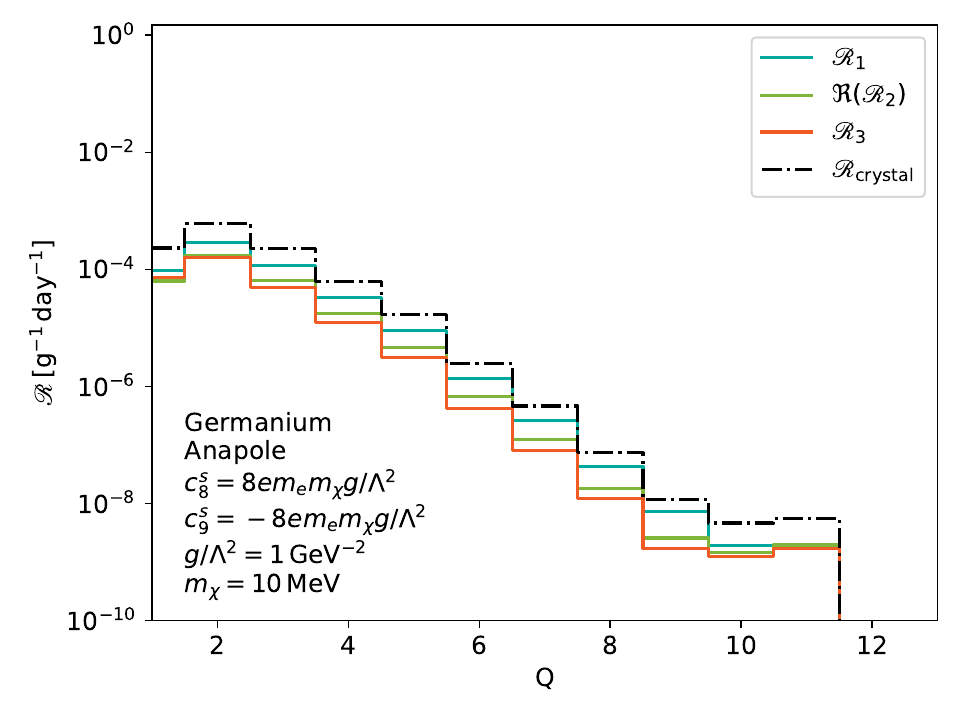}
  \includegraphics[width=0.48\textwidth]{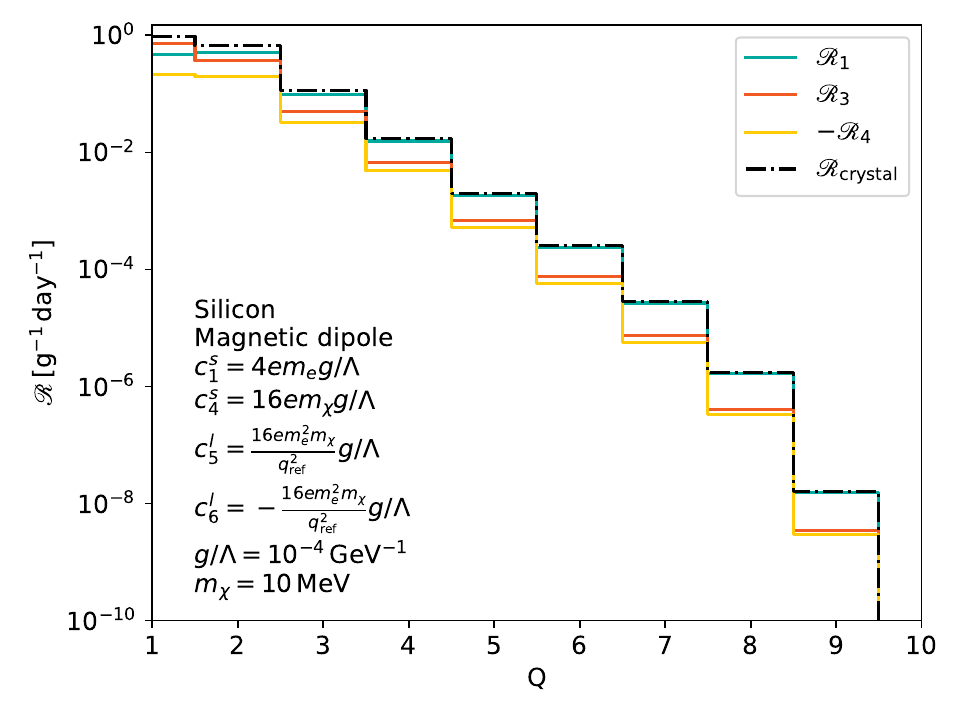}
  \includegraphics[width=0.48\textwidth]{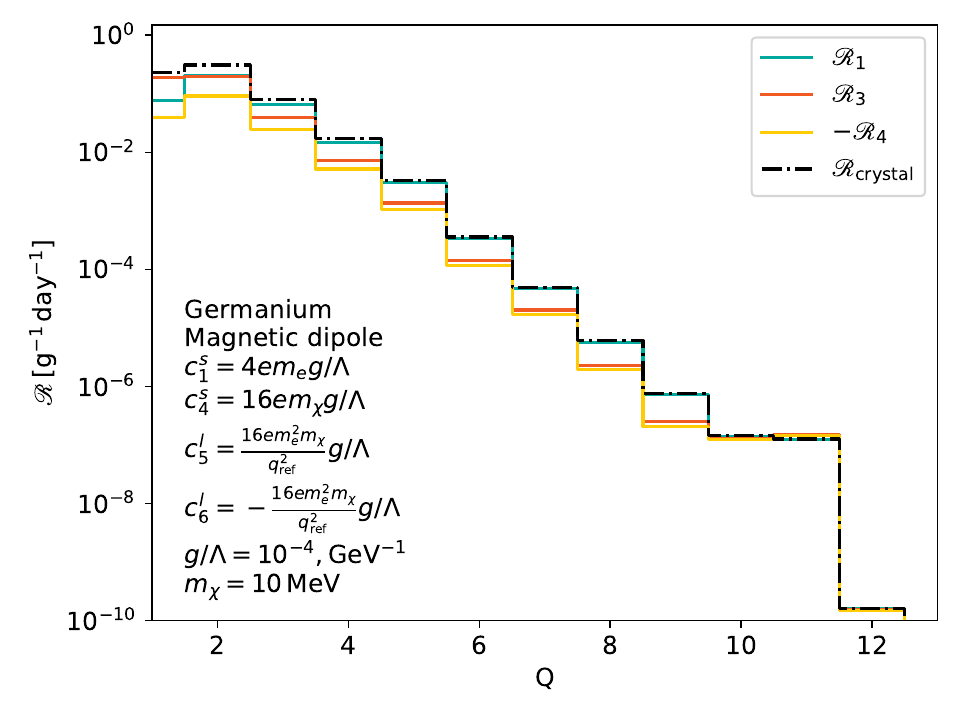}
  \includegraphics[width=0.48\textwidth]{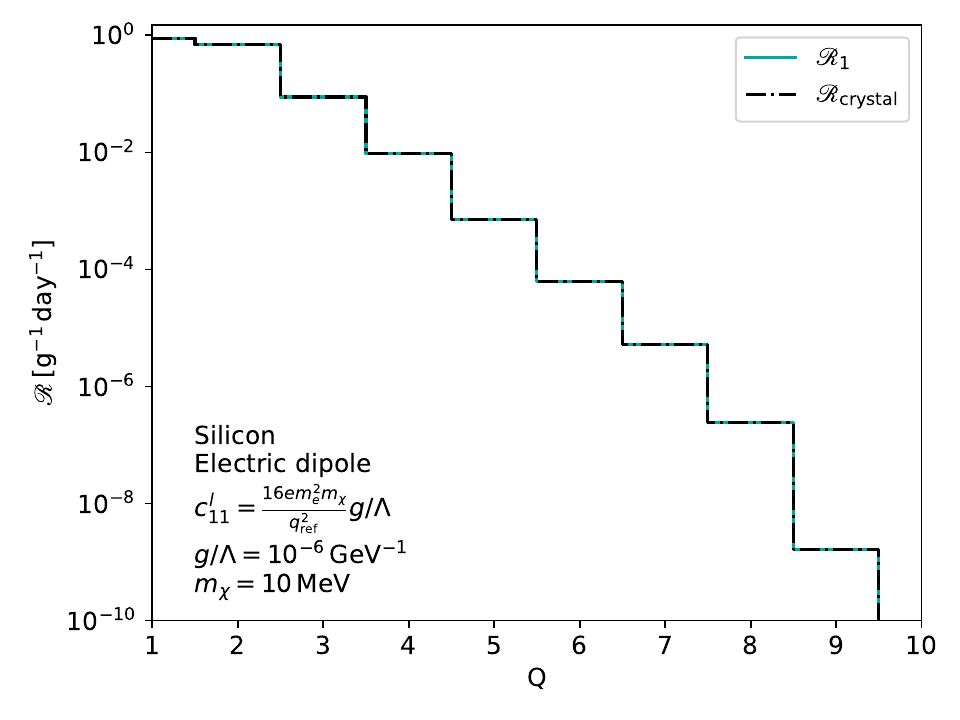}
  \includegraphics[width=0.48\textwidth]{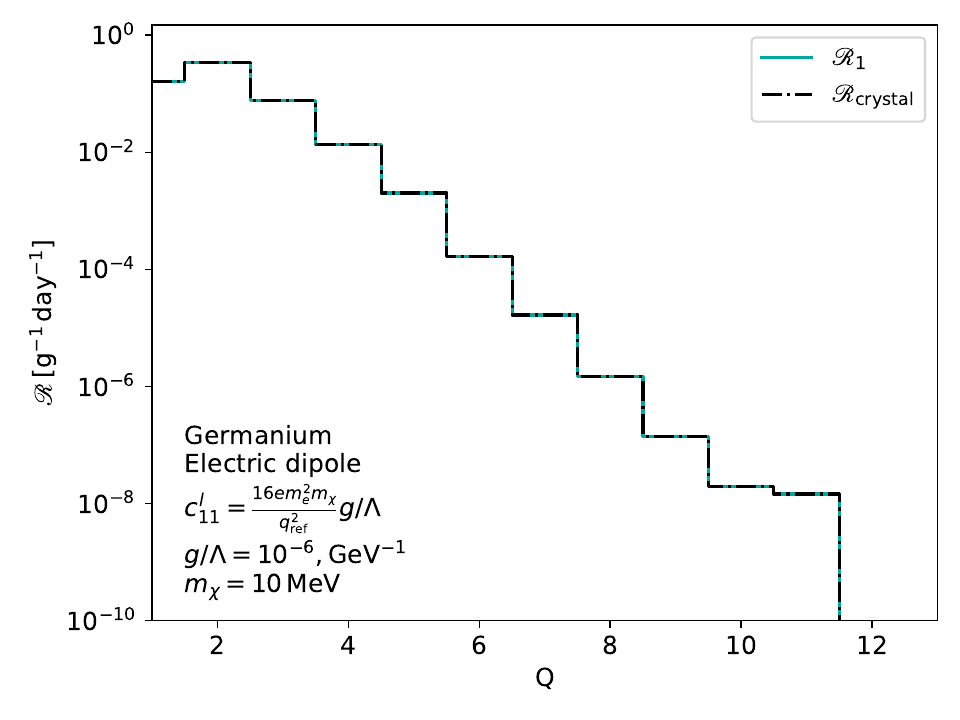}
\caption{Expected excitation rates, $\mathscr{R}$, including the contributions from the individual non-zero responses in each case, as a function of number of electron-hole pairs $Q$. Silicon (germanium) is shown on the left (right), and the excitation rates for the anapole, magnetic dipole and electric dipole interactions are shown in the top, middle and lower panels, respectively.  }
\label{fig:Rate_contribution_poles}
\end{figure*}
\begin{figure}[!h]
\centering
  \includegraphics[width=0.48\textwidth]{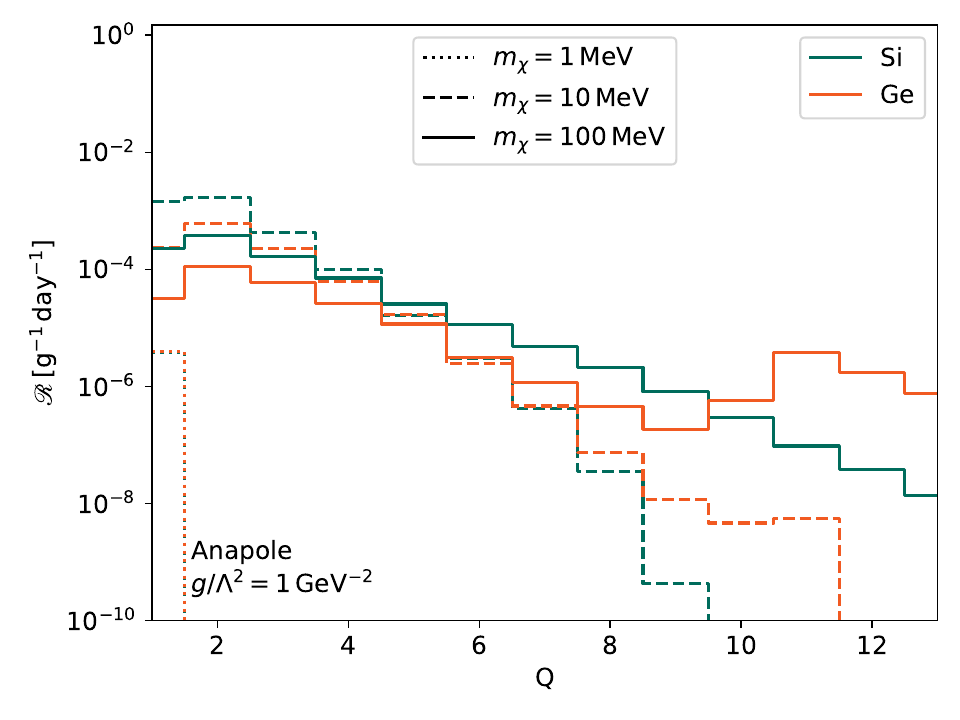}
  \includegraphics[width=0.48\textwidth]{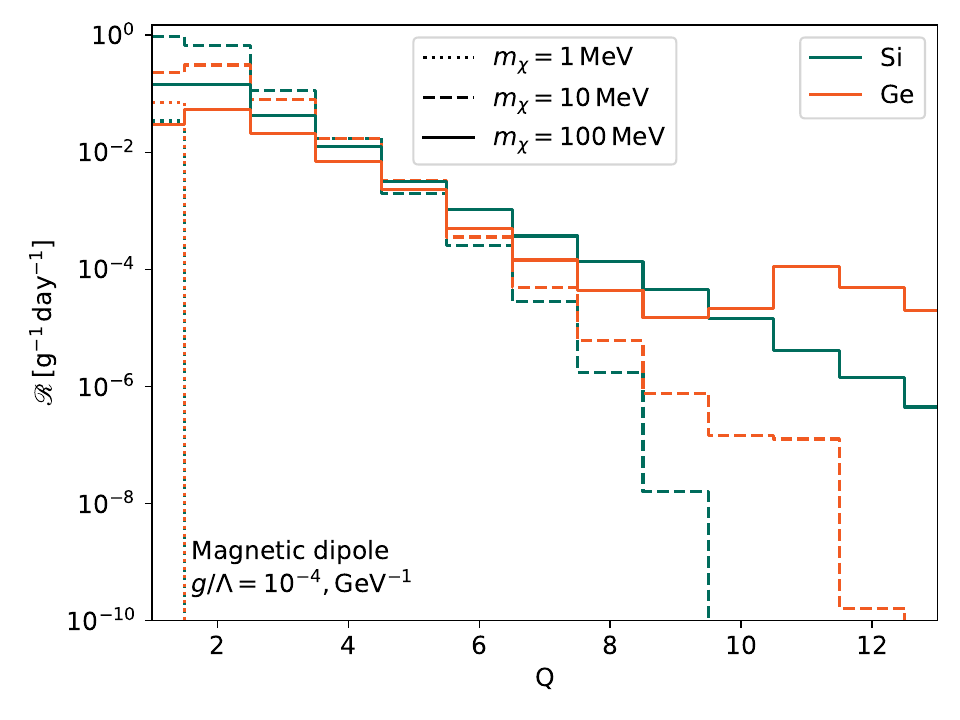}
  \includegraphics[width=0.48\textwidth]{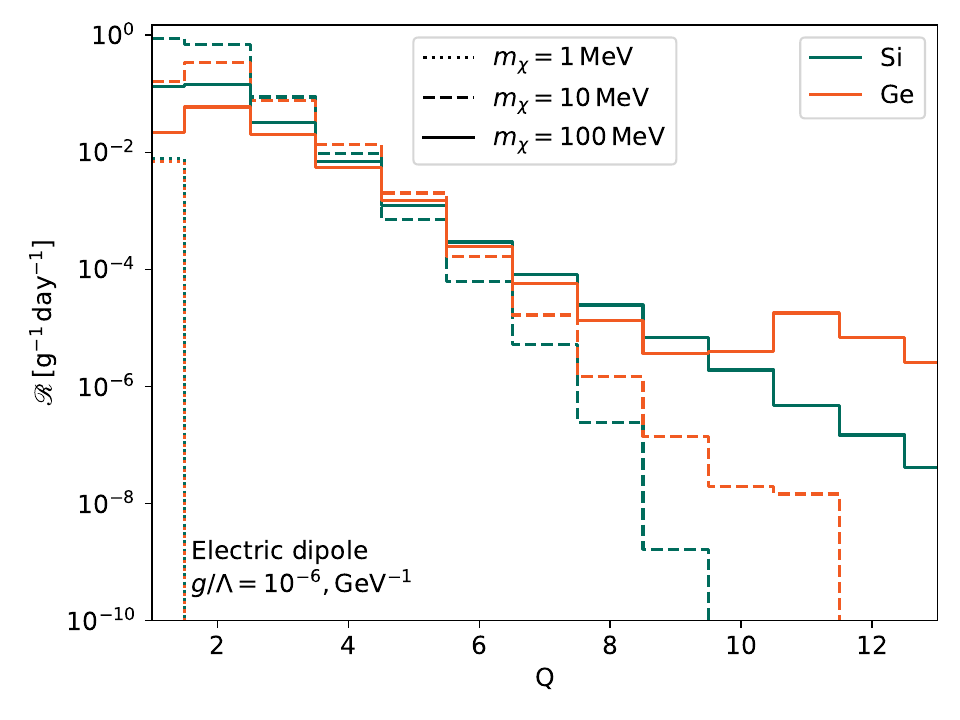}
\caption{Expected excitation rates in silicon and germanium for $m_\chi=1\,\mathrm{MeV}$, $m_\chi=10\,\mathrm{MeV}$ and $m_\chi=100\,\mathrm{MeV}$.}
\label{fig:Multimass_poles}
\end{figure}

\begin{figure*}
\centering
  \includegraphics[width=0.48\textwidth]{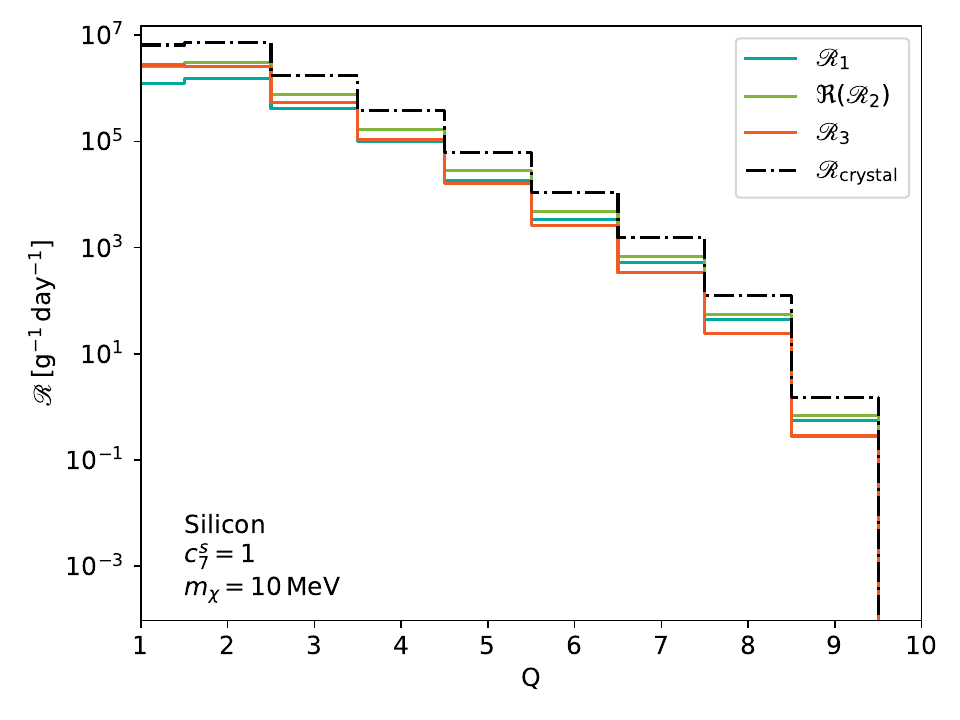}
  \includegraphics[width=0.48\textwidth]{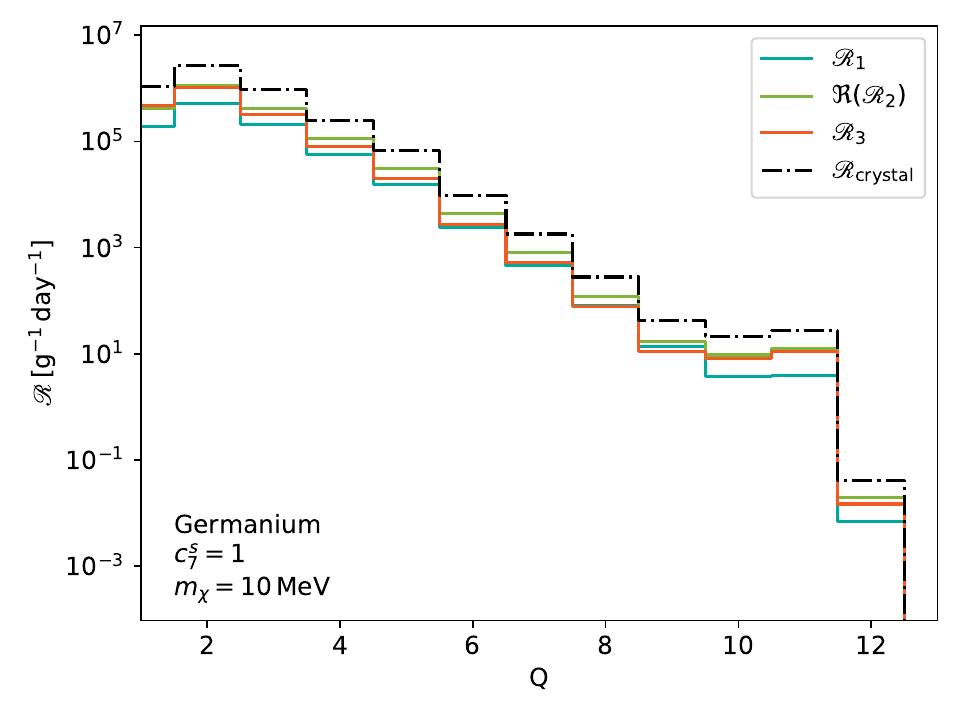}
  \includegraphics[width=0.48\textwidth]{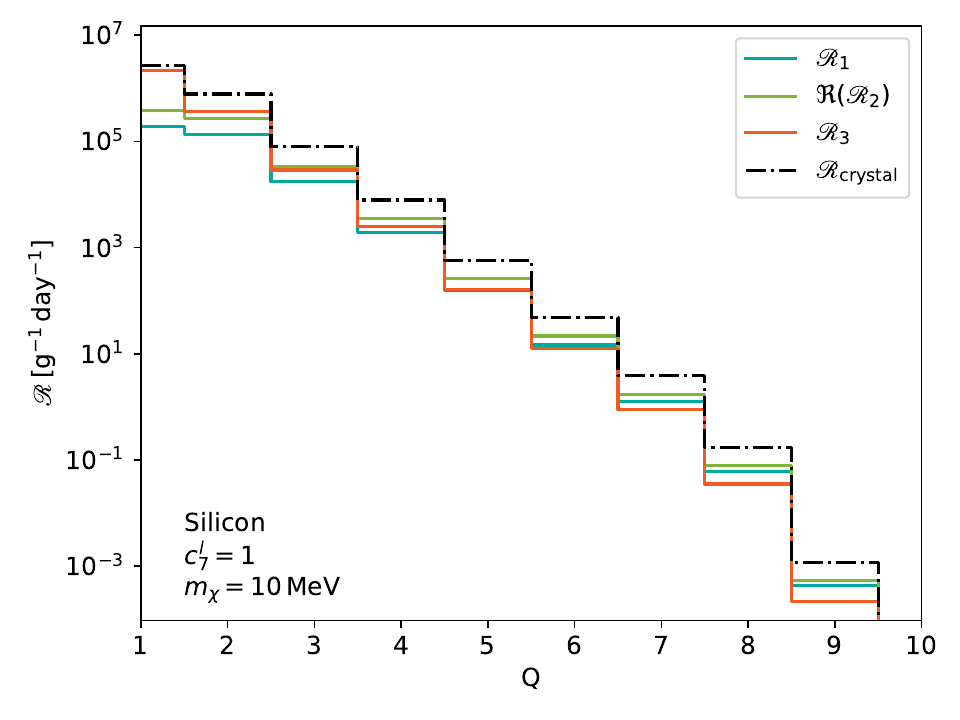}
  \includegraphics[width=0.48\textwidth]{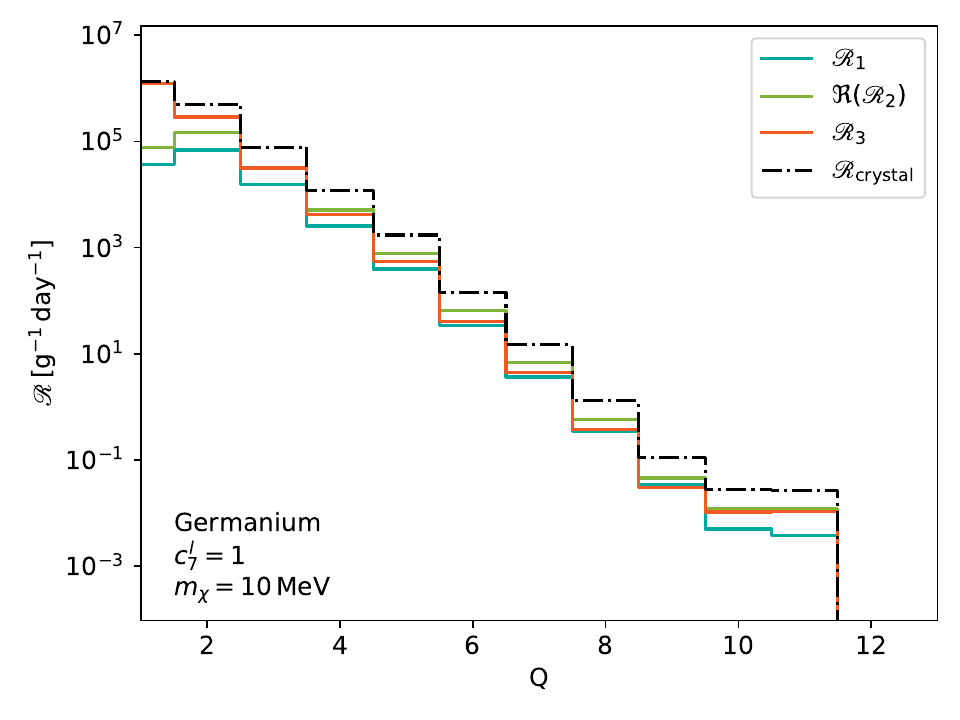}
\caption{Expected excitation rates for $c^s_7=1$ or $c^\ell_7=1$ for silicon (left panels) and germanium (right panels). The contact interactions are shown in the top panels whereas the long-range interactions are shown in the bottom panels. Here, all rate contributions are positive, and we see that $\mathscr{R}_1$, $\Re(\mathscr{R}_2)$ and $\mathscr{R}_3$ give important contributions to the total crystal excitation rate.}
\label{fig:Rate_contribution_c7}
\end{figure*}

\begin{figure*}
\centering
  \includegraphics[width=0.48\textwidth]{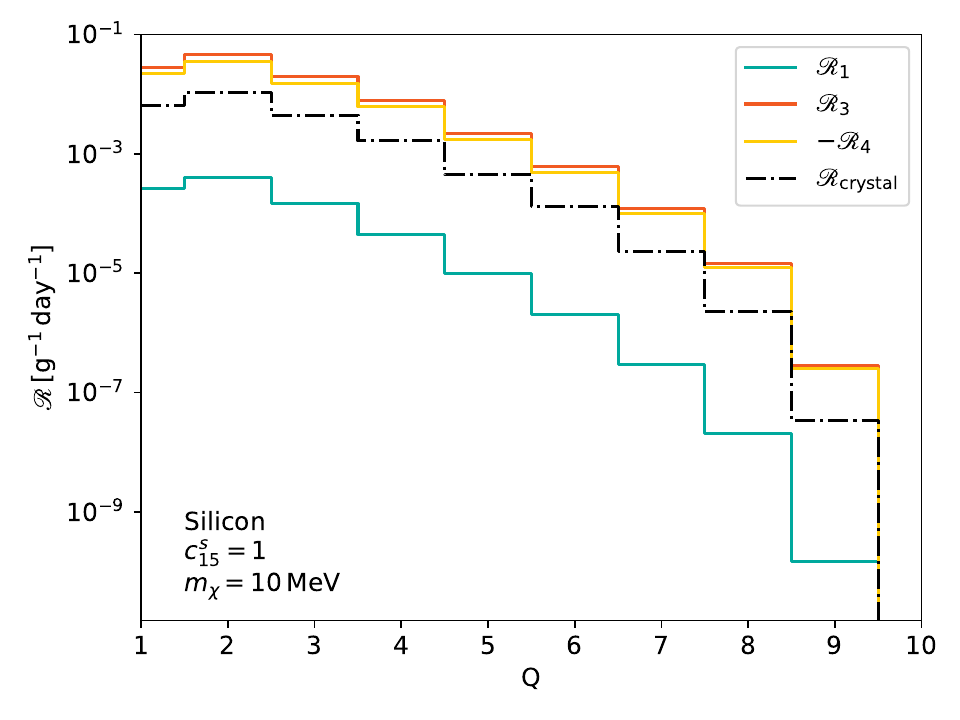}
  \includegraphics[width=0.48\textwidth]{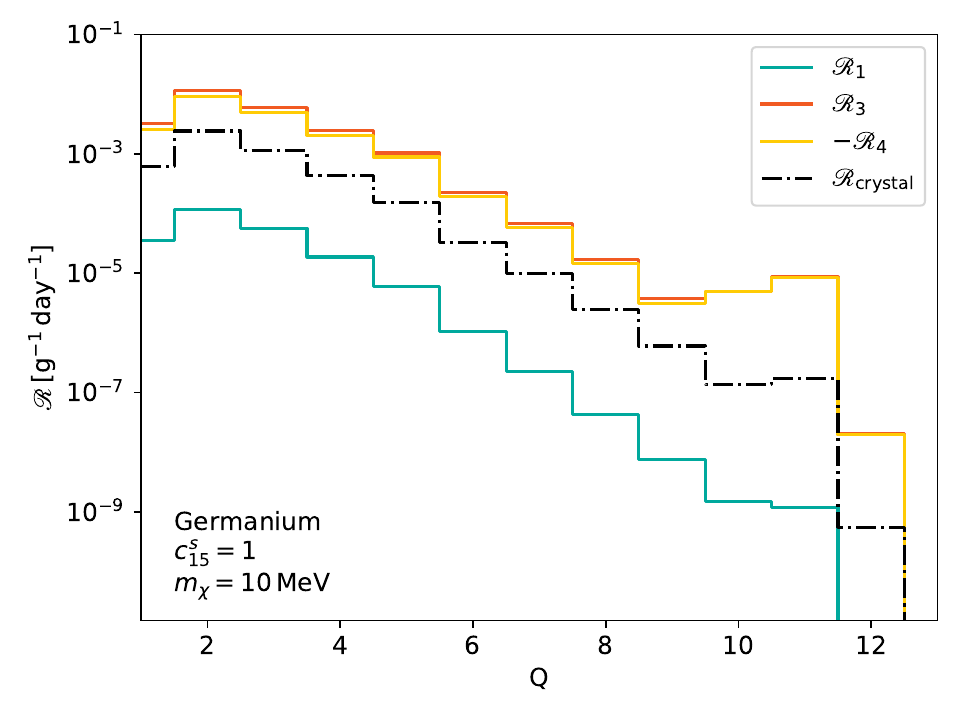}
  \includegraphics[width=0.48\textwidth]{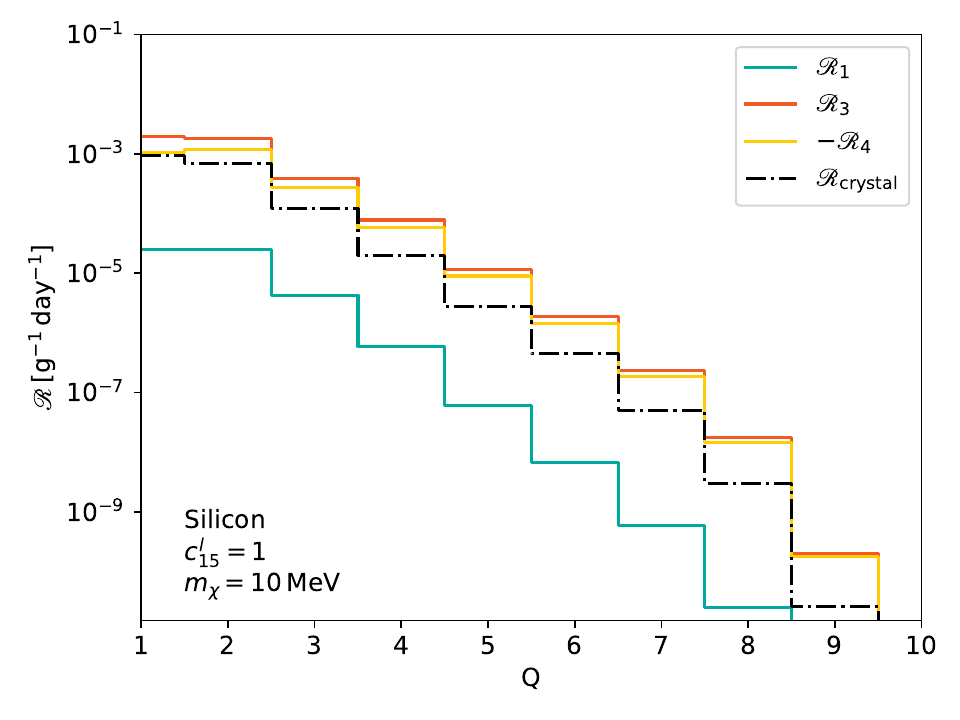}
  \includegraphics[width=0.48\textwidth]{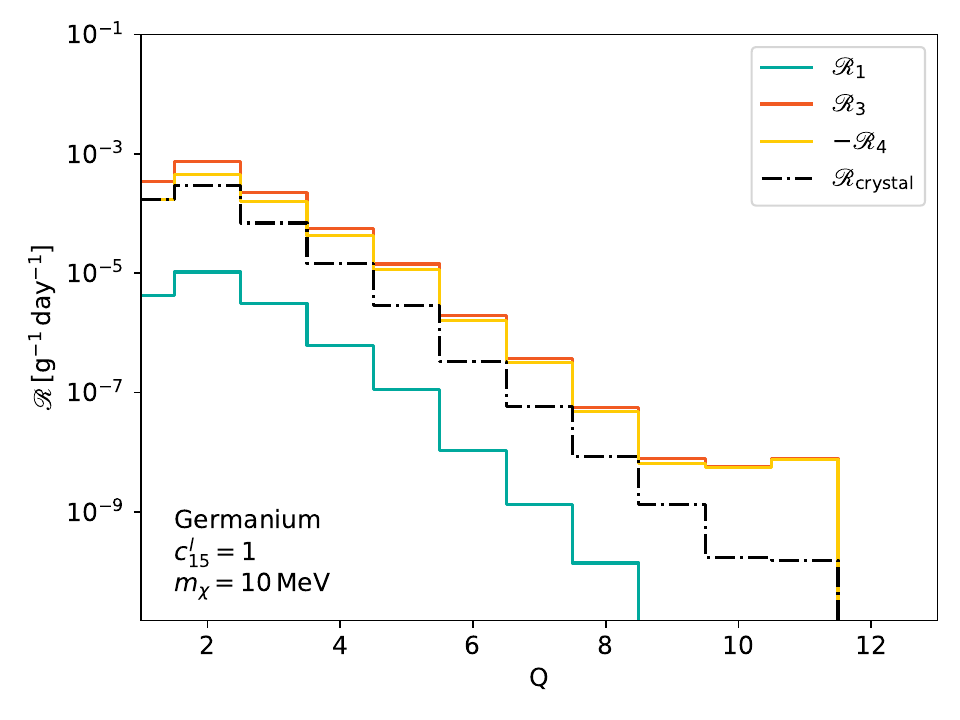}
\caption{Expected excitation rates in silicon (left panels) and germanium (right panels) for $c^s_{15}=1$ or $c^\ell_{15}=1$. The case of short- (long-) range interactions is reported in the top (bottom) panels. In the plot, $-\mathscr{R}_4$ is shown since $\mathscr{R}_4$ is negative, which causes it to cancel with $\mathscr{R}_3$.}
\label{fig:Rate_contribution_c15}
\end{figure*}

\begin{figure}
\centering
  \includegraphics[width=0.48\textwidth]{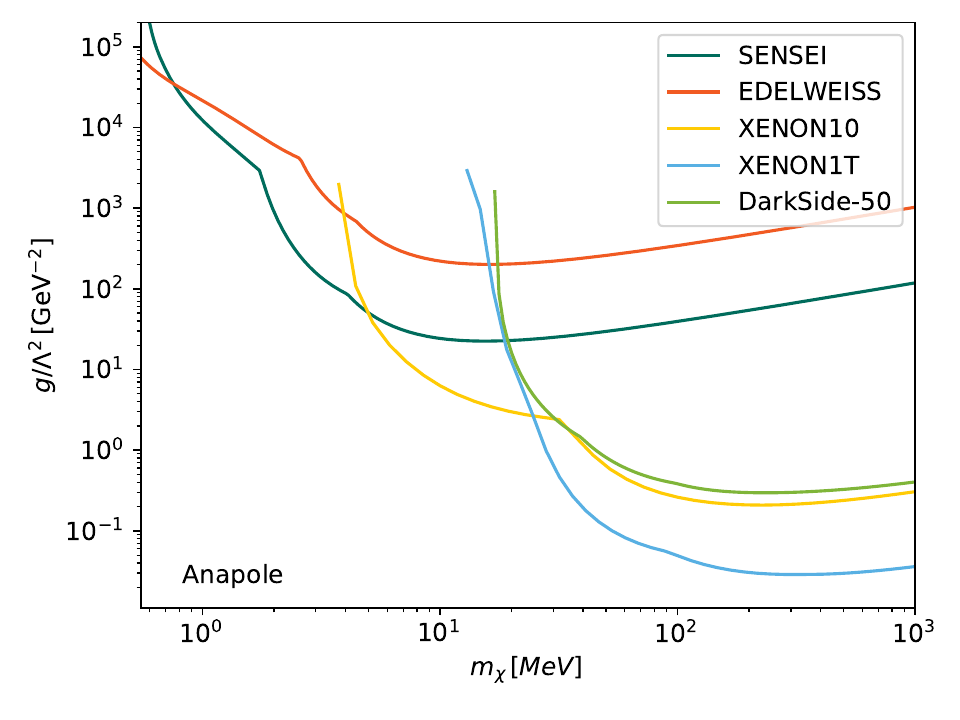}
  \includegraphics[width=0.48\textwidth]{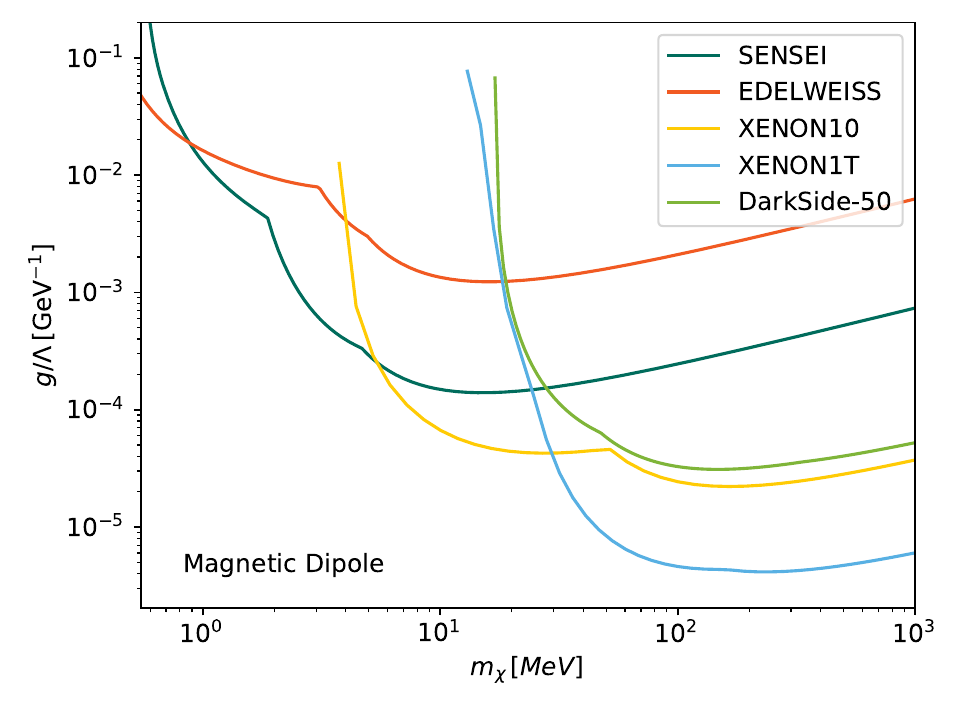}
  \includegraphics[width=0.48\textwidth]{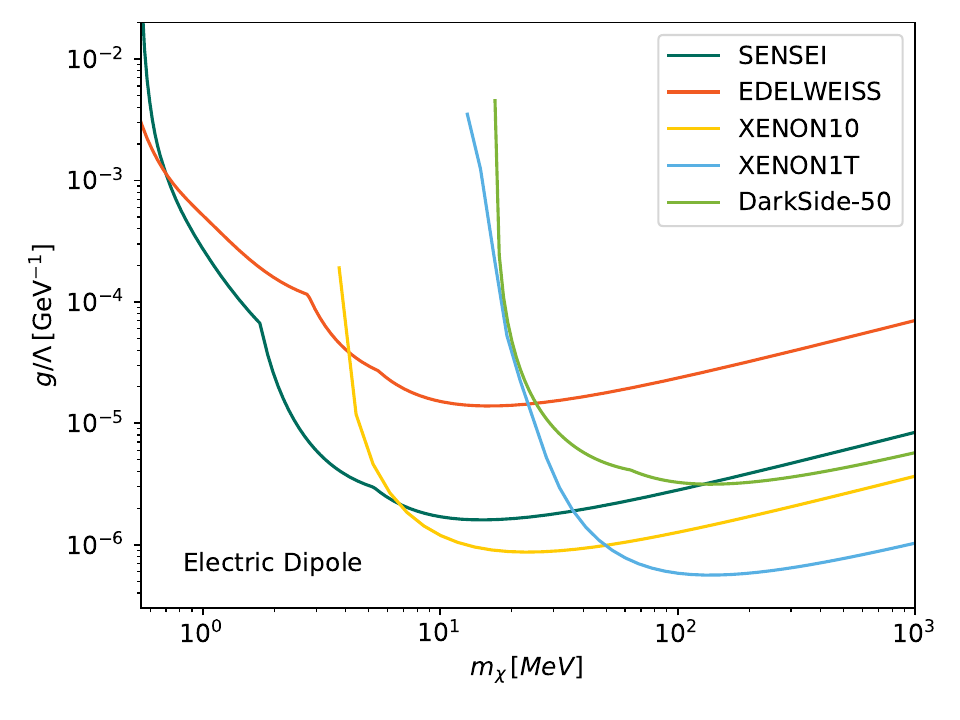}
\caption{90\% C.L. exclusion limits on the DM-electron coupling from data reported by SENSEI@MINOS~\cite{Barak:2020fql} and EDELWEISS~\cite{Arnaud:2020svb} and interpreted within the anapole (top panel), magnetic dipole (central panel) and electric dipole (bottom panel) DM models. For comparison, in each panel we also report the 90\% C.L. exclusion limits found in \cite{Catena:2019gfa} from the null result of experiments operating xenon (XENON10 and XENON1T) and argon (DarkSide-50) detectors.
}
\label{fig:Limits_poles}
\end{figure}

\begin{figure}
\centering
  \includegraphics[width=0.48\textwidth]{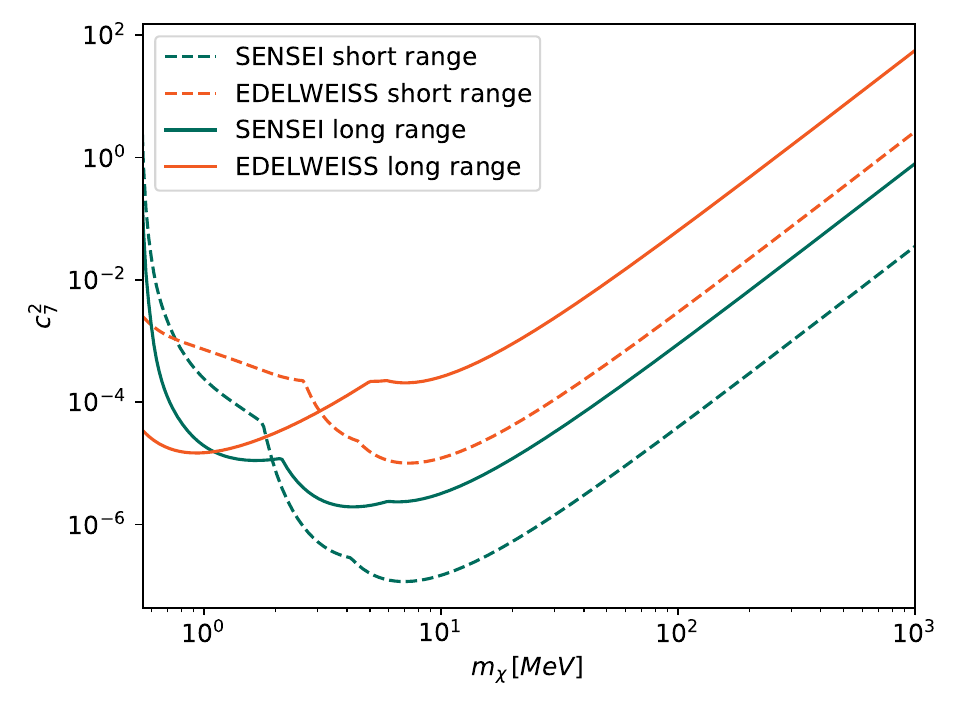}
  \includegraphics[width=0.48\textwidth]{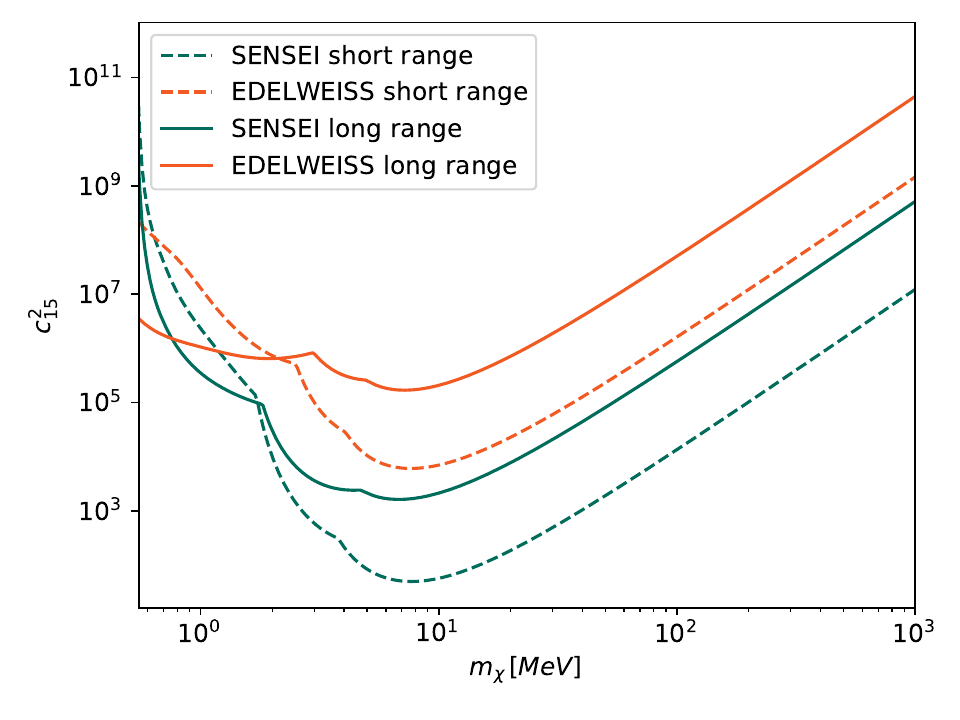}
\caption{Exclusion limits~(90\% C.L.) on $(c^s_7)^2$ and $(c^\ell_7)^2$ (left panel) as well as on $(c^s_{15})^2$ and $(c^\ell_{15})^2$  (right panel).}
\label{fig:Limits_c7_c15}
\end{figure}

\section{Results}
\label{sec:results}
In this section, we numerically evaluate the five response functions in Eq.~(\ref{eq:W_scalar}) focusing on silicon and germanium crystals.~We then use these response functions to compute the expected crystal excitation rates for the anapole, magnetic dipole and electric dipole DM interactions, as well as for specific (linear combinations of) interaction operators in Tab.~\ref{tab:operators}.~Finally, we apply our responses to set 90\% exclusion limits on the strength of such interactions from the null result reported by the EDELWEISS~\cite{Arnaud:2020svb} and SENSEI~\cite{Barak:2020fql} experimental collaborations.  

\subsection{Novel response functions for silicon and germanium crystals}

We display our crystal response functions in the $(q,\Delta E)$ plane, focusing on silicon in Fig.~\ref{fig:Silicon} and on germanium in Fig.~\ref{fig:Germanium}. In each panel of the two figures, the value of the corresponding response function is given by the color bar, with darker colors corresponding to higher values. As described in Appendix~\ref{sec:ff}, our first response function, $\overline{W}_1$,  is equal to $8 \Delta E \alpha m_e^2 /q^3$ times the crystal form factor introduced by Essig et al. in~\cite{Essig:2015cda}  (see our Eq.~(\ref{eq:Comparison_Essig})).~Taking into account this $q$ and $\Delta E$ dependent pre-factor, our first crystal response functions, $\overline{W}_1$, (upper left panel in Figs.~\ref{fig:Silicon} and \ref{fig:Germanium}) are in broad agreement with those given in Fig.~5 of Ref.~\cite{Essig:2015cda}. There are, however, some important differences between our calculations and those in~\cite{Essig:2015cda}. First, the corrected position of our Ge $3d$ levels means that the abrupt increase in $\overline{W}_1$ as a function of $\Delta E$ occurs at a $\Delta E$ value of 30, rather than 25, eV. Second, the higher energy cutoffs that we used allow us to calculate our response functions over a larger energy range.

We now focus on the remaining four response functions, $\overline{W}_2$, $\overline{W}_3$, $\overline{W}_4$ and $\overline{W}_5$, which we compute here for the first time. Since $\overline{W}_2$ is a complex response function, we consider its real and imaginary part separately. Focusing on silicon crystals, the upper right panel of Fig.~\ref{fig:Silicon} shows $\Re(\overline{W}_2)$, while the central left panel reports the corresponding imaginary part, $\Im(\overline{W}_2)$. In the same figure, the central right panel shows $\overline{W}_3$, the lower left panel displays $\overline{W}_4$, while the lower right panel reports $\overline{W}_5$. Fig.~\ref{fig:Silicon} shows the analogous crystal response functions for germanium crystals.~In all panels of  Fig.~\ref{fig:Silicon} and Fig.~\ref{fig:Germanium}, the points below the dashed black line are kinematically inaccessible for DM particles that are gravitationally bound to our galaxy, given our assumptions for the earth and local escape velocity. Note the different color-bar scales in each case. In particular, $\overline{W}_2$ and $\overline{W}_5$ have linear color bars since they can take both positive and negative values. We find that, $\Im\left(\overline{W}_2\right)$ and $\overline{W}_5$ are orders of magnitude smaller than the other responses, and are therefore expected to give a sub-leading contribution to the total crystal excitation rate. In order to understand this result, it is useful to compare our findings with the results of~\cite{Catena:2019gfa} on DM-electron scattering in atoms.

\subsection{Comparison to atomic responses}

Each of the response functions identified here has in principle an atomic analogue. Specifically, we find that our crystal response functions can be mapped on to the atomic response functions found in Ref.~\cite{Catena:2019gfa} if one replaces the crystal Bloch wave functions in Eq.~(\ref{eq:psi}) with the wave functions of electrons in atoms, and the summation/integration over lattice/crystal momenta and the summation over band indices in Eq.~(\ref{eq:W_scalar}) with a summation over the atomic azimuthal and magnetic quantum numbers (compare Eq.~(\ref{eq:W_scalar}) with Eq.~(41) in~\cite{Catena:2019gfa}). This comparison shows that the atomic response functions associated with $\Im\left(\overline{W}_2\right)$ and $\overline{W}_5$ through this mapping are exactly zero. This is the result of $\mathbf{A}'\equiv\sum_{m m'}f_{1\rightarrow 2}\mathbf{f}_{1\rightarrow 2}$ being real and (anti) parallel to $\mathbf{q}$, where the sum is over the initial and final electron magnetic quantum numbers while ``1'' and ``2'' label the initial and final state of the atomic electron, see~\cite{Catena:2019gfa}. Indeed, in the case of atoms $\mathbf{A}'$ is expected to be proportional to $\mathbf{q}$, $\mathbf{q}$ being the only preferred direction in an otherwise spherically symmetric system. Specifically, the electron wave function overlap integral $\mathbf{f}_{1\rightarrow 2}(\mathbf{q})$ (the analogue of $\mathbf{f}_{i,\mathbf{k}\rightarrow i^\prime, \mathbf{k}^\prime}$) is axially symmetric around the direction of $\mathbf{q}$. Similarly, $f_{1\rightarrow 2}(q)$, the analogue of $f_{i,\mathbf{k}\rightarrow i^\prime, \mathbf{k}^\prime}$, is spherically symmetric. In contrast, the spherical symmetry of $f_{i,\mathbf{k}\rightarrow i^\prime, \mathbf{k}^\prime}$ and the axial symmetry of $\mathbf{f}_{i,\mathbf{k}\rightarrow i^\prime, \mathbf{k}^\prime}$ are only approximate in the case of crystals. This explains why, although $\overline{W}_2$ is approximately real and $\overline{W}_5$ is subleading, for semiconductor crystals $\Im\left(\overline{W}_2\right)$ and $\overline{W}_5$ are not exactly zero.
We expect more dramatic departures from axial symmetry in the case of  anisotropic materials, such as 2D materials, or anisotropic 3D Dirac materials. For these systems, we therefore expect larger values for the responses $\Im\left(\overline{W}_2\right)$ and $\overline{W}_5$.

\subsection{Predicted electron excitation rates}
Once the crystal responses have been calculated we can insert them in Eq.~(\ref{eq:R_crystal_2D}) to obtain the expected excitation rates. In Fig.~\ref{fig:Total_Rate_Comparison} we show the expected excitation rates for the $14$ non-relativistic operators of Tab.~\ref{tab:operators} for DM masses of $0.5\,\mathrm{MeV}$, $5\,\mathrm{MeV}$ and $50\,\mathrm{MeV}$ for long- and short-range interactions. 

A notable difference between the predicted excitation rates for silicon and germanium is that silicon has zero excitations for $m_\chi=0.5\,\mathrm{MeV}$, which is due to the band gap of silicon being too large for galaxy-bound DM of this mass to cause excitations. We also see that for most operators the excitation rate is maximum for $m_\chi=5\,\mathrm{MeV}$ showing that both silicon and germanium are well suited to probe DM masses at the $\mathrm{MeV}$-scale.

Here, we focus on the predicted excitation rates for the SENSEI~\cite{Barak:2020fql} and EDELWEISS~\cite{Arnaud:2020svb}  experiments, which employ targets made of silicon and germanium crystals, respectively. Indeed, such experiments do not measure the deposited energy but rather the number of electron hole pairs created by a DM-electron scattering event. In our calculations, we assume a linear relation between the deposited energy and the number of electron hole pairs created in a scattering event, i.e.
\begin{equation}
    Q(\Delta E)=1+\left \lfloor (\Delta E-E_\mathrm{gap})/\epsilon\right\rfloor 
    \label{eq:gap}
\end{equation}
where the floor function $\left \lfloor x\right\rfloor$ rounds $x$ down to the closest integer. The reported values for $E_\mathrm{gap}$ and $\epsilon$ vary somewhat in the literature, and we use the values stated by SENSEI@MINOS~\cite{Barak:2020fql} (EDELWEISS~\cite{Arnaud:2020svb}) for silicon (germanium). They are
\begin{align}
    \epsilon=& 3.8\,\mathrm{eV}\quad (\mathrm{Silicon})\nonumber \\
    \epsilon=& 3.0\,\mathrm{eV}\quad (\mathrm{Germanium})\nonumber \\
    E_\mathrm{gap}=& 1.2\,\mathrm{eV}\quad (\mathrm{Silicon})\nonumber \\
    E_\mathrm{gap}=& 0.67\,\mathrm{eV}\quad (\mathrm{Germanium})\nonumber \,.
\end{align}
The band gap for germanium is considerably lower than that for silicon, which allows germanium target experiments to probe lower masses, as we discussed above.

Having specified the values of $E_{\rm gap}$ and $\varepsilon$ used in our analysis, we can now calculate the expected crystal excitation rate corresponding to a given number of electron-hole pairs $Q=1,2,$~etc...~and to a given target material by setting the boundaries of the $\Delta E$ integration in Eq.~(\ref{eq:R_crystal}) to the values required by Eq.~(\ref{eq:gap}).
In order to illustrate the contributions to the excitation rate from the individual response functions, we find it convenient to rewrite Eq.  \eqref{eq:R_crystal_2D} as 
\begin{equation}
\mathscr{R}_{\textrm{crystal}}=\sum_{l=1}^5 \mathscr{R}_l,
\end{equation}
where
\begin{align}
\mathscr{R}_{l}&=\frac{n_\chi N_\text{cell} }{128\pi m_\chi^2 m_e^2}\int \mathrm{d} (\ln\Delta E)\int \mathrm{d} q \, q \,\widehat{\eta}\left(q, \Delta E_{i\mathbf{k} \rightarrow i'\mathbf{k}'}\right)
\nonumber\\
&\times \Re\left(R_l^*(q,\mathbf{v}) \overline{W}_l(q,\Delta E)\right)\,.
\end{align}
We start by computing the expected excitation rates for the anapole, electric dipole and magnetic dipole interactions using the relation between couplings and interaction scale $\Lambda$ that we derived in Eqs.~\eqref{eq: anapole effective couplings} to \eqref{eq: electric dipole effective couplings}. For these interactions, Fig.~\ref{fig:Rate_contribution_poles} shows the expected excitation rates as a function of $Q$ for silicon and germanium crystals, and
for $m_\chi=10\,\mathrm{MeV}$. The black dashed line gives the total crystal excitation rate while the solid colored lines give contributions from the individual responses. The light blue, light green, dark orange and yellow lines correspond to $\mathscr{R}_1$, $\Re(\mathscr{R}_2)$, $\mathscr{R}_3$ and $\mathscr{R}_4$ respectively. In order to illustrate the dependence of our results on the DM particle mass, in Fig.~\ref{fig:Multimass_poles} we plot the total excitation rate for $1\,\mathrm{MeV}$, $10\,\mathrm{MeV}$ and $100\,\mathrm{MeV}$. Focusing on events producing few electron-hole pairs, i.e.~$Q\sim 1$, we see that the excitation rate is maximum at around $10\,\mathrm{MeV}$, and falls off both for higher and for lower masses. We also see a clear rise in the excitation rate for germanium at $Q=10$ for $m_\chi=100\,\mathrm{MeV}$ caused by the $3d$ electrons. Without the Hubbard-$U$ correction implemented in our DFT calculations, this peak would be located at $Q=9$ (see Appendix~\ref{sec:hubbard}).

Importantly, we find that the crystal excitation rate is not necessarily dominated by the crystal response function $\bar{W}_1$ (the crystal form factor of Essig et al.~\cite{Essig:2015cda}). This is apparent in the case of, e.g. magnetic dipole and anapole DM-electron interactions, but it is also true for other combinations of nonrelativistic effective operators in Tab.~\ref{tab:operators}. 

In Fig.~\ref{fig:Rate_contribution_poles}, we can see the expected excitation rates for different interaction types plotted against the number of electron-hole pairs produced. As expected, we see that these rates drop significantly for larger ionization signals since those require larger energy transfers. We can see that different responses are dominant for different interaction types. Namely that for the case of an anapole interaction, a combination of $\mathscr{R}_1$ and $\mathscr{R}_2$ dominates, for magnetic dipole interaction at low electron-hole-pair production $\mathscr{R}_3$ dominates whereas $\mathscr{R}_1$ dominates elsewhere, and that the electric dipole interaction is dominated by the $\mathscr{R}_1$ interaction.

Figs. \ref{fig:Rate_contribution_c7} and \ref{fig:Rate_contribution_c15} show the crystal excitation rate as a function of the number of excited electron-hole pairs $Q$ for two selected non-relativistic effective operators with coupling constants $c^s_7=1$ or $c^\ell_7=1$  and $c^s_{15}=1$ or $c^\ell_{15}=1$, respectively, and with a DM~mass of $m_\chi=10~\mathrm{MeV}$. In both figures, the rate for silicon (germanium) is given in the left (right) panels and the results for short- (long-) range interactions are given in the top (bottom) panels. In Fig.~\ref{fig:Rate_contribution_c7} all the nonzero response functions give positive contributions of a similar magnitude, whereas in Fig.~\ref{fig:Rate_contribution_c15} the contribution from $\mathscr{R}_4$ almost exactly cancels with that from  $\mathscr{R}_3$, leading to a total crystal excitation rate that is orders of magnitude smaller than the excitation rate produced by $\mathscr{R}_3$ alone. In the case of contact and long-range interactions of type $\mathcal{O}_{15}$ in germanium crystals, we also find a peak at $Q=10$ in the $\mathscr{R}_3$ and $\mathscr{R}_4$ contributions to the total excitation rate (see Fig.~\ref{fig:Rate_contribution_c15}). This peak is due to the DM-induced excitation of $3d$ electrons. Interestingly, the previously mentioned cancellation between $\mathscr{R}_3$ and $\mathscr{R}_4$ washes out this peak, so it is not present in the total excitation rate. Even in this case, however, the $3d$ electrons have an impact on the total crystal excitation rate: they slow down the decrease of the crystal excitation rate for $Q$ above about $10$ electron-hole pairs. In the case of germanium crystals, we find a similar effect also for long- and short-range interactions of type $\mathcal{O}_7$ (see Tab.~\ref{tab:operators} and Fig.~\ref{fig:Rate_contribution_c7}).

\subsection{Exclusion limits}
Once the expected excitation rates have been computed we can compare them to the number of events measured experimentally to place limits on the mass and couplings of the DM particle. As anticipated, here we focus on the results reported by the SENSEI@MINOS \cite{Barak:2020fql} and EDELWEISS \cite{Arnaud:2020svb} experiments. The former operates silicon semiconductor detectors, the latter employs germanium targets. We compare our theoretical predictions with the experimental data by assuming that all events reported by the two experimental collaborations are caused by DM-electron interactions in the detector. By ``event'' we refer to the production of electron-hole pairs as the result of DM-induced electron excitations in the given semiconductor target. In order to compute the expected number of events associated with a given value of $Q$, we multiply the crystal excitation rate in Eq.~(\ref{eq:R_crystal}) by the effective experimental exposure corresponding to that $Q$ value. For the SENSEI@MINOS experiment, we restrict our analysis to $Q=(1,2,3,4)$, and we take the different effective exposures for the four $Q$ values given by the experiment, namely: $(1.38,2.09,9.03,9.10)~$g-day. ~The observed number of events in each of the four $Q$ bins is:~$(758,5,0,0)$, as reported in Tab.~1 and explained at the end of page 4 in~\cite{Barak:2020fql}. For EDELWEISS, we also consider $Q=(1,2,3,4)$, but assume the effective exposures~$(0.04,0.22,1,1)\times 33.4$~g$\times58$h, respectively. The observed number of events in each $Q$ bin is in this case $(5814,44706,2718,227)$, as one can see by digitising Fig.~2 of~\cite{Arnaud:2020svb}.

For each experiment, we calculate 90\% confidence level (C.L.) exclusion limits on the strength of a given DM-electron interaction by requiring that, in each of the four $Q$ bins we consider, the cumulative distribution function of a Poisson probability density function of mean equal to the predicted number of events in that bin is at least equal to $0.1$ if evaluated at the observed number of signal events.

Fig.~\ref{fig:Limits_poles} shows our 90\% C.L. exclusion limits on the strength of an anapole, magnetic and electric dipole DM-electron interaction as a function of the DM particle mass. For comparison, in the same figure we also display the 90\% C.L. exclusion limits on these coupling constants from the null results of XENON10~\cite{Angle:2011th,Essig:2012yx}, XENON1T~\cite{Aprile:2019xxb}, and DarkSide-50~\cite{Agnes:2018oej} as obtained in \cite{Catena:2019gfa}. We find that SENSEI@MINOS provides the strongest constraints on DM masses below $5$~MeV, with EDELWEISS taking over for sub-MeV DM masses.

Using the same procedure we also calculate the 90\% C.L. exclusion limits on the strength of the interactions $\mathcal{O}_{7}$ and $\mathcal{O}_{15}$ that we used as benchmarks in the previous subsection. The resulting constraints on $c^s_7$ and $c^\ell_7$, as well as on $c^s_{15}$ and $c^\ell_{15}$ are given in Fig.~\ref{fig:Limits_c7_c15}. The dashed (solid) lines correspond to short- (long-) range interactions. The dark green line gives the 90\% C.L. exclusion limit from SENSEI@MINOS, while the orange line gives the 90\% C.L. exclusion limit from EDELWEISS. Due to its larger exposure, SENSEI@MINOS generically produces the strongest constraints above about 1~MeV.~Below this threshold, however, the lower band-gap of germanium implies that the strongest constraints arise from EDELWEISS.

\section{Summary}
\label{sec:summary}
We performed a comprehensive and model-independent study of interactions between DM~particles of our galactic halo and electrons bound in crystals.
By modelling the DM-electron interactions in crystals using an effective theory approach, we identified the most general amplitude for DM-electron scatterings and the general responses by the crystal to these interactions.
Our effective approach allows predictions of scattering rates in crystals for virtually all DM~models, it applies e.g. to the anapole, magnetic dipole and electric dipole DM-electron interaction models.
In particular our study focuses on short- and long-range DM-electron interactions with silicon and germanium crystals which are currently used as targets in DM direct detection experiments.

This led us to discover that there at most five ways a crystal can respond to an external probe (not necessarily a DM particle) in the limit of non-relativistic short- and long-range interactions of the most general type. We identified these five independent crystal responses and expressed them in terms of electron wave function overlap integrals. By performing state-of-the-art DFT calculations, we evaluated the five responses, focusing on silicon and germanium crystals, as these are used in operating DM direct detection experiments such as SENSEI and EDELWEISS.   

We applied our crystal response functions to predict the rate of DM-induced electron excitations in silicon and germanium crystals. We performed this calculation for a set of 14 non-relativistic interaction operators (of short- and long-range type) and for specific models, such as the anapole, magnetic dipole and electric dipole models. 

As a second important application of our novel crystal response functions, we computed the 90\% C.L. exclusion limits on the strength with which DM can couple to electrons in crystals by comparing our predicted crystal excitations rates with data collected at the SENSEI@MINOS and EDELWEISS experiments. We performed this calculation for the already mentioned set of non-relativistic DM-electron interactions, as well as for the anapole, magnetic dipole and electric dipole DM models. We compared these limits to constraints arising from different DM direct detection experiments and identified the range of masses where silicon or germanium detectors are expected to set the most stringent bounds on DM-electron interactions.

Within the field of DM direct detection, our novel crystal responses will enable the scientific community to perform predictions for DM particle models that were not tractable before. This is for example the case for the anapole and magnetic dipole interaction models, as they generate crystal responses that were not known previously. Furthermore, our crystal response functions  will allow the community to calculate at which statistical significance a given DM-electron coupling is excluded by the null result of a DM experiment using germanium or silicon semiconductor detectors. Finally, they will allow us to interpret a discovery in a next generation DM direct detection experiment within a range of DM models that can not be covered by the current most widely used approach to data analysis, which is based on the use of a single crystal form factor. 

On a more speculative level, our work paves the way for a new line of research lying at the interface of astroparticle and condensed matter physics: the study of yet hidden material properties that are encoded in the response of materials to external probes sharing the same interaction properties as the elusive galactic DM component.

\acknowledgements{NAS and MM were supported by the ETH Zurich, and by the European Research Council (ERC) under the European Union’s Horizon 2020 research and innovation programme project HERO Grant Agreement No. 810451. RC acknowledges support from an individual research grant from the Swedish Research Council, dnr. 2018-05029. TE was supported by the Knut \& Alice Wallenberg Foundation. The research presented in this paper made use of the following software packages, libraries, and tools: Matplotlib~\cite{Hunter:2007} NumPy~\cite{numpy}, SciPy~\cite{scipy2019}, WebPlotDigitizer~\cite{webplotdigitizer}, Wolfram Mathematica~\cite{Mathematica}. The computations were enabled by resources provided by the Swedish National Infrastructure for  Computing  (SNIC)  at  the  National  Supercomputer Centre  (NSC)  and  the  Chalmers  Centre  for  Computational Science and Engineering (C3SE).}

\appendix
\begin{widetext}
\section{Hubbard $U$ correction for the germanium $3d$-states}
\label{sec:hubbard}

In this appendix we illustrate the effect of the Hubbard $U$ correction. In Fig.~\ref{fig:W_4_comparison} we see how $\overline{W}_4$ is affected by the Hubbard $U$ correction. In particular, we see the high value region produced by the $3d$ electrons is shifted from $25\,\mathrm{eV}$ to $30\,\mathrm{eV}$. In
Fig.~\ref{fig:Hubbard_rate_comparison} we see the impact this has on the expected excitation rate for the anapole, electric dipole and magnetic dipole interactions. For $Q$-values of $9$ and $10$ the the excitation rate differs with orders of magnitude.
\begin{figure*}[h]
    \centering
    \includegraphics[width=0.48\textwidth]{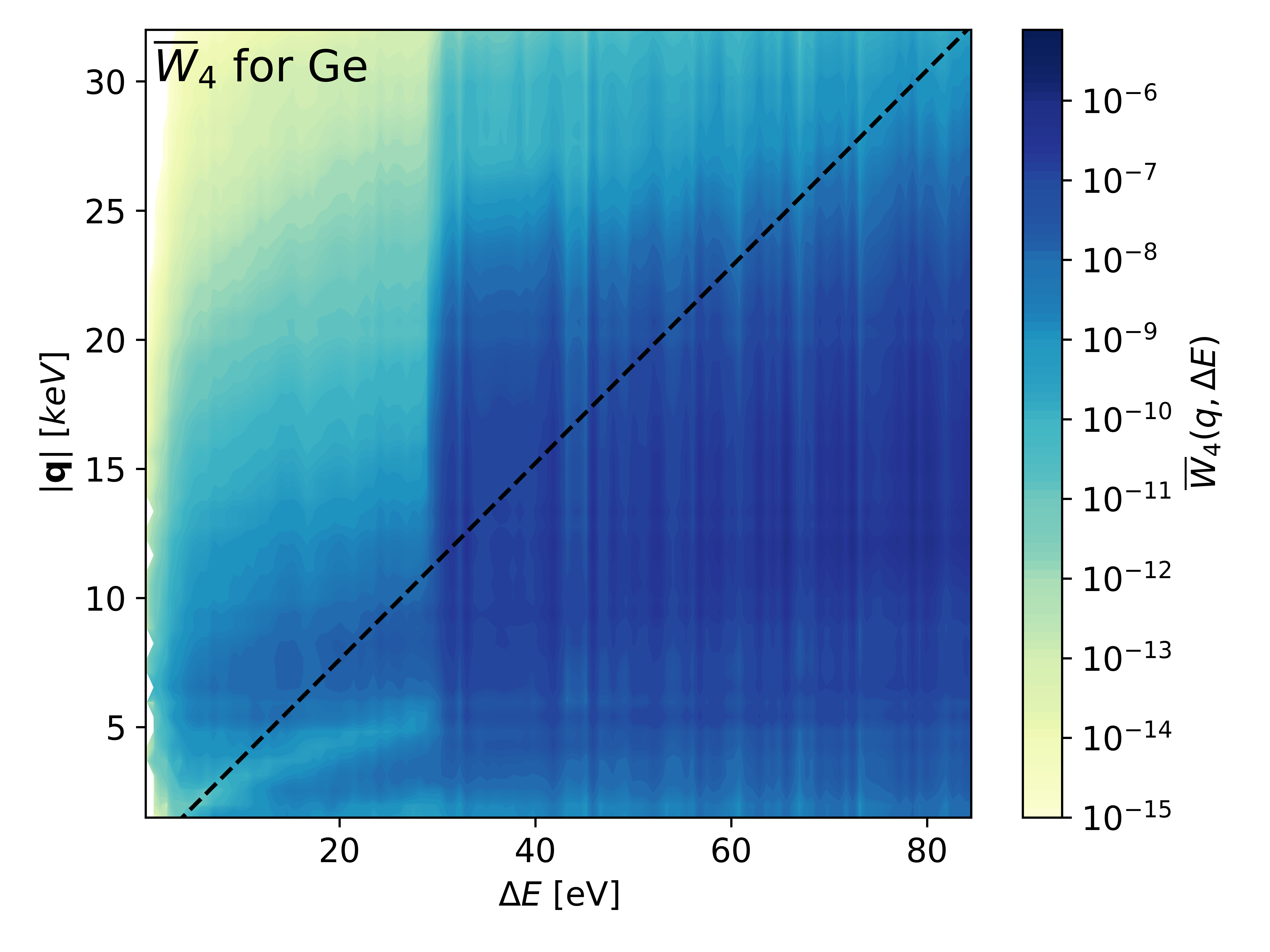}
    \includegraphics[width=0.48\textwidth]{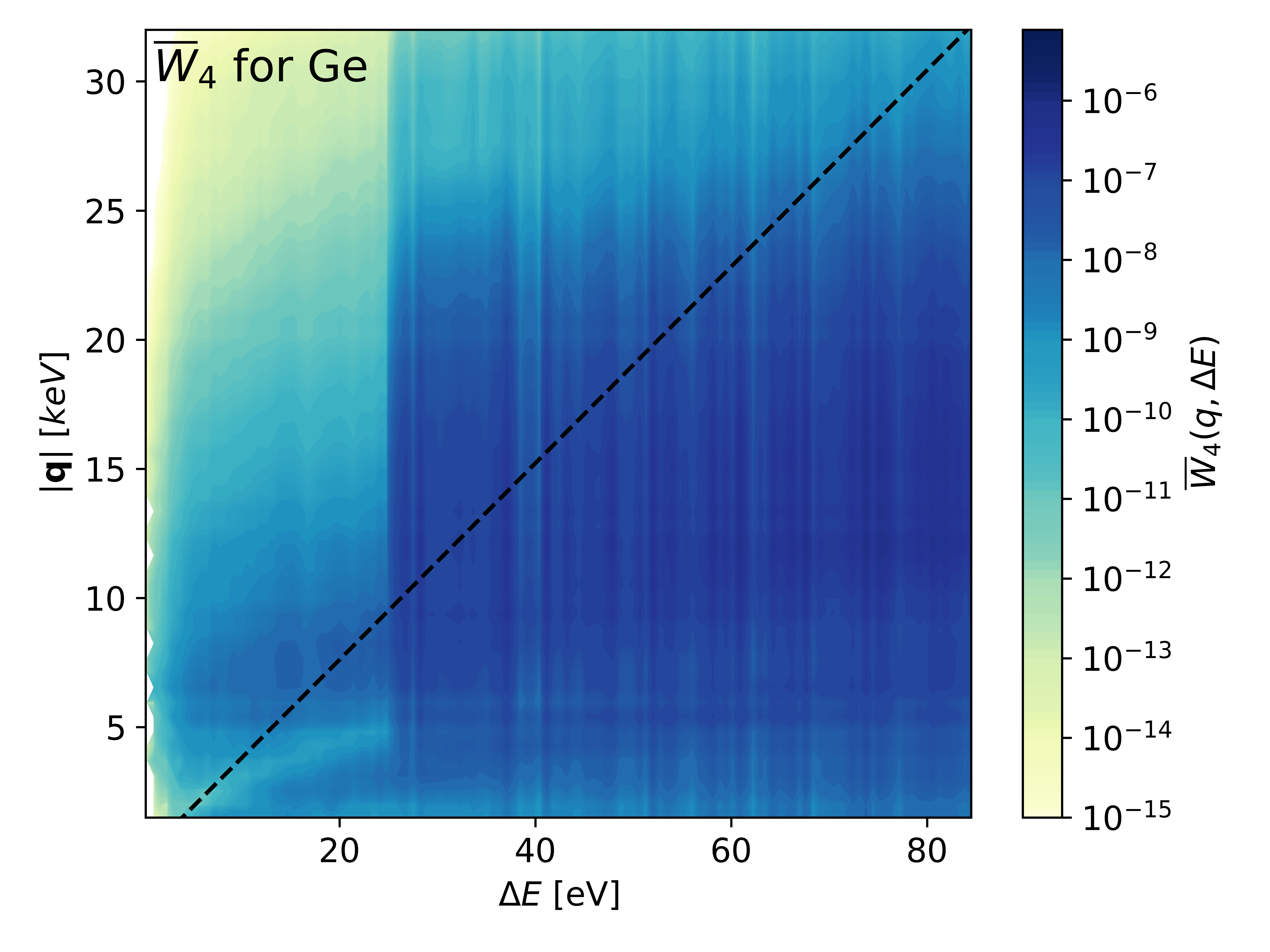}
    \caption{$\overline{W}_4$ calculated with (left) and without (right) the Hubbard $U$ correction applied to the germanium $3d$ states. The $U$ correction shifts the calculated position of the $3d$ bands from about $25\,\mathrm{eV}$ to about $30\,\mathrm{eV}$ below the Fermi energy, and we see a corresponding shift in the increase in intensity onset of $\overline{W}_4$ over the entire $q$ range.}
    \label{fig:W_4_comparison}
\end{figure*}
\begin{figure*}[h]
    \centering
    \includegraphics[width=0.32\textwidth]{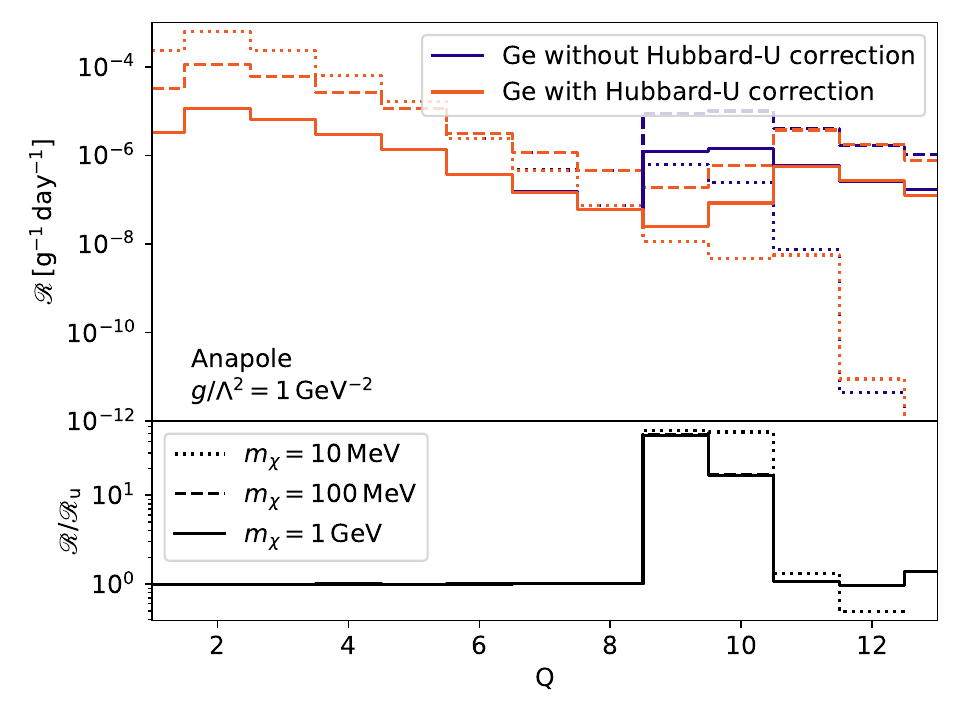}
    \includegraphics[width=0.32\textwidth]{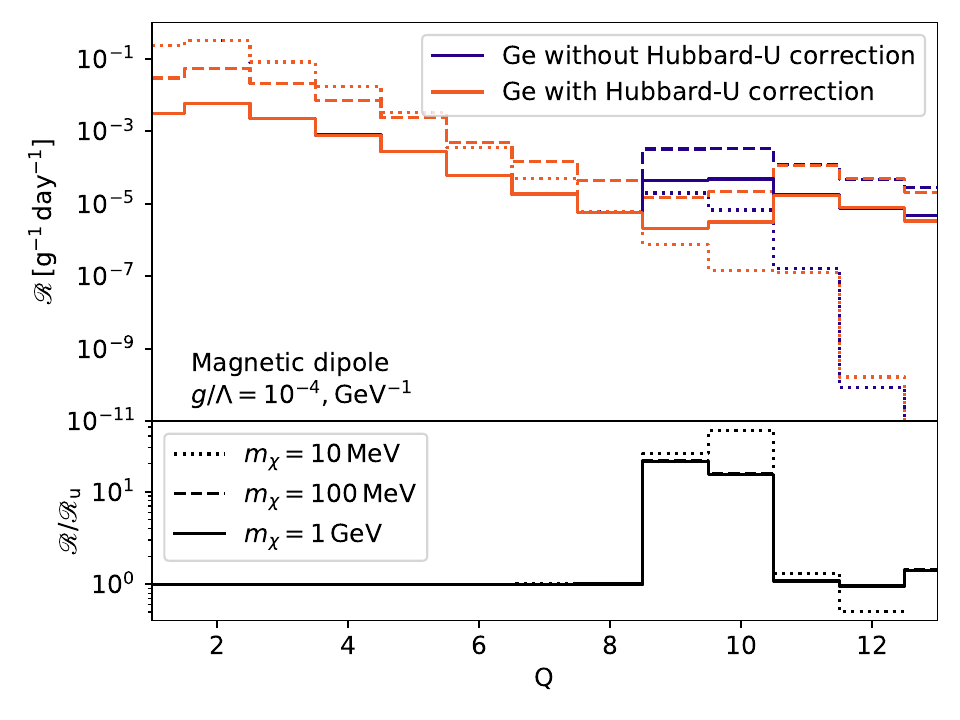}
    \includegraphics[width=0.32\textwidth]{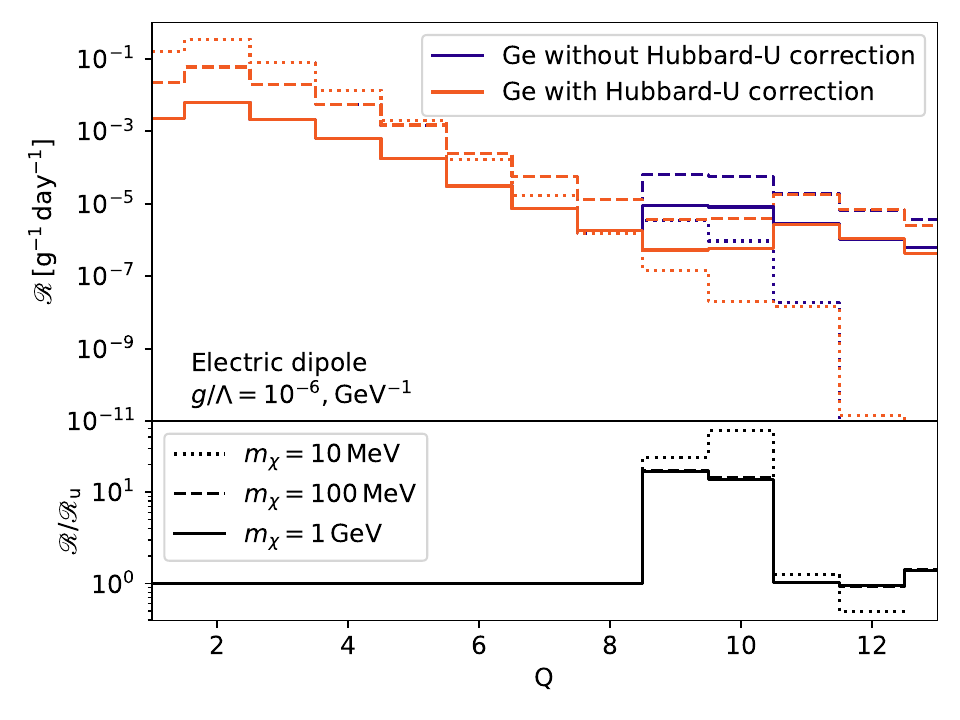}
    \caption{Calculated crystal excitation rates for germanium with and without the Hubbard $U$ correction applied to the $3d$ band. We see that, in the absence of the correction, the lower binding energy, by $\sim5\,\mathrm{eV}$, of the $3d$ electrons leads to excitation rates that are orders of magnitude larger for $9$ and $10$ electron-hole pairs.}
    \label{fig:Hubbard_rate_comparison}
\end{figure*}
\section{Derivation of $\overline{\left|\mathcal{M}^\prime_{i\mathbf{k}\rightarrow i^\prime \mathbf{k}^\prime}\right|^2}$, $f^\prime_{i\mathbf{k}\rightarrow i^\prime \mathbf{k}^\prime}$ and $\mathbf{f}_{i\mathbf{k}\rightarrow i^\prime \mathbf{k}^\prime}$}
\label{sec:R}
In this appendix we cover the gap between Eqs.~ \eqref{eq:M2}-\eqref{eq:fvec} and Eqs.~ \eqref{eq:M2_prime}-\eqref{eq:fvec_prime}. The electron wave-function in a crystal is given on Bloch form as 
\begin{equation}
\psi_{i\mathbf{k}}(\mathbf{x})=\frac{1}{\sqrt{V}}\sum_\mathbf{G}{u_i(\mathbf{k}+\mathbf{G})e^{i(\mathbf{k}+\mathbf{G})\cdot \mathbf{x}}}\, ,
\end{equation}
where $V=N_\text{cell}V_\text{cell}$ is the volume of the crystal. Inserting this in Eqs.~\eqref{eq:f} and \eqref{eq:fvec} gives 
\begin{align}
f_{i,\mathbf{k}\rightarrow i^\prime, \mathbf{k}^\prime}=&\frac{1}{V} \int \mathrm{d}^3x \sum_{\mathbf{G}^\prime} u_{i^\prime,\mathbf{k}^\prime}^* e^{-i(\mathbf{k}^\prime+\mathbf{G}^\prime)\cdot \mathbf{x}} e^{i\mathbf{x}\cdot\mathbf{q}}\sum_{\mathbf{G}} u_{i,\mathbf{k}} e^{i(\mathbf{k}+\mathbf{G})\cdot \mathbf{x}}\nonumber\\
=&\sum_{\mathbf{G} \mathbf{G}^\prime}\frac{u_{i^\prime,\mathbf{k}^\prime}^* u_{i,\mathbf{k}}}{V}\int \mathrm{d}^3x e^{i(\mathbf{k}+\mathbf{G}+\mathbf{q}-\mathbf{k}^\prime-\mathbf{G}^\prime)\cdot \mathbf{x}}\nonumber\\
=&\sum_{\mathbf{G} \mathbf{G}^\prime}\frac{u_{i^\prime,\mathbf{k}^\prime}^* u_{i,\mathbf{k}}}{V} (2\pi)^3\delta^3(\mathbf{k}+\mathbf{G}+\mathbf{q}-\mathbf{k}^\prime-\mathbf{G}^\prime)
\end{align}
and
\begin{align}
\mathbf{f}_{i,\mathbf{k}\rightarrow i^\prime, \mathbf{k}^\prime}&=\int \mathrm{d}^3x \psi_2^*(\mathbf{x})e^{i\mathbf{x}\cdot\mathbf{q}}\frac{i\nabla_\mathbf{x}}{m_e}\psi_1(\mathbf{x})\nonumber\\
&=\frac{1}{V} \int \mathrm{d}^3x \sum_{\mathbf{G}^\prime} u_{i^\prime,\mathbf{k}^\prime}^* e^{-i(\mathbf{k}^\prime+\mathbf{G}^\prime)\cdot \mathbf{x}} e^{i\mathbf{x}\cdot\mathbf{q}} \frac{i\nabla_\mathbf{x}}{m_e} \sum_{\mathbf{G}} u_{i,\mathbf{k}} e^{i(\mathbf{k}+\mathbf{G})\cdot \mathbf{x}}\nonumber\\
&=\sum_{\mathbf{G} \mathbf{G}^\prime}\frac{u_{i^\prime,\mathbf{k}^\prime}^* u_{i,\mathbf{k}}}{V}\int \mathrm{d}^3x e^{-i(\mathbf{k}^\prime+\mathbf{G}^\prime)\cdot \mathbf{x}} e^{i\mathbf{x}\cdot\mathbf{q}} \frac{i\nabla_\mathbf{x}}{m_e} e^{i(\mathbf{k}+\mathbf{G})\cdot \mathbf{x}}\nonumber\\
&=-\sum_{\mathbf{G} \mathbf{G}^\prime}\frac{u_{i^\prime,\mathbf{k}^\prime}^* u_{i,\mathbf{k}}}{m_e V}(\mathbf{k}+\mathbf{G})\int \mathrm{d}^3x e^{i(\mathbf{k}+\mathbf{G}+\mathbf{q}-\mathbf{k}^\prime-\mathbf{G}^\prime)\cdot \mathbf{x}}\nonumber\\
&=-\sum_{\mathbf{G} \mathbf{G}^\prime}\frac{u_{i^\prime,\mathbf{k}^\prime}^* u_{i,\mathbf{k}}}{m_e V}(\mathbf{k}+\mathbf{G}) (2\pi)^3\delta^3(\mathbf{k}+\mathbf{G}+\mathbf{q}-\mathbf{k}^\prime-\mathbf{G}^\prime)
\end{align}
respectively, where $u_{i,\mathbf{k}}\equiv u_i(\mathbf{k}+\mathbf{G})$ and $u_{i^\prime,\mathbf{k}^\prime}\equiv u_{i^\prime}(\mathbf{k}^\prime+\mathbf{G}^\prime)$. We now compute $\left|f_{i,\mathbf{k}\rightarrow i^\prime, \mathbf{k}^\prime}\right|^2$, $f_{i,\mathbf{k}\rightarrow i^\prime, \mathbf{k}^\prime}\mathbf{f}_{i,\mathbf{k}\rightarrow i^\prime, \mathbf{k}^\prime}^*$ and $\left(\mathbf{f}_{i,\mathbf{k}\rightarrow i^\prime, \mathbf{k}^\prime}\cdot \mathbf{w}\right) \left(\mathbf{f}_{i,\mathbf{k}\rightarrow i^\prime, \mathbf{k}^\prime}^*\cdot \mathbf{w}^\prime\right)$ with $\mathbf{w}$ and $\mathbf{w}^\prime$ being arbitrary 3-vectors.~For $\left|f_{i,\mathbf{k}\rightarrow i^\prime, \mathbf{k}^\prime}\right|^2$ we find, 
\begin{align}
\left|f_{i,\mathbf{k}\rightarrow i^\prime, \mathbf{k}^\prime}\right|^2=&\sum_{\mathbf{F} \mathbf{F}^\prime}\frac{u_{i^\prime,\mathbf{k}^\prime}^* u_{i,\mathbf{k}}}{V} (2\pi)^3\delta^3(\mathbf{k}+\mathbf{F}+\mathbf{q}-\mathbf{k}^\prime-\mathbf{F}^\prime)\nonumber\\
&\times \left( \sum_{\mathbf{G} \mathbf{G}^\prime}\frac{u_{i^\prime,\mathbf{k}^\prime}^* u_{i,\mathbf{k}}}{V} (2\pi)^3\delta^3(\mathbf{k}+\mathbf{G}+\mathbf{q}-\mathbf{k}^\prime-\mathbf{G}^\prime)\right)^*\nonumber\\
=&\sum_{\Delta\mathbf{F} \Delta\mathbf{G}} \frac{(2\pi)^6}{V^2}\delta^3(\mathbf{k}+\mathbf{q}-\mathbf{k}^\prime-\Delta\mathbf{F})
\delta^3(\mathbf{k}+\mathbf{q}-\mathbf{k}^\prime-\Delta\mathbf{G})\nonumber\\ 
&\times\left(\sum_{\mathbf{F}}u_{i^\prime}^*\left(\mathbf{k}^\prime+\mathbf{F}+\Delta\mathbf{F}\right) u_i \left(\mathbf{k}+\mathbf{F}\right)\right)\left(\sum_{\mathbf{G}}u_{i^\prime}^*\left(\mathbf{k}^\prime+\mathbf{G}+\Delta\mathbf{G}\right) u_i\left(\mathbf{k}+\mathbf{G}\right)\right)^*\, ,
\end{align}
where $\Delta\mathbf{G}\equiv\mathbf{G}^\prime-\mathbf{G}$ and $\Delta\mathbf{F}\equiv\mathbf{F}^\prime-\mathbf{F}$. From the delta functions we see that the terms in the double sum is only non-zero when $\Delta\mathbf{G}=\Delta\mathbf{F}$. The sums over $\mathbf{F}$ and $\mathbf{G}$ are identical except for the labeling, i.e. $\sum_{\mathbf{G}}u_{i^\prime}^*\left(\mathbf{k}^\prime+\mathbf{G}+\Delta\mathbf{G}\right) u_i\left(\mathbf{k}+\mathbf{G}\right)=\sum_{\mathbf{F}}u_{i^\prime}^*\left(\mathbf{k}^\prime+\mathbf{F}+\Delta\mathbf{G}\right) u_i\left(\mathbf{k}+\mathbf{F}\right)\equiv  f_{i,\mathbf{k}\rightarrow i^\prime, \mathbf{k}^\prime}^\prime$. This yields
\begin{equation}
\left|f_{i,\mathbf{k}\rightarrow i^\prime, \mathbf{k}^\prime}\right|^2=\sum_{\Delta\mathbf{G}}\frac{(2\pi)^3\delta^3(\mathbf{k}+\mathbf{q}-\mathbf{k}^\prime-\Delta\mathbf{G})}{V}\left|f_{i,\mathbf{k}\rightarrow i^\prime, \mathbf{k}^\prime}^\prime\right|^2\, .
\label{eq:t1}
\end{equation}
The calculation for $f_{i,\mathbf{k}\rightarrow i^\prime, \mathbf{k}^\prime}\mathbf{f}_{i,\mathbf{k}\rightarrow i^\prime, \mathbf{k}^\prime}^*$ is analogous, and produces:
\begin{align}
f_{i,\mathbf{k}\rightarrow i^\prime, \mathbf{k}^\prime}\mathbf{f}_{i,\mathbf{k}\rightarrow i^\prime, \mathbf{k}^\prime}^*=&\left(\sum_{\mathbf{F} \mathbf{F}^\prime}\frac{u_{i^\prime,\mathbf{k}^\prime}^* u_{i,\mathbf{k}}}{V} (2\pi)^3\delta^3(\mathbf{k}+\mathbf{F}+\mathbf{q}-\mathbf{k}^\prime-\mathbf{F}^\prime)\right)\nonumber \\
&\times\left(-\sum_{\mathbf{G} \mathbf{G}^\prime}\frac{u_{i^\prime,\mathbf{k}^\prime}^* u_{i,\mathbf{k}}}{m_e V}(\mathbf{k}+\mathbf{G}) (2\pi)^3\delta^3(\mathbf{k}+\mathbf{G}+\mathbf{q}-\mathbf{k}^\prime-\mathbf{G}^\prime)\right)^*\nonumber\\
=&\sum_{\Delta\mathbf{G}}\frac{(2\pi)^3\delta^3(\mathbf{k}+\mathbf{q}-\mathbf{k}^\prime-\Delta\mathbf{G})}{V}\left(\sum_{\mathbf{F}}u_{i^\prime}^*\left(\mathbf{k}^\prime+\mathbf{F}+\Delta\mathbf{G}\right)u_i\left(\mathbf{k}+\mathbf{F}\right)\right)\nonumber\\
&\times\left(-\frac{1}{m_e}\sum_{\mathbf{G}}u_{i^\prime}^*\left(\mathbf{k}^\prime+\mathbf{G}+\Delta\mathbf{G}\right)\left(\mathbf{k}+\mathbf{G}\right)u_i\left(\mathbf{k}+\mathbf{G}\right)\right)^*\nonumber\\
=&\sum_{\Delta\mathbf{G}}\frac{(2\pi)^3\delta^3(\mathbf{k}+\mathbf{q}-\mathbf{k}^\prime-\Delta\mathbf{G})}{V}\left(f_{i,\mathbf{k}\rightarrow i^\prime, \mathbf{k}^\prime}^\prime\right)\left(\mathbf{f}_{i,\mathbf{k}\rightarrow i^\prime, \mathbf{k}^\prime}^\prime\right)^*\, ,
\label{eq:t2}
\end{align}
where 
\begin{equation}
    \mathbf{f}_{i,\mathbf{k}\rightarrow i^\prime, \mathbf{k}^\prime}^\prime\equiv -\frac{1}{m_e}\sum_{\mathbf{G}}u_{i^\prime}^*\left(\mathbf{k}^\prime+\mathbf{G}+\Delta\mathbf{G}\right)\left(\mathbf{k}+\mathbf{G}\right)u_i\left(\mathbf{k}+\mathbf{G}\right)\, .
\end{equation}
Finally,
\begin{align}
(\mathbf{f}_{i,\mathbf{k}\rightarrow i^\prime, \mathbf{k}^\prime}\cdot \mathbf{w})(\mathbf{f}_{i,\mathbf{k}\rightarrow i^\prime, \mathbf{k}^\prime}^*\cdot \mathbf{w}^\prime)=&\sum_{\Delta\mathbf{G}}\frac{(2\pi)^3\delta^3(\mathbf{k}+\mathbf{q}-\mathbf{k}^\prime-\Delta\mathbf{G})}{V} \left(\frac{-1}{m_e}\sum_{\mathbf{F}}u_{i^\prime}^*\left(\mathbf{k}^\prime+\mathbf{F}+\Delta\mathbf{G}\right)\left(\mathbf{k}+\mathbf{F}\right)u_i \left(\mathbf{k}+\mathbf{F}\right)\right) \cdot \mathbf{w}\nonumber\\
&\times\left(\frac{-1}{m_e}\sum_{\mathbf{G}}u_{i^\prime}^*\left(\mathbf{k}^\prime+\mathbf{G}+\Delta\mathbf{G}\right)\left(\mathbf{k}+\mathbf{G}\right)u_i\left(\mathbf{k}+\mathbf{G}\right)\right)^*\cdot\mathbf{w}^\prime\nonumber\\
=&\sum_{\Delta\mathbf{G}}\frac{(2\pi)^3\delta^3(\mathbf{k}+\mathbf{q}-\mathbf{k}^\prime-\Delta\mathbf{G})}{V}\left(\left(\mathbf{f}_{i,\mathbf{k}\rightarrow i^\prime, \mathbf{k}^\prime}^\prime\right)\cdot\mathbf{w}\right)\left(\left(\mathbf{f}_{i,\mathbf{k}\rightarrow i^\prime, \mathbf{k}^\prime}^\prime\right)^*\cdot\mathbf{w}^\prime\right),
\label{eq:t3}
\end{align}
Inserting Eqs.~(\ref{eq:t1}), (\ref{eq:t2}) and (\ref{eq:t3}) in Eq.~(\ref{eq:M2}), and setting $\mathbf{w}=(\nabla_{\mathbf{\ell}}\mathcal{M})_{\mathbf{\ell}=0}$ and $\mathbf{w}^\prime=(\nabla_{\mathbf{\ell}}\mathcal{M}^*)_{\mathbf{\ell}=0}$, we obtain our Eq.~(\ref{eq:M2final}),
\begin{align}
\overline{\left|\mathcal{M}_{i,\mathbf{k}\rightarrow i^\prime, \mathbf{k}^\prime}\right|^2} =& \overline{\left|\mathcal{M}\right|^2}\left|f_{i,\mathbf{k}\rightarrow i^\prime, \mathbf{k}^\prime}\right|^2+ 2 m_e \overline{\Re\left[\mathcal{M}f_{i,\mathbf{k}\rightarrow i^\prime, \mathbf{k}^\prime}(\nabla_{\mathbf{p}_1}\mathcal{M}^*)_{\mathbf{p}_1=0}\cdot\left(\mathbf{f}_{i,\mathbf{k}\rightarrow i^\prime, \mathbf{k}^\prime}\right)^*\right]} + m_e^2 \overline{\left|(\nabla_{\mathbf{p}_1}\mathcal{M})_{\mathbf{p}_1=0}\cdot \mathbf{f}_{i,\mathbf{k}\rightarrow i^\prime, \mathbf{k}^\prime}\right|^2} \nonumber\\
=&\sum_{\Delta\mathbf{G}}\frac{(2\pi)^3\delta^3(\mathbf{k}+\mathbf{q}-\mathbf{k}^\prime-\Delta\mathbf{G})}{V} \nonumber\\
&\times\left(\overline{\left|\mathcal{M}\right|^2}\left|f_{i,\mathbf{k}\rightarrow i^\prime, \mathbf{k}^\prime}^\prime\right|^2 + 2 m_e \overline{\mathfrak{R}\left[\mathcal{M}f_{i,\mathbf{k}\rightarrow i^\prime, \mathbf{k}^\prime}^\prime(\nabla_{\mathbf{p}_1}\mathcal{M}^*)_{\mathbf{p}_1=0}\cdot\left(\mathbf{f}_{i,\mathbf{k}\rightarrow i^\prime, \mathbf{k}^\prime}^\prime\right)^*\right]} \right.\nonumber\\
& \left. + m_e^2 \overline{\left|(\nabla_{\mathbf{p}_1}\mathcal{M})_{\mathbf{p}_1=0}\cdot \mathbf{f}_{i,\mathbf{k}\rightarrow i^\prime, \mathbf{k}^\prime}^\prime\right|^2} \right)\nonumber\\
\equiv
& \sum_{\Delta\mathbf{G}}\frac{(2\pi)^3\delta^3(\mathbf{k}+\mathbf{q}-\mathbf{k}^\prime-\Delta\mathbf{G})}{V} \times \overline{\left|\mathcal{M}_{i,\mathbf{k}\rightarrow i^\prime, \mathbf{k}^\prime}^\prime\right|^2}\,.
\label{Eq:Crystal_matrixelement}
\end{align}

\section{Dark matter and responses}\label{App:DM_responses}
The first, third and fourth dark matter responses are the same as in \cite{Catena:2019gfa}, and we will simply state them in this appendix.
\subsection{The first dark matter response}
The first dark matter response is produced by the first term in Eq.~\eqref{eq:M2_prime} and is
\begin{align}
    R_1(q,v)&=\overline{| \mathcal{M}(\mathbf{q},\mathbf{v}_{\rm el}^\perp)|^2} \nonumber\\
    &= c_1^2 +\frac{c_3^2}{4}\frac{q^2}{m_e^2} (\vPerpEl)^2 -\frac{c_3^2}{4} \left(\frac{\mathbf{q}}{m_e}\cdot \vPerpEl\right)^2 +\frac{c_7^2}{4}(\vPerpEl)^2 + \frac{c_{10}^2}{4}\frac{q^2}{m_e^2}   \nonumber\\
   &+ \frac{j_\chi(j_\chi+1)}{12}\Bigg\{ 3c_4^2 +c_6^2 \frac{q^4}{m_e^4} +(4c_8^2+2c_{12}^2)(\vPerpEl)^2 +(2c_9^2+4c_{11}^2+2c_4c_6)\frac{q^2}{m_e^2}\nonumber \\
   &+\left(4c_5^2+c_{13}^2+c_{14}^2-2c_{12}c_{15}\right)\frac{q^2}{m_e^2}(\vPerpEl)^2+c_{15}^2\frac{q^4}{m_e^4} \left(\vPerpEl\right)^2\nonumber\\
   &-c_{15}^2 \frac{q^2}{m_e^2} \left(\vPerpEl\cdot \frac{\mathbf{q}}{m_e}\right)^2+\left(-4c_5^2+2c_{13}c_{14}+2c_{12}c_{15}\right)\left(\vPerpEl\cdot\frac{\mathbf{q}}{m_e}\right)^2\Bigg\}\, .
\end{align}
\subsection{Expansion of $2m_e\overline{ \Re \left[ \mathcal{M} \nabla_{\mathbf{k}} \mathcal{M}^*\cdot \mathbf{A}  \right]}$}
The second term of Eq.~\eqref{eq:M2_prime} requires more treatment. In \cite{Catena:2019gfa} we expanded the second term and found
\begin{align}
    2m_e\overline{ \Re \left[ \mathcal{M} \nabla_{\mathbf{k}} \mathcal{M}^*\cdot \mathbf{A}  \right]}&= \left[\frac{|c_3|^2}{2}\left(\left(\frac{\mathbf{q}}{m_e}\cdot \vPerpEl\right) \frac{\mathbf{q}}{m_e}-\frac{q^2}{m_e^2}\vPerpEl\right) - \frac{|c_7|^2}{2}\vPerpEl\right]\cdot\Re(\mathbf{A})\nonumber\\
    &+\frac{1}{2}\left( \frac{\mathbf{q}}{m_e}\times\vPerpEl \right)\cdot\Im((c_3c_7^*+c_3^*c_7)\mathbf{A}) +\frac{1}{2}\frac{\mathbf{q}}{m_e}\cdot\Im(c_7^*c_{10}\mathbf{A}) \nonumber\\
 &+\frac{j_\chi(j_\chi+1)}{6}\Bigg\{ \bigg[\left(4|c_5|^2+|c_{15}|^2\frac{q^2}{m_e^2}\right)\left(\left(\frac{\mathbf{q}}{m_e}\cdot \vPerpEl\right) \frac{\mathbf{q}}{m_e}-\frac{q^2}{m_e^2}\vPerpEl\right)\nonumber\\
 &-\left(4|c_8|^2+2|c_{12}|^2+(|c_{13}|^2+|c_{14}|^2)\frac{q^2}{m_e^2}\right)\vPerpEl\bigg]\cdot\Re(\mathbf{A})-\frac{\mathbf{q}}{m_e}\cdot\Im(c_4c_{13}^*\mathbf{A})\nonumber\\
 &-\frac{\mathbf{q}}{m_e}\cdot\Im(c_4c_{14}^*\mathbf{A})+4 \left( \frac{\mathbf{q}}{m_e}\times\vPerpEl \right)\cdot\Im((c_5c_8^*+c_5^*c_8)\mathbf{A}) - \frac{q^2}{m_e^2} \frac{\mathbf{q}}{m_e}\cdot\Im(c_6c_{13}^*\mathbf{A})\nonumber\\
 &- \frac{q^2}{m_e^2} \frac{\mathbf{q}}{m_e}\cdot\Im(c_6c_{14}^*\mathbf{A}) + 2\frac{\mathbf{q}}{m_e}\cdot\Im(c_9c_{12}^*\mathbf{A})+ 4\frac{\mathbf{q}}{m_e}\cdot\Im(c_{11}c_{8}^*\mathbf{A})\nonumber\\
 &-\left( \frac{\mathbf{q}}{m_e}\times\vPerpEl \right)\cdot\Im((c_{12}c_{13}^*+c_{12}^*c_{13})\mathbf{A})+\left( \frac{\mathbf{q}}{m_e}\times\vPerpEl \right)\cdot\Im((c_{12}c_{14}^*+c_{12}^*c_{14})\mathbf{A})\nonumber\\
 &-\left(\left(\frac{\mathbf{q}}{m_e}\cdot \vPerpEl\right) \frac{\mathbf{q}}{m_e}-\frac{q^2}{m_e^2}\vPerpEl\right)\cdot\Re((c_{12}c_{15}^*+c_{12}^*c_{15})\mathbf{A})\nonumber\\
 &-\left(\frac{\mathbf{q}}{m_e}\cdot \vPerpEl\right)\frac{\mathbf{q}}{m_e}\cdot\Re((c_{13}c_{14}^*+c_{13}^*c_{14})\mathbf{A})\nonumber\\
 &-\frac{q^2}{m_e^2}\left( \frac{\mathbf{q}}{m_e}\times\vPerpEl \right)\cdot\Im((c_{14}c_{15}^*+c_{14}^*c_{15})\mathbf{A})\Bigg\}
\end{align}
where $\mathbf{A}\equiv f_{i,\mathbf{k}\rightarrow i^\prime, \mathbf{k}^\prime}^\prime \left(\mathbf{f}_{i,\mathbf{k}\rightarrow i^\prime, \mathbf{k}^\prime}^\prime\right)^*$ is a complex 3-vector. In the case of atoms we found that $\mathbf{A}$ is both real and antiparallel to $\mathbf{q}$. Neither of these simplifications apply to the crystal. In order to separate $R(\mathbf{q},\mathbf{v},c_i)$ from $W(\mathbf{q},\Delta E)$, we rewrite the above equation by expressing it in terms of $\Re(\mathbf{A})$ and $\Im(\mathbf{A})$, so we use that $\Im (c\mathbf{A})=\Im(c)\Re(\mathbf{A})+\Re(c)\Im(\mathbf{A})$ and $\Re(c\mathbf{A})=\Re(c)\Re(\mathbf{A})-\Im(c)\Im(\mathbf{A})$. We also rewrite the terms proportional to $\left(\frac{\mathbf{q}}{m_e}\times\vPerpEl\right)\cdot \mathbf{A}=-\vPerpEl \cdot \left( \frac{\mathbf{q}}{m_e}\times \mathbf{A}\right)$ yielding
\begin{align}
    2m_e\overline{ \Re \left[ \mathcal{M} \nabla_{\mathbf{k}} \mathcal{M}^*\cdot \mathbf{A}  \right]}&= \left[\frac{|c_3|^2}{2}\left(\left(\frac{\mathbf{q}}{m_e}\cdot \vPerpEl\right) \frac{\mathbf{q}}{m_e}-\frac{q^2}{m_e^2}\vPerpEl\right) - \frac{|c_7|^2}{2}\vPerpEl\right]\cdot\Re(\mathbf{A}) \nonumber\\
    &-\frac{1}{2} \vPerpEl \cdot\left(\frac{\mathbf{q}}{m_e}\times(\Im(c_3c_7^*+c_3^*c_7)\Re(\mathbf{A}) + \Re(c_3c_7^*+c_3^*c_7)\Im(\mathbf{A}))\right)\nonumber\\
 &+\frac{1}{2}\frac{\mathbf{q}}{m_e}\cdot(\Im(c_7^*c_{10})\Re(\mathbf{A})+\Re(c_7^*c_{10})\Im(\mathbf{A}))\nonumber \\
 &+\frac{j_\chi(j_\chi+1)}{6}\Bigg\{ \bigg[\left(4|c_5|^2+|c_{15}|^2\frac{q^2}{m_e^2}\right)\left(\left(\frac{\mathbf{q}}{m_e}\cdot \vPerpEl\right) \frac{\mathbf{q}}{m_e}-\frac{q^2}{m_e^2}\vPerpEl\right)\nonumber\\
 &-\left(4|c_8|^2+2|c_{12}|^2+(|c_{13}|^2+|c_{14}|^2)\frac{q^2}{m_e^2}\right)\vPerpEl\bigg]\cdot\Re(\mathbf{A})\nonumber\\
 &-\frac{\mathbf{q}}{m_e}\cdot(\Im(c_4c_{13}^*)\Re(\mathbf{A})+\Re(c_4c_{13}^*)\Im(\mathbf{A}))-\frac{\mathbf{q}}{m_e}\cdot(\Im(c_4c_{14}^*)\Re(\mathbf{A})+\Re(c_4c_{14}^*)\Im(\mathbf{A}))\nonumber\\
 &-4 \vPerpEl \cdot\left( \frac{\mathbf{q}}{m_e}\times(\Im(c_5c_8^*+c_5^*c_8)\Re(\mathbf{A})+\Re(c_5c_8^*+c_5^*c_8)\Im(\mathbf{A}))\right) \nonumber \\
 &- \frac{q^2}{m_e^2} \frac{\mathbf{q}}{m_e}\cdot(\Im(c_6c_{13}^*)\Re(\mathbf{A})+\Re(c_6c_{13}^*)\Im(\mathbf{A}))\nonumber\\
 &- \frac{q^2}{m_e^2} \frac{\mathbf{q}}{m_e}\cdot(\Im(c_6c_{14}^*)\Re(\mathbf{A})+\Re(c_6c_{14}^*)\Im(\mathbf{A})) \nonumber \\
 &+ 2\frac{\mathbf{q}}{m_e}\cdot(\Im(c_9c_{12}^*)\Re(\mathbf{A})+\Re(c_9c_{12}^*)\Im(\mathbf{A}))
 + 4\frac{\mathbf{q}}{m_e}\cdot(\Im(c_{11}c_{8}^*)\Re(\mathbf{A})+\Re(c_{11}c_{8}^*)\Im(\mathbf{A}))\nonumber\\
 &+\vPerpEl \cdot\left( \frac{\mathbf{q}}{m_e}\times(\Im(c_{12}c_{13}^*+c_{12}^*c_{13})\Re(\mathbf{A})+\Re(c_{12}c_{13}^*+c_{12}^*c_{13})\Im(\mathbf{A}))\right)\nonumber\\
 &-\vPerpEl \cdot\left( \frac{\mathbf{q}}{m_e}\times(\Im(c_{12}c_{14}^*+c_{12}^*c_{14})\Re(\mathbf{A})+\Re(c_{12}c_{14}^*+c_{12}^*c_{14})\Im(\mathbf{A}))\right)\nonumber\\
 &-\left(\left(\frac{\mathbf{q}}{m_e}\cdot \vPerpEl\right) \frac{\mathbf{q}}{m_e}-\frac{q^2}{m_e^2}\vPerpEl\right)\cdot(\Re(c_{12}c_{15}^*+c_{12}^*c_{15})\Re(\mathbf{A})-\Im(c_{12}c_{15}^*+c_{12}^*c_{15})\Im(\mathbf{A}))\nonumber\\
 &-\left(\frac{\mathbf{q}}{m_e}\cdot \vPerpEl\right)\frac{\mathbf{q}}{m_e}\cdot(\Re(c_{13}c_{14}^*+c_{13}^*c_{14})\Re(\mathbf{A})-\Im(c_{13}c_{14}^*+c_{13}^*c_{14})\Im(\mathbf{A}))\nonumber\\
 &+\frac{q^2}{m_e^2}\vPerpEl \cdot\left( \frac{\mathbf{q}}{m_e}\times(\Im(c_{14}c_{15}^*+c_{14}^*c_{15})\Re(\mathbf{A})+\Re(c_{14}c_{15}^*+c_{14}^*c_{15})\Im(\mathbf{A})\right)\Bigg\}.
\end{align}
For real couplings this reduces to
\begin{align}
    2m_e\overline{ \Re \left[ \mathcal{M} \nabla_{\mathbf{k}} \mathcal{M}^*\cdot \mathbf{A}  \right]}&=\frac{\mathbf{q}}{m_e}\cdot\Re(\mathbf{A})\left[ \frac{c_3^2}{2}\left(\frac{\mathbf{q}}{m_e}\cdot \vPerpEl\right) \right. \nonumber\\
    & \left. +\frac{j_\chi(j_\chi+1)}{6}\left(\frac{\mathbf{q}}{m_e}\cdot \vPerpEl\right)\Bigg( 4c_5^2-2\left(c_{12}c_{15}+c_{13}c_{14}\right)+c_{15}^2\frac{q^2}{m_e^2} \Bigg) \right]\nonumber\\
    &+ \frac{\mathbf{q}}{m_e}\cdot\Im(\mathbf{A})\left[ \frac{c_7c_{10}}{2} \right. \nonumber\\
    & \left.+\frac{j_\chi(j_\chi+1)}{6}\Bigg( 4 c_8 c_{11} + 2c_{9}c_{12} -c_4 c_{13} -c_4 c_{14} - \left( c_6 c_{13} + c_6 c_{14} \right) \frac{q^2}{m_e^2} \Bigg) \right] \nonumber \\
    &+ \vPerpEl \cdot \Re(\mathbf{A})\left[ -\frac{c_3^2}{2}\frac{q^2}{m_e^2} - \frac{c_7^2}{2} 
    \right. \nonumber\\
    &\left. +\frac{j_\chi(j_\chi+1)}{6}\Bigg\{ -c_8^2-c_{12}^2 + \left(-4c_5^2 + 2c_{12}c_{15} - c_{13}^2 - c_{14}^2\right)\frac{q^2}{m_e^2} - c_{15}^2 \frac{q^4}{m_e^4} \Bigg\}\right]\nonumber\\
    &+\vPerpEl\cdot\left( \frac{\mathbf{q}}{m_e} \times \Im(\mathbf{A}) \right)\left[ -c_3 c_7 +\frac{j_\chi(j_\chi+1)}{6}\Bigg\{ -8c_5c_8 +2c_{12}c_{13} -2 c_{12}c_{14} + 2 c_{14}c_{15}\frac{q^2}{m_e^2} \Bigg\} \right]\, . \label{eq:2nd_term_expanded}
\end{align}
From this we can define the second scalar overlap integral as
\begin{equation}
B_2=\frac{\mathbf{q}}{m_e}\cdot\mathbf{A}=\frac{\mathbf{q}}{m_e}\cdot f_{i,\mathbf{k}\rightarrow i^\prime, \mathbf{k}^\prime}^\prime \left(\mathbf{f}_{i,\mathbf{k}\rightarrow i^\prime, \mathbf{k}^\prime}^\prime\right)^*\, .
\end{equation} 
In addition to this we also find two complex vectorial overlap integrals, which we can call 
\begin{equation}
    \mathbf{B}_6=\mathbf{A}=f_{i,\mathbf{k}\rightarrow i^\prime, \mathbf{k}^\prime}^\prime \left(\mathbf{f}_{i,\mathbf{k}\rightarrow i^\prime, \mathbf{k}^\prime}^\prime\right)^*\, 
\end{equation}
and 
\begin{equation}
    \mathbf{B}_7=\frac{\mathbf{q}}{m_e}\times\mathbf{A}=\frac{\mathbf{q}}{m_e}\times f_{i,\mathbf{k}\rightarrow i^\prime, \mathbf{k}^\prime}^\prime \left(\mathbf{f}_{i,\mathbf{k}\rightarrow i^\prime, \mathbf{k}^\prime}^\prime\right)^*\,.
\end{equation}
\subsection{Velocity averaging of $\mathbf{B}_6$ and $\mathbf{B}_7$}
\label{App:Velocity_averaging}
For the scope of this paper we have treated the velocity distribution as isotropic, and we can use this to further simplify Eq.~\eqref{eq:2nd_term_expanded}. We start by decomposing $\mathbf{v}$ and $\mathbf{A}$ in a component parallel to $\mathbf{q}$ and a component perpendicular to $\mathbf{q}$,
\begin{equation}
    \mathbf{v}=\frac{(\mathbf{v}\cdot\mathbf{q})\mathbf{q}}{q^2}+\frac{\left|\mathbf{v}\times\mathbf{q}\right|\hat{\mathbf{n}}_v^\perp}{q}
\end{equation}
and
\begin{equation}
    \mathbf{A}=\frac{(\mathbf{A}\cdot\mathbf{q})\mathbf{q}}{q^2}+\frac{\left|\mathbf{A}\times\mathbf{q}\right|\hat{\mathbf{n}}_A^\perp}{q}\,,
\end{equation}
where $\hat{\mathbf{n}}_v^\perp$ and $\hat{\mathbf{n}}_A^\perp$ are unit vectors in the plane perpendicular to $\mathbf{q}$. Inserting this decomposition in the terms proportional to $\vPerpEl\cdot \mathbf{A}$ in Eq.~\eqref{eq:Mazi} we find something proportional to
\begin{align}
    \frac{1}{2\pi}\int_0^{2\pi} \mathrm{d}\phi \, \mathbf{\vPerpEl}\cdot\mathbf{A}\Bigg|_{\cos\theta=\xi}=&
     \frac{1}{2\pi}\int_0^{2\pi} \mathrm{d}\phi\, \mathbf{v}\cdot\mathbf{A} -\frac{m_e}{2\mu} \frac{\mathbf{q}}{m_e} \cdot \mathbf{A}\Bigg|_{\cos\theta=\xi}\nonumber \\
     =& \frac{1}{2\pi}\int_0^{2\pi} \mathrm{d}\phi \, \frac{v\cos\theta m_e}{q} \frac{\mathbf{q}}{m_e} \cdot \mathbf{A} + \frac{|\mathbf{v}\times\mathbf{q}||\mathbf{A}\times\mathbf{q}| }{q^2} \hat{\mathbf{n}}_v^\perp\cdot\hat{\mathbf{n}}_A^\perp -\frac{m_e}{2\mu} \frac{\mathbf{q}}{m_e} \cdot \mathbf{A}\Bigg|_{\cos\theta=\xi}
\end{align}
Without loss of generality we can now choose our coordinates such that $\hat{\mathbf{n}}_v^\perp\cdot\hat{\mathbf{n}}_A^\perp=\cos \phi$. We then have
\begin{align}
    \frac{1}{2\pi}\int_0^{2\pi} \mathrm{d}\phi \, \mathbf{\vPerpEl}\cdot\mathbf{A}\Bigg|_{\cos\theta=\xi}=&\frac{1}{2\pi}\int_0^{2\pi} \mathrm{d}\phi \, \frac{v\cos\theta m_e}{q} \frac{\mathbf{q}}{m_e} \cdot \mathbf{A} + \frac{|\mathbf{v}\times\mathbf{q}||\mathbf{A}\times\mathbf{q}| }{q^2} \cos\phi -\frac{m_e}{2\mu} \frac{\mathbf{q}}{m_e} \cdot \mathbf{A}\Bigg|_{\cos\theta=\xi}\nonumber \\
    =& \frac{\mathbf{q}}{m_e} \cdot \mathbf{A}\left[\frac{v\xi m_e}{q}  -\frac{m_e}{2\mu} \right]\nonumber \\
    =& \frac{\mathbf{q}}{m_e} \cdot \mathbf{A}\left[\frac{m_e}{2m_\chi} + \frac{m_e \Delta E}{q^2}  -\frac{m_e}{2\mu} \right]\nonumber \\
    =& \frac{\mathbf{q}}{m_e} \cdot \mathbf{A}\left[\frac{m_e}{2m_\chi} + \frac{m_e \Delta E}{q^2}  -\frac{m_e+m_\chi}{2m_\chi} \right] \nonumber \\
    =&\frac{\mathbf{q}}{m_e} \cdot \mathbf{A}\left[\frac{m_e \Delta E}{q^2}  -\frac{1}{2} \right]\nonumber \\
    =& \frac{\mathbf{q}}{m_e} \cdot \mathbf{A} \left(\frac{q}{m_e}\right)^{-2} \frac{\mathbf{q}}{m_e} \cdot \vPerpEl\, ,
\end{align}
which is the same as what was found in the case of atoms. For $\frac{\mathbf{q}}{m_e}\times\mathbf{A}$ we find
\begin{align}
    \frac{1}{2\pi}\int_0^{2\pi} \mathrm{d}\phi \, \mathbf{\vPerpEl}\cdot \left(\frac{\mathbf{q}}{m_e}\times\mathbf{A}\right) \Bigg|_{\cos\theta=\xi}=&
     \frac{1}{2\pi}\int_0^{2\pi} \mathrm{d}\phi\, \mathbf{v}\cdot\left(\frac{\mathbf{q}}{m_e}\times\mathbf{A}\right) -\frac{m_e}{2\mu} \frac{\mathbf{q}}{m_e} \cdot \left(\frac{\mathbf{q}}{m_e}\times\mathbf{A}\right)\Bigg|_{\cos\theta=\xi}\nonumber \\
     =& \frac{1}{2\pi}\int_0^{2\pi} \mathrm{d}\phi \, \frac{v\cos\theta m_e}{q} \frac{\mathbf{q}}{m_e} \cdot \left(\frac{\mathbf{q}}{m_e}\times\mathbf{A}\right) + \frac{|\mathbf{v}\times\mathbf{q}|}{q} \hat{\mathbf{n}}_v^\perp\cdot\left(\frac{\mathbf{q}}{m_e}\times\mathbf{A}\right) \Bigg|_{\cos\theta=\xi} \nonumber \\
     =&  \frac{1}{2\pi}\int_0^{2\pi} \mathrm{d}\phi \, \frac{|\mathbf{v}\times\mathbf{q}|}{q} \left|\frac{\mathbf{q}}{m_e}\times\mathbf{A}\right|\cos\phi \Bigg|_{\cos\theta=\xi} \nonumber \\
     =& 0
\end{align}
This result follows from our treatment of the velocity distribution as isotropic. Intuitively one can understand our treatment as not considering directionality. We expect directional effects such as a daily modulation to be small in silicon and germanium crystals justifying our treatment of the velocity distribution. In less isotropic materials such as 2D materials we expect these directional effects to be important and we will not in these cases be able to absorb $\mathbf{B}_6$ and $\mathbf{B}_7$ (fully) in the other responses. 
\subsection{The second to fifth dark matter response}
Having averaged over the azimuthal angle we arrive at the same dark matter response as we obtained in \cite{Catena:2019gfa} for $R_2$
\begin{align}
    &\Re(R_2)=  \left(\frac{\Delta E m_e}{q^2} - \frac{1}{2}\right)\left[-\frac{c_7^2}{2} 
 -\frac{j_\chi(j_\chi+1)}{6}\Bigg\{\left( 4c_8^2+2c_{12}^2\right)  +\left(
 c_{13}+c_{14}\right)^2\frac{q^2}{m_e^2}  \Bigg\}\right]\,,
\end{align}
and
\begin{align}
    &\Im(R_2)= \frac{1}{2}c_7c_{10}+\frac{j_\chi(j_\chi+1)}{6}\Bigg\{ -c_4c_{13}-c_4c_{14}+ 2c_9c_{12}
 + 4c_{11}c_{8} -\left( c_6c_{13}+  c_6c_{14} \right)\frac{q^2}{m_e^2}\Bigg\}\,.
\end{align}
The third and fourth DM responses were found in \cite{Catena:2019gfa} to be 
\begin{align}
   R_3(q,v) &= \frac{c_3^2}{4}\frac{q^2}{m_e^2}+\frac{c_7^2}{4}+\frac{j_\chi(j_\chi+1)}{12}\Bigg\{4c_8^2+2c_{12}^2+\left(4c_5^2+c_{13}^2+c_{14}^2-2c_{12}c_{15}\right)\frac{q^2}{m_e^2}+c_{15}^2\frac{q^4}{m_e^4}\Bigg\}\, ,\\
   R_4(q,v) &= -\frac{c_3^2}{4}+\frac{j_\chi(j_\chi+1)}{12}\Bigg\{-4c_5^2-c_{15}^2\frac{q^2}{m_e^2}+2c_{12}c_{15}+2c_{13}c_{14}\Bigg\}\, .
\end{align}
Finally, we find the fifth response to be
\begin{equation}
R_5(q,v)=\frac{j_\chi(j_\chi+1)}{6}\Bigg\{ 4c_3c_7+4c_5c_8 - c_{12}c_{13}+c_{12}c_{14}-4c_{14}c_{15}\frac{q^2}{m_e^2}\Bigg\}.
\end{equation}

\section{Comparison with the crystal form factor by Essig et al.}
\label{sec:ff}
In this appendix we compare our first crystal response function to the crystal form factor of Ref.~\cite{Essig:2015cda}. In our notation, Eq.~(3.17) in \cite{Essig:2015cda} reads
\begin{align}
    \left| f_\mathrm{crystal}(q,\Delta E) \right|^2 &= \frac{2\pi^2(\alpha m_e^2 V_\text{cell})^{-1}}{\Delta E}\sum_{i i^\prime}\int_\mathrm{BZ} \frac{V_\text{cell}\dd^3\;k}{(2\pi)^3}\frac{V_\text{cell}\dd^3\;k^\prime}{(2\pi)^3} 
    \times \Delta E \delta(\Delta E -E_{i\mathbf{k}}+E_{i'\mathbf{k}^\prime})\nonumber\\ &\times\sum_{\Delta\mathbf{G}}q
    \delta(\left|\mathbf{k}-\Delta\mathbf{G}-\mathbf{k}^\prime\right|-q) \left| f_{i,\mathbf{k}\rightarrow i^\prime, \mathbf{k}^\prime}^\prime \right|^2\,.
\end{align}
This can be compared to Eq.~(\ref{eq:W_scalar_2D_2}) to obtain 
\begin{equation}
    \left|f_{\text {crystal}}\left(q, \Delta E\right)\right|^{2}= \frac{q^3}{8\Delta E\alpha m_e^2}\overline{W}_1(q,\Delta E)\, . \label{eq:Comparison_Essig}
\end{equation}

\end{widetext}

\bibliography{ref,ref2,bibliography,Nicola}

\end{document}